\documentclass[a4paper,11pt]{article}
\usepackage{jheppub} % for details on the use of the package, please see the JINST-author-manual
\usepackage{lineno}
\usepackage{comment}
\usepackage{orcidlink}
\newcommand{\spth}{$SU(4)/Sp(4)$ }
\newcommand{\soth}{$SU(4)/SO(4)$ }

\arxivnumber{} % if you have one

\title{\boldmath QCD-like theories at next-to-next-to-leading order with $N_F=2$ non-degenerate fermions}

 \author[a]{Johan Bijnens\,\orcidlink{0000-0002-1618-2844}}
\author[b]{and Daniil Krichevskiy\,\orcidlink{0000-0002-5355-8839}}

\affiliation[a]{Division of Particle and Nuclear Physics, Department of Physics, Lund University,
Box 118, \\SE 221 00 Lund, Sweden}
\affiliation[b]{Department of Mathematics and Physics, 
University of Stavanger\\
Kristine Bonnevies vei 22, 4021 Stavanger, Norway
}

\emailAdd{johan.bijnens@fysik.lu.se}
\emailAdd{daniil.krichevskiy@uis.no}

\abstract{QCD-like theories with $N_F=2$ fermion flavours in real and pseudoreal representations are studied within Chiral Perturbation Theory. For the pseudoreal symmetry-breaking pattern $SU(4)/Sp(4)$, the reduced NLO Lagrangian is derived. The NLO and NNLO corrections to the pion masses, decay constants, and vacuum condensates are calculated for non-degenerate fermion masses, extending previous results obtained for degenerate masses and for the non-degenerate case at NLO. Using the available spectroscopic and scattering lattice data for the $Sp(N_c=4)$ gauge theory with two fermion flavours, fits of the NLO low-energy constants are performed at NNLO precision. It is found that higher-order corrections are important for reproducing lattice observables and have a significant impact on the phenomenology of strongly interacting pionic dark matter, particularly in the regime of big $M_\pi/F_\pi$.}
\keywords{Spontaneous Symmetry Breaking, Chiral Lagrangian, Models for Dark Matter}

\begin{document}
\maketitle
\flushbottom

\begin{comment}
\section{To do list}
\begin{itemize}
  \item some red notes are left
\item check all the formulas
\item formatting of long formulas
 \item Is footnote 4 correct?
 \item Is footnote 10 fine?
\end{itemize}
\end{comment}

\section{Introduction}\label{sec:intro}

The Standard Model (SM) of particle physics is in excellent agreement with experimental results across a wide range of energy scales and phenomena. However, several important issues remain unresolved within the SM, such as the dark matter problem \cite{Balazs:2024uyj} and the electroweak hierarchy problem \cite{Peskin:2025lsg}. One possible approach to addressing these problems is to introduce beyond-the-Standard-Model (BSM) physics in the form of a strongly coupled QCD-like dark sector. In the context of dark matter, possible solutions are provided, for example, by theories of dark baryons \cite{Cline:2021itd} and dark pions. In particular, significant attention has been devoted to the strongly interacting massive particle (SIMP) paradigm \cite{Hochberg:2014dra, Hochberg:2014kqa, Kamada:2022zwb} and its generalizations including vector states \cite{Bernreuther:2023kcg, Berlin:2018tvf, Choi:2018iit}. In the context of the hierarchy problem, strongly coupled theories such as technicolor and composite Goldstone Higgs models have been proposed. Technicolor models explain electroweak symmetry breaking through the condensation of new strongly interacting fermions \cite{ Peskin:1980gc, Preskill:1980mz, Dimopoulos:1979sp}, while composite Goldstone Higgs models interpret the Higgs boson as a pseudo-Nambu–Goldstone boson arising from a new strong sector; see, e.g., \cite{Hill:2002ap, Cacciapaglia:2020kgq} for reviews.

A common feature of QCD-like sectors is spontaneous chiral symmetry breaking through the formation of a chiral condensate in the infrared (IR) regime. This results in the appearance of (pseudo-)Goldstone bosons, which we will hereafter refer to as pions, in analogy with real-world QCD, whose low-energy interactions can be described within the framework of Chiral Perturbation Theory (ChPT). The pattern of symmetry breaking depends on the number of fermion flavours and on their representation under the gauge group. ChPT provides a universal effective description of different symmetry-breaking patterns, with the structure of the low-energy Lagrangian determined solely by the symmetry-breaking pattern~\cite{Callan:1969sn,Coleman:1969sm,Weinberg:1978kz,Gasser:1983yg,Gasser:1984gg,Scherer:2012xha}.

For $N_F$ fermions in complex, real, and pseudoreal representations, the corresponding symmetry-breaking patterns are $SU(N_F)_L\times SU(N_F)_R/SU(N_F)_V$, $SU(2N_F)/SO(2N_F)$, $SU(2N_F)/Sp(2N_F)$ respectively. 
The low-energy description of the $SU(4)/Sp(4)$ and $SU(4)/SO(4)$ symmetry breaking theories relevant for this paper goes back a long time, see e.g. \cite{Smilga:1994tb,Splittorff:2001fy}. A comprehensive leading-order (LO) discussion of these cases can be found in \cite{Kogut:2000ek}.\footnote{QCD-like theories with quarks in (pseudo)real representations do not suffer from the sign problem~\cite{Nagata:2021ugx}, making lattice simulations at finite chemical potential feasible, unlike in real-world QCD.}
QCD-like theories with fermions in pseudoreal or real representations have recently attracted considerable attention; see \cite{Kulkarni:2022bvh} and \cite{Pomper:2024otb} for detailed discussions of the pseudoreal and real cases for $N_F=2$, respectively.

The validity of the EFT expansion is controlled by the pseudo-Goldstone mass scale, implying that higher-order effects become increasingly relevant in BSM scenarios with comparatively heavy pseudo-Goldstone states. In this context, both next-to-leading (NLO) and next-to-next-to-leading (NNLO) corrections can have a sizable phenomenological impact, as demonstrated, for instance, in studies of SIMP dark matter~\cite{Hansen:2015yaa}. In addition, incorporating higher-order contributions is important for reliably relating the chiral regime of nearly massless fermions, relevant for composite Higgs constructions, to the finite-fermion-mass region probed in lattice simulations.

From both phenomenological and lattice perspectives, it is important to consider the case of non-degenerate fermion masses. In particular, mass splittings can substantially enlarge the viable parameter space of SIMP dark matter models~\cite{Hochberg:2014kqa}, while lattice simulations which employ non-degenerate fermion masses cannot be accurately described within the fully degenerate framework. The NLO analysis of the non-degenerate case with $N_F=2$ fermions in real and pseudoreal representations was performed in \cite{Kolesova:2025ghl}. 
The NNLO analysis in the degenerate case can be found in \cite{Bijnens:2009qm,Bijnens:2011fm}. In the present paper, we extend this analysis to NNLO and compute the pseudo-Goldstone masses, decay constants, and vacuum expectation values at this order. In addition, for the  \spth theory, we derive the reduced form of the general NLO $SU(2N_F)/Sp(2N_F)$ Lagrangian which contains a smaller number of independent terms.

The NLO and NNLO Lagrangians contain Wilson coefficients, which in the present context are referred to as low-energy constants (LECs). At least a subset of these LECs can be determined by fitting to lattice data. Using the expressions derived in this work, we perform fits to the $Sp(N_c=4)$ lattice results of \cite{Kulkarni:2022bvh} and \cite{Dengler:2024maq}. Our analysis extends and improves upon the NLO study of \cite{Kolesova:2025ghl}, where the available precision was insufficient to reproduce the lattice data for the split decay constants. Including NNLO corrections allows us to accurately capture the observed features of the lattice results.

The structure of this paper is as follows. In section~\ref{secEFT}, we discuss the construction of EFT. In section~\ref{SecReduction}, we present the reduction of the \spth\ Lagrangian. The calculation of the NNLO masses and decay constants is carried out in sections~\ref{secMasses} and \ref{secDecay}, respectively, while the NNLO condensates are discussed in section~\ref{secCondensates}. In section~\ref{sec:fit}, we perform a fit of the low-energy constants (LECs) for the $Sp(N_c=4)$, $N_F=2$ theory and discuss applications to strongly interacting pionic dark matter. The NNLO expressions for the \soth case are lengthy and are therefore collected in the appendices.

\section{Effective theory}\label{secEFT}
In this section, we review the QCD-like gauge theories and the corresponding low-energy effective theory employed in their description. Our presentation follows Refs.~\cite{Kolesova:2025ghl,Bijnens:2009qm}. To keep the discussion self-contained, we summarize here the relevant setup. 
\subsection{UV theory}\label{subsecUV}

Consider a gauge theory containing $N_F$ left-handed and $N_F$ right-handed fermion flavors, corresponding altogether to $2N_F$ Weyl fermionic degrees of freedom, transforming under a real or pseudoreal representation of a gauge group. Typical examples include fermions in the fundamental representation of the gauge groups $SO(N_c)$ and $Sp(N_c)$ correspondingly, where $N_c$ denotes the number of colors.

Including an external (pseudo)scalar source $\mathcal{M}$, the Lagrangian takes the form
\begin{align}\label{eq.EFT1}
\begin{split}
\mathcal{L} =& \overline{q}_{Li} i \gamma^\mu D_\mu q_{Li}
+ \overline{q}_{Ri} i \gamma^\mu D_\mu q_{Ri}
- \overline{q}_{Ri} \mathcal{M}_{ij} q_{Lj}
- \overline{q}_{Li} \mathcal{M}_{ij}^\dagger q_{Rj},
\end{split}
\end{align}
where implicit summation over flavor indices $i,j=1,\ldots,N_F$ is understood. For massive fermions, the expectation value of the source $\mathcal{M}$ may be identified with the fermion mass matrix. The covariant derivative is defined as
\begin{equation}
D_{\mu} q = \partial_{\mu} q - i G_{\mu}^{a} t^{a},
\end{equation}
with $G_\mu^a$ denoting the gauge fields and $t^a$ the Hermitian generators of the gauge group in the representation carried by the fermions. The fields $q_L$ and $q_R$ are vectors in flavor space.\footnote{Color and spinor indices are omitted throughout.}
For fermions transforming in a (pseudo)real representation, there exists a unitary matrix $\epsilon$ that is either symmetric or antisymmetric and satisfies
\begin{align}\label{eq.EFT2}
\begin{split}
-t^{a\ast}=\epsilon^{-1} t^a \epsilon,
\qquad
\epsilon^T = \eta \epsilon,
\end{split}
\end{align}
where $\eta = +1$ for a real representation and $\eta = -1$ for a pseudoreal representation.

Following the construction of~\cite{Kogut:2000ek,Bijnens:2009qm}, one may introduce the field
\begin{align}\label{eq.EFT3}
\begin{split}
\tilde{q}_{R} = \epsilon C \bar{q}_{L}^{\,T},
\end{split}
\end{align}
which transforms as a right-handed spinor under Lorentz transformations and belongs to the same gauge-group representation as $q_L$. Here, $C=i\gamma^2\gamma^0$ is a  matrix acting in Dirac space.

Making use of Eq.~\eqref{eq.EFT2}, integrating by parts, and exploiting the Grassmann nature of the fermion fields, the Lagrangian in Eq.~\eqref{eq.EFT1} can be recast into the form
\begin{align}\label{eq.EFT4}
\begin{split}
\mathcal{L}
=
\bar{\hat{q}}\, i\gamma^\mu D_\mu \hat{q}
-\frac{1}{2}\hat{q}^{T}\hat{\mathcal{M}}^\dagger \epsilon^\dagger C \hat{q}
-\frac{1}{2}\bar{\hat{q}}\, C\epsilon \hat{\mathcal{M}}\, \bar{\hat{q}}^{\hspace{0.1em}T},
\end{split}
\end{align}
where the multiplet $\hat{q}$ contains $2N_F$ flavor components, and the matrix $\hat{\mathcal{M}}$ acts in the enlarged flavor space:
\begin{align}\label{eq.EFT5}
\begin{split}
\hat{q}
=
\begin{pmatrix}
q_R \\
\tilde{q}_R
\end{pmatrix},
\qquad
\hat{\mathcal{M}}
=
\begin{pmatrix}
0 & \eta \mathcal{M} \\
\mathcal{M}^{T} & 0
\end{pmatrix}.
\end{split}
\end{align}
If $\mathcal{M}$ is identified with a fermion mass matrix, then in the limit $\hat{\mathcal{M}}=0$, the classical theory exhibits a global $U(2N_F)$ flavor symmetry under transformations $\hat{q}\to g\hat{q}$.\footnote{This symmetry is larger than in QCD with fermions in a complex representation, where the classical flavor symmetry is given by $U(N_F)_L\times U(N_F)_R$.} At the quantum level, however, the axial anomaly breaks the symmetry down to $SU(2N_F)$.\footnote{It was shown in \cite{Pomper:2024otb} that the non-anomalous symmetry in the real case is, in fact, $\mathbb{Z}_{2} \times SU\left( 2N_{F}\right)$, see discussion in the section \ref{subsecSplitMasses}. However, this does not affect the construction of the EFT.} 

To construct correlation functions involving the currents
\begin{equation}
j_\mu^a = \bar{\hat{q}} \gamma_\mu T^a \hat{q},
\end{equation}
with $T^a$ the Hermitian generators of $SU(2N_F)$, one introduces external vector fields coupled to these currents. The generators are normalized as $\left< T^{a}T^{b}\right>  =\delta^{ab} $, where $\left< \ldots \right> $ denotes a trace in the flavor space. The Lagrangian then becomes
\begin{align}\label{eq.EFT4.1}
\begin{split}
\mathcal{L}
=
\bar{\hat{q}}\, i\gamma^\mu D_\mu \hat{q}
-\frac{1}{2}\hat{q}^{T}\hat{\mathcal{M}}^\dagger \epsilon^\dagger C \hat{q}
-\frac{1}{2}\bar{\hat{q}}\, C\epsilon \hat{\mathcal{M}}\, \bar{\hat{q}}^{\hspace{0.1em}T}
+\bar{\hat{q}} \gamma^\mu V_\mu \hat{q},
\end{split}
\end{align}
where the external source is written as $V_\mu = V_\mu^a T^a$.
The Lagrangian remains invariant under $SU(2N_F)$ transformations provided that the external fields transform accordingly. In fact, the symmetry can be promoted to a local one if the following transformation rules are imposed:
\begin{align}\label{eq.EFT6}
\begin{split}
V_\mu \rightarrow g V_\mu g^\dagger + i g \partial_\mu g^\dagger,
\qquad
\hat{\mathcal{M}} \rightarrow g \hat{\mathcal{M}} g^T,
\qquad
g \in SU(2N_F).
\end{split}
\end{align}

\subsection{IR description}\label{subsecIR}

We assume that at low energies the theory develops a non-vanishing quark condensate,
\begin{align}\label{eq.EFT7}
\begin{split}
\left< \bar{q} q \right>
\equiv
\sum_{i=1}^{N_F}
\left<
\bar{q}_{Li} q_{Ri}
+
\bar{q}_{Ri} q_{Li}
\right>
=
\left<
\frac{1}{2}
\hat{q}^{T} J \epsilon^\ast C \hat{q}
+
\frac{1}{2}
\bar{\hat{q}}\, C \epsilon J \bar{\hat{q}}^{\hspace{0.1em}T}
\right>
\neq 0,
\end{split}
\end{align}
where
\begin{align}\label{eq.EFT8}
\begin{split}
J=
\begin{pmatrix}
0 & \eta \mathbf{1}_{N_F} \\
\mathbf{1}_{N_F} & 0
\end{pmatrix},
\end{split}
\end{align}
and $\mathbf{1}_{N_F}$ denotes the $N_F\times N_F$ identity matrix. The condensate for individual flavor components can then be expressed as
\begin{align}\label{eq.EFT9}
\begin{split}
\left<
\hat{q}^{T}_{i}\epsilon^\ast C \hat{q}_{j}
\right>
=
\frac{\left< \bar{q}_{L} q_{R} \right>}{N_F}
J_{ij}.
\end{split}
\end{align}
The condensate remains invariant only under a subgroup $H\subset G=SU(2N_F)$ satisfying
\begin{align}\label{eq.EFT10}
\begin{split}
J = h^{T} J h,
\qquad
h\in H\subset G=SU(2N_F),
\qquad
\begin{cases}
H=SO(2N_F), & \eta =1,\\
H=Sp(2N_F), & \eta =-1.
\end{cases}
\end{split}
\end{align}
For fermions in a real representation, the Goldstone manifold is  given by $SU(2N_F)/SO(2N_F)$ and contains $N_F(2N_F+1)-1$ Goldstone bosons. In the pseudoreal case, the corresponding coset is $SU(2N_F)/Sp(2N_F)$ with $N_F(2N_F-1)-1$ Goldstone modes. In the  case discussed in this paper, $N_F=2$, this yields $N_\pi=9$ pions for real representations and $N_\pi=5$ for pseudoreal ones.

Using Eq.~\eqref{eq.EFT9}, the generators of the group $G$ can be decomposed into broken generators $X^a$ and unbroken generators $Q^a$ (see Appendix~\ref{appendixSU4gen} for their explicit form in the $N_F=2$ case):
\begin{align}\label{eq.EFT10b}
\begin{split}
JQ^a + Q^{aT}J = 0,
\qquad
JX^a - X^{aT}J = 0.
\end{split}
\end{align}
The global symmetry is explicitly broken in the presence of fermion masses. For mass-degenerate fermions with mass $m$, corresponding to $\mathcal{M}=m\mathbf{1}_{N_F}$ and $\hat{\mathcal{M}}=mJ$, the generators are separated into the same broken and unbroken subsets.

Within the CCWZ construction~\cite{Coleman:1969sm,Callan:1969sn}, the coset space is parametrized by the matrix
\begin{align}\label{eq.EFT12}
\begin{split}
u
=
e^{\frac{i}{\sqrt{2}F}\pi^a X^a}
\in G/H,
\end{split}
\end{align}
where $F$ denotes the leading-order pion decay constant and $\pi^a$ are the pseudoscalar Goldstone fields, commonly referred to as dark pions.

Under a global transformation $g\in G$, the field $u$ transforms nonlinearly according to
\begin{equation}
u \rightarrow g u h^\dagger,
\end{equation}
where $h\in H$ generally depends nonlinearly on both $u$ and $g$. Using the transformation law in Eq.~\eqref{eq.EFT6}, one can construct the object
\begin{align}\label{eq.EFT13}
\begin{split}
u^\dagger (\partial_\mu - iV_\mu)u
\equiv
\Gamma_\mu - \frac{i}{2}u_\mu,
\qquad
\Gamma_\mu = \Gamma_\mu^a Q^a,
\qquad
u_\mu = u_\mu^a X^a,
\end{split}
\end{align}
which belongs to the Lie algebra of $G$ and can therefore be decomposed into its broken and unbroken components. The quantity $u_\mu$ serves as one of the basic building blocks of the chiral effective theory.

More generally, an arbitrary matrix $F$ may be decomposed into components transforming as unbroken and broken generators, respectively:
\begin{align}
\bar{F}
=
\frac{1}{2}
\left(
F - JF^T J^T
\right),
\qquad
\tilde{F}
=
\frac{1}{2}
\left(
F + JF^T J^T
\right),
\end{align}
which satisfy
\begin{equation}
\bar{F}J = -J\bar{F}^T,
\qquad
\tilde{F}J = J\tilde{F}^T.
\end{equation}
The field $u_\mu$ can then be written explicitly as
\begin{align}\label{eq.EFT14}
\begin{split}
u_\mu
=
i\left[
u^\dagger(\partial_\mu - iV_\mu)u
-
u(\partial_\mu + iJV_\mu^T J^T)u^\dagger
\right].
\end{split}
\end{align}
Another  ingredient of the  theory is provided by the spurion fields $\chi_\pm$,
\begin{align}
\chi_\pm
=
u^\dagger \chi J^T u^\dagger
\pm
uJ\chi^\dagger u,
\label{eqEFT16}
\end{align}
where $\chi = 2B_0\hat{\mathcal{M}}$. The low-energy constant $B_0$ has dimensions of mass and is  related to the quark condensate.

The quantities defined in Eqs.~\eqref{eq.EFT14}--\eqref{eqEFT16} transform homogeneously under $G$,
\begin{equation}
u_\mu \rightarrow h u_\mu h^\dagger,
\qquad
\chi_\pm \rightarrow h \chi_\pm h^\dagger.
\end{equation}
If the vector fields $V_\mu^a$ are promoted to dynamical degrees of freedom, one must also include operators constructed from the field-strength tensor
\begin{align}
V_{\mu\nu}
\equiv V_{\mu\nu}^aT^a =
\partial_\mu V_\nu
-
\partial_\nu V_\mu
-
i[V_\mu,V_\nu],
\end{align}
which transforms under $G$ as
\begin{equation}
V_{\mu\nu}\rightarrow gV_{\mu\nu}g^\dagger.
\end{equation}
To preserve a close analogy with the complex-representation case of QCD in the Standard Model (see~\cite{Bijnens:2009qm} for the details), one formally introduces the fields
\begin{equation}
l_\mu = -JV_\mu^T J^T,
\qquad
r_\mu = V_\mu,
\end{equation}
with corresponding field-strength tensors  $l_{\mu\nu}=-JV_{\mu\nu}^T J^T$ and $r_{\mu\nu}=V_{\mu\nu}$.\footnote{In this notation, the combinations $v_\mu=r_\mu+l_\mu$ and $a_\mu=r_\mu-l_\mu$ correspond to currents associated with unbroken and broken generators, respectively.}
One may then define
\begin{align}
f_{\pm\mu\nu}
=
ul_{\mu\nu}u^\dagger
\pm
u^\dagger r_{\mu\nu}u,
\end{align}
which transform homogeneously under $G$,\footnote{Our definition of $f_{\pm\mu\nu}$ differs slightly from that in~\cite{Bijnens:2009qm} (see their equations (46) and (51)); the transpose acting on $u$ is omitted there due to a typographical error.}
\begin{equation}
f_{\pm\mu\nu}\rightarrow hf_{\pm\mu\nu}h^\dagger.
\end{equation}
In addition we have
\begin{align}
  \Gamma_\mu\to h\Gamma_\mu h^\dagger +h\partial_\mu h^\dagger 
\end{align}
allowing to construct a covariant derivative.

Note that the Lagrangian \eqref{eq.EFT1} is symmetric under a transformation which interchanges $q_L$ and $q_R$ which we call intrinsic parity.
On the level of the Lagrangian \eqref{eq.EFT4.1} it is realized as 
\begin{equation}
    \hat{q} \rightarrow \epsilon CJ\bar{\hat{q} }^{T},
\end{equation}
provided that the external scalar and vector sources transform as follows:
\begin{align}
\hat{\mathcal{M} } \rightarrow \eta J\hat{\mathcal{M} }^{\dagger } J^{T},\  \quad V_{\mu }\rightarrow -JV^{T}_{\mu }J^{T}.
\end{align}
Assuming $u\to u^\dagger$, $\chi_+$ and $f_{+\mu\nu}$ are even, while $\chi_-$ and $f_{-\mu\nu}$ are odd under this transformation.

\subsection{Symmetry in the case of non-degenerate fermion masses}\label{subsecSplitMasses}
If one treats the source $\hat{\mathcal{M}}$ as the fermion mass matrix for $N_F=2$, then in the non-degenerate case it takes the form
\begin{align}\label{eqQuarkMassMatrix_4x4}
\mathcal{\hat{M} } =\left( \begin{matrix}0&0&\eta m_{u}&0\\ 0&0&0&\eta m_{d}\\ m_{u}&0&0&0\\ 0&m_{d}&0&0\end{matrix} \right), 
\end{align}
where $m_u$ and $m_d$ denote the masses of the two fermions. Non-zero quark masses explicitly break the chiral symmetry. If the two masses were equal, the pattern of explicit symmetry breaking would coincide with the pattern of spontaneous symmetry breaking induced by the non-zero condensate. In the non-degenerate case, however, the symmetry is broken further. The details can be found in \cite{Kulkarni:2022bvh,Cacciapaglia:2020kgq,Pomper:2024otb}.

In the pseudoreal case, the flavour symmetry $SU(4)$ is broken down to $SU(2)\times SU(2)$. The five pions decompose into a singlet and a four-plet transforming as $(\frac{1}{2},\frac{1}{2})$ under the unbroken symmetry. In the real case, it was shown in \cite{Pomper:2024otb} that the non-anomalous flavour symmetry is in fact $\mathbb{Z}_{2} \ltimes SU(4)$, due to the existence of a non-anomalous transformation that acts as charge conjugation on a single Dirac fermion. This symmetry is broken down to $O(2)\times O(2)$. The nine pions decompose into a four-plet, two doublets, and a singlet. The mass and decay-constant splittings calculated below reflect the multiplet structure described above.

For further convenience, we define the following combination which will appear in the NLO and NNLO expressions:
\begin{align}\label{eqquarkmassratioR}
    R_q=\frac{m_{u}-m_{d}}{m_{d}+m_{u}}.
\end{align}

\subsection{Power counting}

At energies well below the cutoff scale $4\pi F_\pi$, where $F_\pi$ is the pion decay constant, the dynamics of the Goldstone bosons can be described within a systematic chiral expansion. The chiral Lagrangian is organized as
\begin{align}\label{eq.EFT17}
\begin{split}
\mathcal{L}_{\text{ChPT}}
=
\sum_{k=1}^{\infty}
\mathcal{L}_{\text{N}^{(k-1)}\text{LO}}[\pi,\chi,V^\mu],
\end{split}
\end{align}
where each term contains only operators with the corresponding chiral dimension according to the power counting. The chiral expansion is organized in powers of soft momenta. Thus, a derivative (or external momentum) counts as
\begin{equation}
\mathrm{deg} \ \partial_\mu = 1 .
\end{equation}
Explicit symmetry breaking by fermion masses generates nonzero pion masses. Assigning chiral dimension $\mathrm{deg}  \  m_q=2$ implies $\mathrm{deg}\ M_\pi=1$, consistent with $M_\pi\propto\sqrt{m_q}$. Similarly, since the vector fields enter through the covariant derivative in Eq.~\eqref{eq.EFT14}, one assigns $\mathrm{deg} \ V_\mu=1$. Consequently,
\begin{equation}
\mathrm{deg} \ \chi_\pm = 2,
\qquad
\mathrm{deg}\ u_\mu = 1.
\end{equation}
For a fixed $k$, Lagrangian $\mathcal{L}_{\text{N}^{(k-1)}\text{LO}}$ contains terms of the degree $2k$ and thus can be called an $\mathcal{O}(p^{2k})$ Lagrangian. 

The expansion of physical observables is organized analogously:
\begin{equation}\label{eqChExp}
\mathcal{O}_{\text{phys}}
=
\mathcal{O}_{\text{LO}}
+
\mathcal{O}_{\text{NLO}}
+
\mathcal{O}_{\text{NNLO}}
+\dots,
\end{equation}
where the expansion parameter is
\begin{equation}
\frac{p^2}{(4\pi F_\pi)^2}
\sim
\frac{M_\pi^2}{(4\pi F_\pi)^2}
\lesssim 1.
\end{equation}
Throughout this work, $M_\pi \equiv M_{\text{phys}}$ and $F_\pi \equiv F_{\text{phys}}$ denote the physical pion mass and decay constant, respectively.
A given Feynman diagram $\Gamma$ scales as $\mathcal{O}(p^{\mathrm{deg}\,\Gamma})$, with chiral dimension
\begin{align}\label{eq.EFT18}
\begin{split}
\mathrm{deg}\,\Gamma
=
2 + (n-2)N_L
+
\sum_{i=1}^{\infty}
2(i-1)N_{2i},
\end{split}
\end{align}
where $n$ is the spacetime dimension, $N_L$ the number of loops, and $N_{2i}$ the number of vertices from $\mathcal{L}_{\text{N}^{(i-1)}\text{LO}}$. Diagrams with dimension $2k$ contribute to $\mathcal{O}_{\text{N}^{(k-1)}\text{LO}}$.

\subsubsection{Lagrangians}

At each order in the chiral expansion, the Lagrangian contains all operators consistent with chiral symmetry, Lorentz invariance and intrinsic parity. According to the power counting, the LO Lagrangian is given by
\begin{align}\label{eq.EFT15}
\mathcal{L}_\text{LO} =\frac{F^2}{4} \left< u_{\mu }u^{\mu }+\chi_{{}+} \right>\,.
\end{align}
The NLO Lagrangian takes the form
\begin{align}\label{eq.EFT16}
  \mathcal{L}_\text{NLO} &=L_{0}\left< u^{\mu }u^{\nu }u_{\mu }u_{\nu }\right>  +L_{1}\left< u^{\mu }u_{\mu }\right>  \left< u^{\nu }u_{\nu }\right>  +L_{2}\left< u^{\mu }u^{\nu }\right>  \left< u_{\mu }u_{\nu }\right>  +L_{3}\left< u^{\mu }u_{\mu }u^{\nu }u_{\nu }\right>  \nonumber\\ &+L_{4}\left< u^{\mu }u_{\mu }\right>  \left< \chi_{+} \right>  +L_{5}\left< u^{\mu }u_{\mu }\chi_{+} \right>  +L_{6}\left< \chi_{+} \right>^{2}  +L_{7}\left< \chi_{-} \right>^{2}  +\frac{1}{2} L_{8}\left< \chi^{2}_{+} +\chi^{2}_{-} \right> \nonumber\\ &-iL_{9}\left< f_{+\mu \nu }u^{\mu }u^{\nu }\right>  +\frac{1}{4} L_{10}\left< f^{2}_{+}-f^{2}_{-}\right>  +H_{1}\left< l_{\mu \nu }l^{\mu \nu }+r_{\mu \nu }r^{\mu \nu }\right> + H_{2}\left< \chi \chi^{\dagger } \right>,
\end{align}
The Lagrangians in Eqs.~\eqref{eq.EFT15} and \eqref{eq.EFT16} apply to theories with real, pseudoreal, and complex fermion representations. In the complex case, this requires embedding the $N_F\times N_F$ flavor matrices into the corresponding $2N_F\times2N_F$ matrix structure.
The NNLO  Lagrangian has been discussed in detail for the complex case in Ref.~\cite{Bijnens:1999sh} and contains 115 operators with corresponding NNLO LECs denoted by $K_i$. The structure of the $\mathcal{L}_\text{NNLO}$ in the (pseudo)real case is analogous, although some operators may become redundant. We do not present the full expression here and instead refer the reader to the literature mentioned above.

To absorb the divergences originating from diagrams with $\mathrm{deg}\,\Gamma=4$, the renormalized LECs are defined in the $\text{ChPT-}\overline{\text{MS}}$ scheme as 
\begin{equation}\label{eq.EFT17a}
    \begin{split}
L_{i}=\left( \mu c\right)^{-2\varepsilon }  \left( -\frac{\Gamma_{i} }{32\pi^{2} \varepsilon } +L^{r}_{i}\left( \mu \right)  \right)  ,
\end{split}
\end{equation}
where $2\varepsilon=4-n$ and
\begin{equation}\label{eq.EFT18a}
\begin{split}
\log c=-\frac{1}{2} \left( \log 4\pi -\gamma +1\right)  ,
\end{split}
\end{equation}
while $\gamma_E=-\Gamma'(1)$ is the Euler--Mascheroni constant. The coefficients $\Gamma_i$ for complex, real, and pseudoreal fermion representations are given in Ref.~\cite{Bijnens:2009qm} and are summarized in Table~\ref{tab:LECs_table}.
Diagrams with $\mathrm{deg}\,\Gamma=6$ generate both local divergences of the form
\begin{equation}
\frac{1}{n-4},
\qquad
\frac{1}{(n-4)^2},
\end{equation}
and nonlocal divergences proportional to
\begin{equation}
\frac{1}{n-4}\log\left(\frac{M^2}{\mu^2}\right).
\end{equation}
The nonlocal divergences cannot be absorbed into the NNLO LECs, since the LECs parameterize only local counterterms. Therefore, these divergences must cancel in the sum of all diagrams~\cite{Bijnens:1999hw}. We verified this cancellation explicitly in all calculations presented in this work.
The remaining local divergences are absorbed through the renormalization of the NNLO LECs~\cite{Amoros:1999dp}:
\begin{align}
K_{i}=\left( \mu c\right)^{-4\varepsilon }  \left( \frac{\gamma_{2,i} }{\varepsilon^{2} } +\frac{\gamma_{1,i} }{\varepsilon } +K^{r}_{i}\left( \mu \right)  \right),  
\end{align}
where $\gamma_{1,i}$ and $\gamma_{2,i}$ are NNLO renormalization coefficients.\footnote{We do not provide explicit values for $\gamma_{1,i}$ and $\gamma_{2,i}$, since in practice we renormalize only particular linear combinations of the NNLO LECs due to the large operator degeneracy at this order.}

In the case of the \spth theory, all pions have the same LO mass $M$, even when the fermions have different masses. This allows all loop integrals to be expressed in terms of a single function
\begin{align}
    \overline{A} \left( M^{2}\right)  =-\pi_{16} M^{2}\log \frac{M^{2}}{\mu^{2}},\quad \pi_{16}=\frac{1}{16\pi^2}.
\end{align}
In contrast, in the case of the \soth theory, the LO masses are different, and the reduction of all diagrams to $\overline{A} \left( M^{2}\right)$ is not possible. In this case, the functions $H^F(m_1^2,m_2^2,m_3^2,p^2)$, $H^F_1(m_1^2,m_2^2,m_3^2,p^2)$, $H^F_{21}(m_1^2,m_2^2,m_3^2,p^2)$, and their derivatives with respect to $p^2$ appear from the sunset diagrams (diagram (h) in figure \ref{fig:SelfEnergy}). In general, these functions cannot be expressed in terms of elementary functions, although they are implemented numerically in the CHIRON package \cite{Bijnens:2014gsa}. A detailed discussion of the relevant loop integrals can be found in \cite{Amoros:1999dp}.

\renewcommand{\arraystretch}{1.5}
\begin{table}[h]
\centering
\begin{tabular}{c c c c c c c c c c c c c c}
\hline
Theory & $L_0$ & $L_1$ & $L_2$ & $L_3$ & $L_4$ & $L_5$ & $L_6$ & $L_7$ & $L_8$ & $L_9$ & $L_{10}$& $H_1$ & $H_2$ \\
\hline
$SU(4)/Sp(4)$ & $-\frac{1}{24}$ & $\frac{1}{32}$ & $\frac{1}{16}$ & $\frac{1}{6}$ & $\frac{1}{16}$ & $\frac{1}{4}$ & $\frac{5}{128}$ & $0$ & $0$&$\frac{1}{2}$ & $-\frac{1}{2}$& $-\frac{3}{4}$ & $0$ \\
$SU(4)/SO(4)$ & $\frac{1}{8}$ & $\frac{1}{32}$ & $\frac{1}{16}$ & $0$ & $\frac{1}{16}$ & $\frac{1}{4}$ & $\frac{5}{128}$ & $0$ & $\frac{1}{8}$ &  $\frac{3}{2}$&  $-\frac{3}{2}$& $-\frac{3}{4}$ & $\frac{1}{4}$ \\
\hline
\end{tabular}
\caption{Coefficients $\Gamma_i$ for the NLO LECs of $SU(4)/Sp(4)$ and $SU(4)/SO(4)$ theories reproduced from \cite{Bijnens:2009qm}.}
\label{tab:LECs_table}
\end{table}
\renewcommand{\arraystretch}{1.0}

\section{Reduction of the \spth Lagrangian}\label{SecReduction}
It turns out that, in the case of \spth theory, only certain linear combinations of the LECs $L_i^r$ appear in the final formulas, as already noted in \cite{Kolesova:2025ghl}. This is a consequence of the redundancy of the Lagrangian \eqref{eq.EFT16}. In this section, we present a reduced set of independent terms.

Firstly, let us note that in this special case\footnote{The relation may be checked explicitly using the generators listed in Appendix~\ref{appendixSU4gen}.}
\begin{equation}
    \left< X^{a}X^{b}X^{c}X^{d}\right>  =\frac{1}{4} \left( \delta^{ab} \delta^{cd} -\delta^{ac} \delta^{bd} +\delta^{ad} \delta^{bc} \right).
\end{equation}
Then given that $u_\mu$ belongs to the broken part of $SU(4)$, i.e. $u_\mu=u_\mu^aX^a$, one can write
\begin{align}
    \left< u^{\mu }u^{\nu }u_{\mu }u_{\nu }\right>  &=\frac{1}{2} \left< u^{\mu }u^{\nu }\right>  \left< u_{\mu }u_{\nu }\right>  -\frac{1}{4} \left< u^{\mu }u_{\mu }\right>  \left< u^{\nu }u_{\nu }\right>, \\
    \left< u^{\mu }u_{\mu }u^{\nu }u_{\nu }\right> & =\frac{1}{4} \left< u^{\mu }u_{\mu }\right>  \left< u^{\nu }u_{\nu }\right> .
\end{align}
Secondly, we note that for \spth the following relation holds:
\begin{align}
\label{eq:XaXbSP4}
    \left\{ X^{a},X^{b}\right\}  =\frac{1}{2} \left< X^{a}X^{b}\right>  \mathbf{1}_{4} =\frac{\delta^{ab} }{2} \mathbf{1}_{4}. 
\end{align}
Consequently, the totally symmetric group invariant $d^{abc}$ is zero when it contains at least two unbroken generators:
\begin{align}
    d^{a bc}=\frac{1}{2} \left< \left\{ X^{a},X^{b}\right\}  T^{c}\right>  =0.
\end{align}
This means that
\begin{align}
    \left< u^{\mu }u_{\mu }\chi_{+} \right>  =u^{a}_{\mu }u^{\mu ,b}\left< X^{a}X^{b}\chi_{+} \right>  =\frac{\delta^{ab} }{4} u^{a}_{\mu }u^{\mu ,b}\left< \chi_{+} \right>  =\frac{1}{4} \left< u^{\mu }u_{\mu }\right>\left<\chi_{+} \right> , 
\end{align}
where the following relation was used. 
\begin{align}
    T^{a}T^{b}=\frac{1}{4} \delta^{ab} \mathbf{1}_{4} +d^{abc}T^{c}+\frac{i}{2} f^{abc}T^{c},
\end{align}
The antisymmetric part cancels because of contraction with symmetric $u^{a}_{\mu }u^{\mu ,b}$. 

Finally, let us introduce an auxiliary parameter $\alpha$ such that $u(\alpha)=e^{\frac{i\alpha}{\sqrt{2} F} \pi^{a} X^{a}}$. We define a function
\begin{align}\label{eqReduction2}
   f(\alpha )=\left< \chi^{2}_{+} +\chi^{2}_{-} \right>  -\frac{1}{2} \left< \chi_{+} \right>^{2}  -\frac{1}{2} \left< \chi_{-} \right>^{2},
\end{align}
where $\chi_\pm$ depend on $\alpha$ via $u(\alpha)$. $f(0)$ does not depend on the pion fields as it depends only on
\begin{align}
    \chi_{\pm } (0)=\chi J^{T}\pm J\chi^{\dagger }, 
\end{align}
In case of  \eqref{eqQuarkMassMatrix_4x4}, its value is $f(0)=-64B_0 m_u m_d$.  
The derivative of the function turns out to be identically zero:
\begin{align}\label{eqReduction1}
    \frac{df\left( \alpha \right)  }{d\alpha } =\frac{\sqrt[]{2} i}{F} \pi^{a} \Big( \left< X^{a}\chi_{-} \right>  \left< \chi_{+} \right>  +\left< X^{a}\chi_{+} \right>  \left< \chi_{-} \right>  -2\left< X^{a}\left\{ \chi_{+} ,\chi_{-} \right\}  \right>  \Big)=0.
\end{align}
This equality is proven by a direct calculation which relies on the  fact that the matrix $ \tilde{\chi } =u^{T}\chi^{\dagger } u$ is antisymmetric provided that the matrix $\chi$ is antisymmetric, as well as $\left< X^{a}\right> =0$. Equation \eqref{eqReduction1} means that $f(\alpha)$ is constant and the value for the physically relevant value of $\alpha=1$ is therefore known:
\begin{align}
    f\left( 1\right) =f(0).
\end{align}
We deduce that the terms in the \eqref{eqReduction2} are not independent and, moreover, this expression does not depend on the pion fields $\pi^a$. Thus, the term $\left< \chi^{2}_{+} +\chi^{2}_{-} \right>  $ can be replaced in the Lagrangian \eqref{eq.EFT16} by a linear combination of $\left< \chi_{+} \right>^{2}  $, $\left< \chi_{-} \right>^{2}  $ and the contact term $f(0)$. The latter can be shown to be invariant under infinitesimal transformation $g \in G=SU(4)$, using only the transformation properties of $\chi$ together with the requirement that $\chi$ is antisymmetric.

Taking into account the above we can rewrite the Lagrangian \eqref{eq.EFT16}:
\begin{align}\label{eqReduction3}
     \mathcal{L}_\text{NLO} &= l_{1}\left< u^{\mu }u_{\mu }\right>  \left< u^{\nu }u_{\nu }\right>  +l_{2}\left< u^{\mu }u^{\nu }\right>  \left< u_{\mu }u_{\nu }\right>  +l_{3}\left< u^{\mu }u_{\mu }\right>  \left< \chi_{+} \right>   +l_{4}\left< \chi_{+} \right>^{2}  +l_{5}\left< \chi_{-} \right>^{2} \nonumber  \\ &-il_{6}\left< f_{+\mu \nu }u^{\mu }u^{\nu }\right>  +\frac{1}{4} l_{7}\left< f^{2}_{+}-f^{2}_{-}\right>  +h_{1}\left< l_{\mu \nu }l^{\mu \nu }+r_{\mu \nu }r^{\mu \nu }\right> + h_{2}\left< \chi \chi^{\dagger } \right> \nonumber \\&
     +h_3\left( 2\left< \chi J^{T}\chi J^{T}\right>  +2\left< \chi^{\dagger } J\chi^{\dagger } J\right>  -\frac{1}{2} \left< \chi J^{T}+J\chi^{\dagger } \right>^{2}  -\frac{1}{2} \left< \chi J^{T}-J\chi^{\dagger } \right>^{2}  \right) ,
\end{align}

where the  new LECs were defined:
\begin{align}
     l_{1}&=-\frac{1}{4} L_{0}+L_{1}+\frac{1}{4} L_{3},\\
      l_{2}&=\frac{1}{2} L_{0}+L_{2},\\
      l_{3}&=L_{4}+\frac{1}{4} L_{5},\\
      l_{4}&=L_{6}+\frac{1}{4} L_{8},\\
    l_{5}&=L_{7}+\frac{1}{4} L_{8},\\
    l_6&=L_9,\\
    l_7&=L_{10},\\
    h_1&=H_1,\\
    h_2&=H_2,\\
    h_3&=\frac{L_8}{2}.
\end{align}
The new LECs should be renormalized with the new renormalization coefficients $\gamma_i$ which are provided in the table \ref{tab:LECsNew_table}:
\begin{align}
    l_{i}=\left( \mu c\right)^{-2\varepsilon }  \left( -\frac{\gamma_{i} }{32\pi^{2} \varepsilon } +l^{r}_{i}\left( \mu \right)  \right) .
\end{align}

The NNLO Lagrangian of \cite{Bijnens:1999sh} is also expected to be reduced because of the relation (\ref{eq:XaXbSP4}) in addition to the trace reflection already mentioned in \cite{Bijnens:2009qm}.

\renewcommand{\arraystretch}{1.5}
\begin{table}[h]
\centering
\begin{tabular}{c c c c c c c c c c c }
\hline
Theory & $l_1$ & $l_2$ & $l_3$ & $l_4$ & $l_5$ & $l_6$ & $l_7$ & $h_1$ & $h_2$ & $h_3$  \\
\hline
$SU(4)/Sp(4)$ & $\frac{1}{12}$  & $\frac{1}{24}$ &$\frac{1}{8}$  & $\frac{5}{128}$ & $0$ &$\frac{1}{2}$ & $-\frac{1}{2}$ & $-\frac{3}{4}$ & $0$ &  $0$ \\
\hline
\end{tabular}
\caption{Coefficients $\gamma_i$ for the NLO LECs of $SU(4)/Sp(4)$.}
\label{tab:LECsNew_table}
\end{table}
\renewcommand{\arraystretch}{1.0}

\section{Masses at NNLO}\label{secMasses}

\begin{figure}[t]
    \centering
    \includegraphics[width=0.6\textwidth]{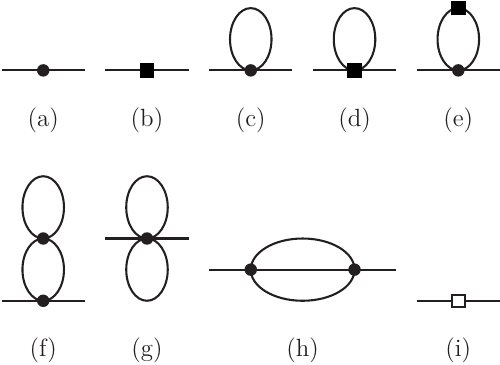}
    \caption{Diagrams which contribute to the pion mass at LO (a), NLO (b,c) and NNLO (d-i).
    The black circle vertex comes from $\mathcal{L}_{\text{LO}}$, black box vertex comes from $\mathcal{L}_{\text{NLO}}$ and the white box vertex comes from $\mathcal{L}_{\text{NNLO}}$.}
    \label{fig:SelfEnergy}
\end{figure}

The physical mass of the  $k$-th pion is given by a pole of a full propagator of $k$-th field \cite{Scherer:2012xha}:
\begin{equation}\label{eq.Sp4.32}
\begin{split}
     i\bigtriangleup_{k} =\frac{i}{p^{2}-
     M_{LO,k}^2+\Sigma_{k} \left( p^{2}\right)  +i\epsilon },
\end{split}
\end{equation}
where the self-energy $i\Sigma_{k}$ can be found as a sum of all one-particle irreducible diagrams at the given order of perturbation theory. The  %physical 
pion masses at the corresponding order are then defined from the equation
\begin{equation}\label{eq.Sp4.33}
\begin{split}
    M^{2}_{\pi ,k}-M^2_{\text{LO},k}+\Sigma_{k} \left(p^2 = M^{2}_{\pi ,k}\right)  =0.
\end{split}
\end{equation}
This equation is solved order by order:
\begin{align}
  M^2_{\text{NLO},k}& = -\Sigma_{\text{NLO},k}( M^2_{\text{LO},k}),\\
  M^2_{\text{NNLO},k}& = -\Sigma_{\text{NNLO},k}( M^2_{\text{LO},k}) + \Sigma_{\text{NLO},k}( M^2_{\text{LO},k})\frac{d}{dp^{2}} \Sigma_{\text{NLO},k}( M^2_{\text{LO},k}).
\end{align}
The diagrams which contribute to the pion mass are shown in figure \ref{fig:SelfEnergy}.

\subsection{\spth}

In the \spth theory, the LO  masses are equal for all the pion species:\footnote{
For notational simplicity, pion-species indices are omitted whenever no confusion can arise. 
The low-energy coefficients are generally different for different pion species, and their explicit values are given in the corresponding tables.
}
\begin{align}\label{massSpLO}
    M_{LO}^2=M^2 = B_0(m_u+m_d) 
\end{align}
The NLO and NNLO contributions read
\begin{align}\label{massSpNLO}
    M^{2}_{\text{NLO} }&=\frac{M^{2}}{F^{2}} \left(a_{M}\overline{A}  \left( M^{2}\right)  +b_{M}M^{2} + \alpha_M R_q^2 M^{2}\right),
\end{align}
\begin{align}\label{massSpNNLO}
    M^{2}_{\text{NNLO} }&=\frac{M^{2}}{F^{4}} \Bigg( c_{M}\overline{A} \left( M^{2}\right)^{2}  +M^{2}\overline{A} \left( M^{2}\right)  \left( d_{M}+\pi_{16} e_{M}\right)  +M^{4}\left( f_{M}+g_{M}\pi_{16} +h_{M}\pi^{2}_{16} \right)  \nonumber\\&+M^{2}R_q^{2}\left( M^{2}\beta_{M}+M^{2}\pi_{16} \gamma_{M}+\overline{A} \left( M^{2}\right)  \delta_{M}\right)  \Bigg). 
\end{align}
See table \ref{tab:MassSP4} for the values of coefficients for different pion species. The observed mass splitting is in agreement with the multiplet structure discussed in \ref{subsecSplitMasses}.
We note that the following linear combinations of the NNLO LECs are introduced:
\begin{align}
    r_{M,0}^r &= -16 \big(2 K_{17}^r + 8 K_{18}^r + K_{19}^r +  4 K_{20}^r + 4 K_{21}^r + 16 K_{22}^r + 
    K_{23}^r - 3 K_{25}^r - 12 K_{26}^r \nonumber\\&- 48 K_{27}^r - 2 K_{39}^r - 8 K_{40}^r\big),\label{rM0} \\
    r_{M,1}^r &= -16 (K_{19}^r + 4 K_{21}^r - K_{23}^r - 3 K_{25}^r - 4 K_{26}^r),\label{rM1}\\
    r_{M,2}^r &= -16 \big(K_{19}^r + 4 K_{21}^r + K_{23}^r + 4 K_{24}^r - 9 K_{25}^r - 12 K_{26}^r - 
    6 K_{39}^r - 8 K_{40}^r - 16 K_{41}^r  \nonumber \\&- 32 K_{42}^r\big)\label{rM2}.
\end{align}

\begin{table}[h]
\centering
\begin{tabular}{|c|c|}
\hline
Pion species & All\\
\hline
$a_M$ & $-\frac{3}{4}$\\
\hline
$b_M$ &$-32\left(l_3^r - 2l_4^r\right)$\\
\hline
$c_M$ & $ \frac{69}{32}$\\
\hline
$d_M$ & $8\left(11l_1^r + 5l_2^r - l_3^r + 2l_4^r\right)$\\
\hline
$e_M$ &$ \frac{3}{2}$\\
\hline
$f_M$ & $ r_{M,0}^r +1024\,l_3^r\left(l_3^r - 2l_4^r\right)$ \\
\hline
$g_M$ &$ 4\left(l_1^r + 3\left(l_2^r - 2l_3^r + 4l_4^r\right)\right)$\\
\hline
$h_M$ & $\frac{5}{6}$\\
\hline
Pion species &  $\pi_1$, $\pi_2$, $\pi_4$, $\pi_5$ \\
\hline
$\alpha_M$ & $0$\\
\hline
$\beta_M$ & $r_{M,1}^r$ \\
\hline
$\gamma_M$ & $16l_5^r$ \\
\hline
$\delta_M$ & $-16l_5^r$ \\
\hline
Pion species & $\pi_3$ \\
\hline
$\alpha_M$ & $64l_5^r $ \\
\hline
$\beta_M$ & $r_{M,2}^r -2048\,l_3^r l_5^r $ \\
\hline
$\gamma_M$ & $-16l_5^r $ \\
\hline
$\delta_M$ & $80l_5^r$ \\
\hline
\end{tabular}
\caption{NLO and NNLO coefficients for the mass of \spth theory.}
\label{tab:MassSP4}
\end{table}

%table with old LECs
\begin{comment}
\begin{table}[h]
\centering
\begin{tabular}{|c|c|}
\hline
Pion species & All\\
\hline
$a_M$ & $-\frac{3}{4}$\\
\hline
$b_M$ & $16L^r_8 + 64L^r_6 - 8L^r_5 - 32L^r_4$\\
\hline
$c_M$ & $ \frac{69}{32}$\\
\hline
$d_M$ & $4L^r_8 + 16L^r_6 - 2L^r_5 - 8L^r_4 + 22L^r_3 + 40L^r_2 + 88L^r_1 - 2L^r_0$\\
\hline
$e_M$ &$ \frac{3}{2}$\\
\hline
$f_M$ & $ r_{M,0}^r - 128L^r_5L^r_8 - 512L^r_5L^r_6 + 64(L^r_5)^2 - 512L^r_4L^r_8 - 2048L^r_4L^r_6 + 512L^r_4L^r_5 + 1024(L^r_4)^2$ \\
\hline
$g_M$ &$ 12L^r_8 + 48L^r_6 - 6L^r_5 - 24L^r_4 + L^r_3 + 12L^r_2 + 4L^r_1 + 5L^r_0$\\
\hline
$h_M$ & $\frac{5}{6}$\\
\hline
Pion species &  $\pi_1$, $\pi_2$, $\pi_4$, $\pi_5$ \\
\hline
$\alpha_M$ & $0$\\
\hline
$\beta_M$ & $r_{M,1}^r$ \\
\hline
$\gamma_M$ & $4L^r_8 + 16L^r_7$ \\
\hline
$\delta_M$ & $- 4L^r_8 - 16L^r_7$ \\
\hline
Pion species & $\pi_3$ \\
\hline
$\alpha_M$ & $16L_r^8 + 64L_r^7 $ \\
\hline
$\beta_M$ & $r_{M,2}^r - 128L_r^5L^r_8 - 512L_r^5L_r^7 - 512L^r_4L^r_8 - 2048L^r_4L^r_7 $ \\
\hline
$\gamma_M$ & $- 4L^r_8 - 16L^r_7 $ \\
\hline
$\delta_M$ & $20L_r^8 + 80L_r^7$ \\
\hline
\end{tabular}
\caption{NLO and NNLO coefficients for the mass of \spth theory.}
\label{tab:MassSP4}
\end{table}
\end{comment}

\subsection{\soth}

For the \soth theory with non-degenerate quark masses, the nine pions split into three groups with equal LO masses:
\begin{align}
M^2_1&\equiv M^{2}_{\text{LO},1}=M^{2}_{\text{LO},2}=M^{2}_{\text{LO},3}=M^{2}_{\text{LO},4}=M^{2}_{\text{LO},5}=B_{0}\left(m_{u}+m_{d}\right), \label{NLOso.1} \\
M^2_{6}&\equiv M^{2}_{\text{LO},6}=M^{2}_{\text{LO},7}=2B_{0}m_{u},\label{NLOso.2} \\ 
 M^2_{8}&\equiv M^{2}_{\text{LO},8}=M^{2}_{\text{LO},9}=2B_{0}m_{d}\label{NLOso.3}.  
\end{align}
The NLO contribution reads
\begin{align}
    M^{2}_{\text{NLO} }&=\frac{M^{2}_{\text{LO} }}{F^{2}} \Big( a_{M}\overline{A} \left( M^{2}_{1}\right)  +b_{M}M^{2}_{1}+\alpha_{M} R_q^{2}M^{2}_{1}+\zeta_{M} R_qM^{2}_{1} \nonumber \\&+\left( \overline{A} \left( M^{2}_{6}\right)  \left( 1+R_q\right)  +\overline{A} \left( M^{2}_{8}\right)  \left( 1-R_q\right)  \right)  \varkappa_{M} \Big).
\end{align}
As in the case of \spth theory, the observed mass splitting at NLO is in agreement with the multiplet structure.
The NNLO expressions are given in the appendix \ref{appSO4Masses}.

\begin{table}[h]
\centering
\begin{tabular}{|c|c|}
\hline
Pion species& All\\
\hline
$b_M$ & $16L^r_8 + 64L^r_6 - 8L^r_5 - 32L^r_4$ \\
\hline
Pion species &  $\pi_1,\pi_2,\pi_4,\pi_5$ \\
\hline
$a_M$&  $\frac{1}{4} $\\
\hline
$\alpha_M$& 0  \\
\hline
$\zeta_M$& 0  \\
\hline
$\varkappa_M$& 0  \\
\hline
Pion species &  $\pi_3,$ \\
\hline
$a_M$& $-\frac{3}{4}$  \\
\hline
$\alpha_M$& $16L^r_8 + 64L^r_7$  \\
\hline
$\zeta_M$&  $0$ \\
\hline
$\varkappa_M$&   $\frac{1}{2}$ \\
\hline
Pion species &  $\pi_6,\pi_7$ \\
\hline
$a_M$& $\frac{1}{4}$  \\
\hline
$\alpha_M$&  $0$ \\
\hline
$\zeta_M$&   $ 16L^r_8 - 8L^r_5 $\\
\hline
$\varkappa_M$&   $0$\\
\hline
Pion species &  $\pi_8,\pi_9$ \\
\hline
$a_M$&  $\frac{1}{4}$ \\
\hline
$\alpha_M$& $0$  \\
\hline
$\zeta_M$&  $-16L^r_8 + 8L^r_5$ \\
\hline
$\varkappa_M$& $0$  \\
\hline
\end{tabular}
\caption{NLO coefficients for the mass of \soth theory.}
\label{tab:MassSO4}
\end{table}

\section{Decay constants at NNLO}\label{secDecay}

The physical pseudoscalar decay constants $F_{\pi,k}$ are defined through the matrix elements of the axial currents corresponding to broken generators \cite{Amoros:1999dp}:
\begin{equation}\label{eqdecay1}
\langle 0|A^{k}_{\mu }(0)|\pi^{k}(p)\rangle
=i\sqrt{2}\, p_{\mu }F_{\pi,k},
\qquad k=1,\ldots,N_{\pi }.
\end{equation}
Inserting a complete set of one-particle states and using \eqref{eqdecay1}, the current--pion correlator can be related to the decay constant as
\begin{align}
G_{\mu }(p)
=
\int d^{4}x\, e^{ipx}
\left<0\left|
T\left\{
A^{k}_{\mu }(x)\,\pi^{k}(0)
\right\}
\right|0\right>=
\frac{\sqrt{2Z_{k}}\,F_{\pi ,k}}
{p^{2}-M^{2}_{\pi,k}+i\epsilon}\,
p_{\mu },
\end{align}
where $Z_k$ is the pion wave-function renormalization constant. The correlator $G_\mu(p)$ can also be written as the product of the external pion propagator and the amputated vertex function $\Gamma_\mu(p)$:
\begin{align}
G_{\mu }(p)
=
\frac{iZ_{k}}
{p^{2}-M^{2}_{\pi,k}+i\epsilon}\,
\Gamma_{\mu }(p).
\end{align}
Comparing the two expressions gives
\begin{align}
i\sqrt{Z_{k}}\,\Gamma_{\mu }(p)
=
\sqrt{2}\,F_{\pi,k}\,p_{\mu }.
\end{align}
To compute the contribution of the current insertion to $\Gamma_\mu(p)$, one takes a functional derivative of the low-energy generating functional $Z_{\mathrm{ChPT}}[u,\chi,V^\mu]$ with respect to the external source $V_\mu^k$ coupled to the broken current $A_\mu^k$, and subsequently sets $V_\mu^k=0$. This corresponds to evaluating diagrams with one external $V_\mu^k$ leg.\footnote{Some details of the external source method in ChPT can be found in \cite{Bijnens:2006zp} or \cite{Scherer:2012xha}.} The diagrams contributing to this calculation up to NNLO are identical to those for the mass calculation (see figure~\ref{fig:SelfEnergy}), except that one pion leg is replaced by an external vector leg.  We denote the corresponding two-point vertex function with external pion $\pi_k$ and source $V_\mu^k$ legs (with the overall factor of $p^\mu$ removed) by $\sqrt{2}\mathcal{M}_{\pi_k V_k}$.

The wave function renormalization of the pion calculated as a residue of the full pion propagator. 
It is related to the self-energy correction via \cite{Peskin:1995ev}
\begin{equation}\label{eqdecay3}
\begin{split}
Z_k =\left( 1-\frac{d\Sigma_k }{dp^{2}} \left(M^{2}_{\pi ,k}\right)  \right)^{-1},
\end{split}
\end{equation}
where the chiral expansion is assumed:
\begin{equation}\label{eqdecay4}
\Sigma_{k} (p^{2})=\Sigma_{\text{LO} ,k} (p^{2})+\Sigma_{N\text{LO} ,k} (p^{2})+\Sigma_{NN\text{LO} ,k} (p^{2}),
\end{equation}
where $\Sigma_{\text{LO}}=0$.
Then using the chiral expansion of mass, we get
\begin{align}\label{eqdecay5}
    \frac{d\Sigma_{k} }{dp^{2}} \left( M^{2}_{\pi ,k}\right)  =\Sigma^{\prime }_{\text{NLO} ,k} + \Sigma^{\prime }_{\text{NNLO} ,k} ,
\end{align}
with
\begin{align}\label{eqdecay5.1}
    \Sigma^{\prime }_{\text{NLO} ,k} &=\frac{d\Sigma_{\text{NLO} ,k} }{dp^{2}} \left( M^{2}_{\text{LO} ,k}\right),\\
    \Sigma^{\prime }_{\text{NNLO} ,k} &=M^{2}_{\text{NLO} ,k}\frac{d^{2}\Sigma_{\text{NLO} ,k} }{d\left( p^{2}\right)^{2}  } \left( M^{2}_{\text{LO} ,k}\right)  +\frac{d\Sigma_{\text{N} \text{NLO} ,k} }{dp^{2}} \left( M^{2}_{\text{LO} ,k}\right).
\end{align}
Finally, we have up to the NNLO:
\begin{align}\label{eqdecay6}
 F_{\text{LO} ,k}&=\mathcal{M}_{\text{LO} ,\pi_{k} V_{k}} ,\\
 F_{\text{NLO} ,k}&=\mathcal{M}_{\text{NLO} ,\pi_{k} V_{k}} -\frac{1}{2} \mathcal{M}_{\text{LO} ,\pi_{k} V_{k}}\Sigma^{\prime }_{\text{NLO},k}  , \\
  F_{\text{NNLO} ,k}&=\mathcal{M}_{\text{NNLO} ,\pi_{k} V_{k}} -\frac{1}{2}\mathcal{M}_{\text{NLO} ,\pi_{k} V_{k}}\Sigma^{\prime }_{\text{NLO},k}\nonumber \\&+\frac{1}{8} \mathcal{M}_{\text{LO} ,\pi_{k} V_{k}} \left( 3\left( \Sigma^{\prime }_{\text{NLO},k } \right)^{2}  -4\Sigma^{\prime }_{\text{NNLO},k } \right).
\end{align}

\subsection{\spth}

In the \spth theory, the LO and NLO decay constants  are equal for all the pion species:
\begin{align}\label{decaySpLO}
    F_{LO}=F,
\end{align}
\begin{align}\label{decaySpNLO}
 F_{\text{NLO} }=F\left( a_{M}\frac{\overline{A} \left( M^{2}\right)  }{F^{2}} +b_{M}\frac{M^{2}}{F^{2}} \right).
\end{align}
The NNLO contribution reads
\begin{align}\label{decaySpNNLO}
   F_{\text{NNLO} }&=F\Bigg( c_{F}\frac{\bar{A} \left( M^{2}\right)^{2}  }{F^{4}} +M^{2}\frac{\bar{A} \left( M^{2}\right)  }{F^{4}} \left( d_{F}+e_{F}\pi 16\right) \\& +\frac{M^{4}}{F^{4}} \left( f_{F}+g_{F}\pi 16+h_{F}\pi^{2}_{16} \right)  +\frac{M^{2}}{F^{4}} R_q^{2}\left( \gamma_{F}M^2 \pi_{16} +\delta_{F} \bar{A} \left( M^{2}\right) +\rho_FM^2 \right)  \Bigg)  \nonumber.
\end{align}
See table \ref{tab:DecaySP4} for the values of coefficients for different pion species.  The following linear combinations of the NNLO LECs are introduced:

\begin{comment}
\begin{align}
   r^{r}_{F,0}&=8(K^{r}_{19}(1+R^{2})+4K^{r}_{20}+4K^{r}_{21}\left( 1+R^{2}\right)  +16K^{r}_{22}+K^{r}_{23}\left( 1-R^{2}\right)  ),\\
   r^{r}_{F,1}&=8(K^{r}_{19}(1+R^{2})+4K^{r}_{20}+4K^{r}_{21}\left( 1+R^{2}\right)  +16K^{r}_{22}+K^{r}_{23}\left( 1+R^{2}\right) \nonumber \\&  +4K^{r}_{24}R^{2}).
\end{align}
\end{comment}

\begin{align}
   r^r_{F,0}&= 8 (K_{19}^r+4 K_{20}^r+4 K_{21}^r+16 K_{22}^r+K_{23}^r),\label{rF0}\\
   r^r_{F,1}&=8  (K_{19}^r+4 K_{21}^r-K_{23}^r),\label{rF1}\\
   r^r_{F,2}& = 8(K_{19}^r+4 K_{21}^r+K_{23}^r+4 K_{24}^r).\label{rF2}
\end{align}

\begin{table}[h]
\centering
\begin{tabular}{|c|c|}
\hline
Pion species & All\\
\hline
$a_F$ & $1$\\
\hline
$b_F$ & $ 16l_3^r$\\
\hline
$c_F$ & $- \frac{11}{8}$\\
\hline
$d_F$ & $-4\left(11l_1^r + 5l_2^r + 7l_3^r - 16l_4^r\right)$\\
\hline
$e_F$ &$ -\frac{43}{96}$\\
\hline
$f_F$ & $ r_{F,0}^r-128\left(l_3^r\right)^2$ \\
\hline
$g_F$ &$ -2\left(l_1^r + 3l_2^r - 16l_3^r + 32l_4^r\right)$\\
\hline
$h_F$ & $-\frac{5}{96}$\\
\hline
Pion species &  $\pi_1$, $\pi_2$, $\pi_4$, $\pi_5$ \\
\hline
$\gamma_F$ & $ -16l_5^r$ \\
\hline
$\delta_F$ & $ 16l_5^r$ \\
\hline
$\rho_F$ & $r_{F,1}^r$ \\
\hline
Pion species & $\pi_3$ \\
\hline
$\gamma_M$ & $0 $ \\
\hline
$\delta_M$ & $0$ \\
\hline
$\rho_F$ & $r_{F,2}^r$ \\
\hline
\end{tabular}
\caption{NLO and NNLO coefficients for the decay constants of \spth theory.}
\label{tab:DecaySP4}
\end{table}

\begin{comment} %table with old LECs
    
\begin{table}[h]
\centering
\begin{tabular}{|c|c|}
\hline
Pion species & All\\
\hline
$a_F$ & $1$\\
\hline
$b_F$ & $ 4L^r_5 + 16L^r_4$\\
\hline
$c_F$ & $- \frac{11}{8}$\\
\hline
$d_F$ & $16L^r_8 + 64L^r_6 - 7L^r_5 - 28L^r_4 - 11L^r_3 - 20L^r_2 - 44L^r_1 + L^r_0 $\\
\hline
$e_F$ &$ -\frac{43}{96}$\\
\hline
$f_F$ & $ r_{F,0}^r - 8(L^r_5)^2 - 64L^r_4L^r_5 - 128(L^r_4)^2$ \\
\hline
$g_F$ &$  - 16L^r_8 - 64L^r_6 + 8L^r_5 + 32L^r_4 - \frac{1}{2}L^r_3 - 6L^r_2 - 2L^r_1 - \frac{5}{2}L^r_0$\\
\hline
$h_F$ & $-\frac{5}{96}$\\
\hline
Pion species &  $\pi_1$, $\pi_2$, $\pi_4$, $\pi_5$ \\
\hline
$\gamma_F$ & $ - 4L^r_8 - 16L^r_7$ \\
\hline
$\delta_F$ & $ 4L^r_8 + 16L^r_7$ \\
\hline
$\rho_F$ & $r_{F,1}^r$ \\
\hline
Pion species & $\pi_3$ \\
\hline
$\gamma_M$ & $0 $ \\
\hline
$\delta_M$ & $0$ \\
\hline
$\rho_F$ & $r_{F,2}^r$ \\
\hline
\end{tabular}
\caption{NLO and NNLO coefficients for the decay constants of \spth theory.}
\label{tab:DecaySP4}
\end{table}
\end{comment}

\subsection{\soth}

For the \soth theory with non-degenerate quark masses, the LO decay constant is the same for all species:
\begin{align}
    F_\text{LO}=F,
\end{align}
while at NLO and NNLO the decay constants split into four groups. The NLO contribution reads
\begin{align}
 F_{\text{NLO} }=F\left( a_{F}\frac{\overline{A} \left( M^{2}_{1}\right)  }{F^{2}} +b_{F}\frac{M^{2}_{1}}{F^{2}} +\xi_{F} \frac{\overline{A} \left( M^{2}_{6}\right)  }{F^{2}} +\zeta_{F} \frac{\overline{A} \left( M^{2}_{8}\right)  }{F^{2}} +\varkappa_{F} R_q\frac{M^{2}_{1}}{F^{2}} \right)   ,
\end{align}
where the values of coefficients are given in the table \ref{tab:DecaySO4}.
The NNLO expressions are given in the appendix \ref{appSO4Decay}.

\begin{table}[h]
\centering
\begin{tabular}{|c|c|}
\hline
Pion species & All \\
\hline
$b_F$ & $ 4L^r_5 + 16L^r_4 $ \\
\hline
Pion species &  $\pi_1,\pi_2,\pi_4,\pi_5$ \\
\hline
$a_F$&  $\frac{1}{2} $\\
\hline
$\xi_F$& $\frac{1}{4}$  \\
\hline
$\zeta_F$& $\frac{1}{4}$  \\
\hline
$\varkappa_F$& $0$  \\
\hline
Pion species &  $\pi_3,$ \\
\hline
$a_F$&  $ 1$\\
\hline
$\xi_F$& $0$  \\
\hline
$\zeta_F$& $0 $ \\
\hline
$\varkappa_F$& $0$  \\
\hline
Pion species &  $\pi_6,\pi_7$ \\
\hline
$a_F$&  $\frac{1}{2} $\\
\hline
$\xi_F$& $\frac{1}{2}$  \\
\hline
$\zeta_F$& $0$  \\
\hline
$\varkappa_F$& $ 4L^r_5$  \\
\hline
Pion species &  $\pi_8,\pi_9$ \\
\hline
$a_F$&  $\frac{1}{2} $\\
\hline
$\xi_F$& $0$  \\
\hline
$\zeta_F$& $\frac{1}{2}$  \\
\hline
$\varkappa_F$ & $ -4L^r_5$  \\
\hline
\end{tabular}
\caption{NLO coefficients for the decay constants of \soth theory.}
\label{tab:DecaySO4}
\end{table}

\section{Condensates at NNLO}\label{secCondensates}
\begin{figure}[h]
    \centering
    \includegraphics[width=0.5\textwidth]{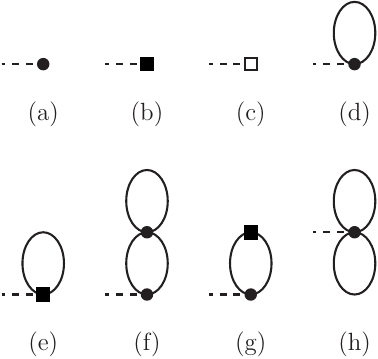}
   \caption{Diagrams which contribute to the quark condensates at LO (a), NLO (b,d) and NNLO (c, e-h).   The black circle vertex comes from $\mathcal{L}_{\text{LO}}$, black box vertex comes from $\mathcal{L}_{\text{NLO}}$ and the white box vertex comes from $\mathcal{L}_{\text{NNLO}}$.
    }
    \label{fig:Cond}
\end{figure}

The quark condensates are calculated using the external scalar source $\chi$ which contains the fermion masses:
\begin{align}
    \left< \bar{u} u\right>  =-\frac{1}{i} \frac{\delta }{\delta m_{u}} Z_{\text{ChPT} }\left[ u,\chi ,V^{\mu }\right]  ,\quad \left< \bar{d} d\right> =-\frac{1}{i} \frac{\delta }{\delta m_{d}} Z_{\text{ChPT} }\left[ u,\chi ,V^{\mu }\right] . 
\end{align}
Feynman diagrams which contribute to the calculation of the vacuum condensates up to NNLO are shown in figure \ref{fig:Cond}, where the dashed line denotes the scalar source leg. The results in this section are in accordance with the later work \cite{Kolesova:2025ghl} and differ by a factor of 2 from the  earlier paper \cite{Bijnens:2009qm}.

\subsection{\spth}
At LO both condensates are equal:
\begin{align}
 \left< \bar{u} u\right>_{\text{LO} }=\left< \bar{d} d\right>_{\text{LO} } =-2B_{0}F^{2}.
\end{align}
At NLO and NNLO the condensates are different. These contributions read:
\begin{align}
    \left< \bar{q} q\right>_{\text{NLO} } & =\left< \bar{q} q\right>_{\text{LO} }  \left( {}a_{V}\frac{\bar{A} \left( M^{2}\right)  }{F^{2}} +b_{V}\frac{M^{2}}{F^{2}} +\alpha_{V} R_q\frac{M^{2}}{F^{2}} \right),\\
    \left< \bar{q} q\right>_{\text{NNLO} } & =\left< \bar{q} q\right>_{\text{LO} }  \Bigg( {}c_{V}\frac{\bar{A} \left( M^{2}\right)^{2}  }{F^{4}} +\frac{M^{2}\bar{A} \left( M^{2}\right)  }{F^{4}} \left( d_{V}+\pi_{16} e_{V}\right)  +\frac{M^{4}}{F^{4}} \left( f_{V}+g_{V}\pi_{16} \right)  \nonumber \\& +\frac{M^{2}}{F^{4}} R_q\Big( \beta_{V} M^{2}+\gamma_{V} R_qM^{2}+\delta_{V} R_qM^{2}\pi_{16} +\zeta_{V} \bar{A} \left( M^{2}\right)  \nonumber \\& +\eta_{V} R_q\bar{A} \left( M^{2}\right)  \Big) \Bigg). 
\end{align}
where $q$ means either $u$ or $d$. 
The values of coefficients are given in the table \ref{tab:CondensatesSP4}. The following linear combinations of the NNLO LECs are introduced:
\begin{align}
    r_{V,0}^r&=48 (K_{25}^r+4 K_{26}^r+16 K_{27}^r),\label{rV0}\\
    r_{V,1}^r&= 64 (3 K_{25}^r + 4K_{26}^r)\label{rV1}.
\end{align}

\begin{table}[h]
\centering
\begin{tabular}{|c|c|}
\hline
Condensate & All \\
\hline
$a_V$ & $\frac{5}{4}$ \\
\hline
$b_V$ & $4\left(h_2^r - 4h_3^r + 16l_4^r\right)$ \\
\hline
$c_V$ & $-\frac{45}{32}$ \\
\hline
$d_V$ & $-120\left(l_3^r - 2l_4^r\right)$ \\
\hline
$e_V$ & $\frac{15}{16}$ \\
\hline
$f_V$ & $r_{V,0}^r$ \\
\hline
$g_V$ & $40\left(l_3^r - 2l_4^r\right)$ \\
\hline
$\gamma_V$ & $\frac{1}{4}r_{V,1}^r$ \\
\hline
$\delta_V$ & $-16l_5^r$ \\
\hline
$\eta_V$ & $16l_5^r$ \\
\hline
Condensate & $\left< \bar{u} u\right> $ \\
\hline
$\alpha_V$ & $4\left(h_2^r + 4h_3^r\right)$ \\
\hline
$\beta_V$ & $\frac{1}{2}r_{V,1}^r$ \\
\hline
$\zeta_V$ & $32l_5^r$ \\
\hline
Condensate &$\left< \bar{d} d\right> $  \\
\hline
$\alpha_V$ & $-4\left(h_2^r + 4h_3^r\right)$ \\
\hline
$\beta_V$ & $-\frac{1}{2}r_{V,1}^r$ \\
\hline
$\zeta_V$ & $-32l_5^r$ \\
\hline
\end{tabular}
\caption{NLO and NNLO coefficients for the  condensates in \spth theory.}
\label{tab:CondensatesSP4}
\end{table}

% old table with old LECs
\begin{comment}
    \begin{table}[h]
\centering
\begin{tabular}{|c|c|}
\hline
Condensate & All \\
\hline
$a_V$ & $\frac{5}{4}$ \\
\hline
$b_V$ & $4H_2^r + 8L_8^r + 64L_6^r$ \\
\hline
$c_V$ & $-\frac{45}{32}$ \\
\hline
$d_V$ & $60L_8^r + 240L_6^r - 30L_5^r - 120L_4^r$ \\
\hline
$e_V$ & $\frac{15}{16}$ \\
\hline
$f_V$ & $r_{V,0}^r$ \\
\hline
$g_V$ & $-20L_8^r - 80L_6^r + 10L_5^r + 40L_4^r$ \\
\hline
$\gamma_V$ & $\frac{1}{4}r_{V,1}^r$ \\
\hline
$\delta_V$ & $-4L_8^r - 16L_7^r$ \\
\hline
$\eta_V$ & $4L_8^r + 16L_7^r$ \\
\hline
Condensate & $\left< \bar{u} u\right> $ \\
\hline
$\alpha_V$ & $4H_2^r + 8L_8^r$ \\
\hline
$\beta_V$ & $\frac{1}{2}r_{V,1}^r$ \\
\hline
$\zeta_V$ & $8L_8^r + 32L_7^r$ \\
\hline
Condensate &$\left< \bar{d} d\right> $  \\
\hline
$\alpha_V$ & $-4H_2^r - 8L_8^r$ \\
\hline
$\beta_V$ & $-\frac{1}{2}r_{V,1}^r$ \\
\hline
$\zeta_V$ & $-8L_8^r - 32L_7^r$ \\
\hline
\end{tabular}
\caption{NLO and NNLO coefficients for the  condensates in \spth theory.}
\label{tab:CondensatesSP4}
\end{table}
\end{comment}

\subsection{\soth}

At LO both condensates are equal: 
\begin{align}
 \left< \bar{u} u\right>_{\text{LO} }=\left< \bar{d} d\right>_{\text{LO} } =-2B_{0}F^{2}.
\end{align}
At NLO and NNLO  the condensates are different. The NLO contribution reads:
\begin{align}
    \left< \bar{q} q\right>_{\text{NLO} } & =\left< \bar{q} q\right>_{\text{LO} }  \left(a_{V}\frac{\bar{A} \left( M^{2}_{1}\right)  }{F^{2}} +b_{V}\frac{M^{2}_{1}}{F^{2}} +\alpha_{V} R_q\frac{M^{2}_{1}}{F^{2}} +\xi_{V} \frac{\bar{A} \left( M^{2}_{6}\right)  }{F^{2}} +\varkappa_{V} \frac{\bar{A} \left( M^{2}_{8}\right)  }{F^{2}}  \right), 
\end{align}
where as before $q$ means either $u$ or $d$. The values of coefficients are given in the table \ref{tab:CondensatesSO4}. The NNLO result is given in the appendix \ref{appSO4Vacuum}.

\begin{table}[h]
\centering
\begin{tabular}{|c|c|}
\hline
Condensate & All \\
\hline
$a_V$ & $\frac{5}{4}$ \\
\hline
$b_V$ & $4H_2^r + 8L_8^r + 64L_6^r$ \\
\hline
Condensate & $\left< \bar{u} u\right> $ \\
\hline
$\alpha_V$ & $4H_2^r + 8L_8^r$ \\
\hline
$\xi_V$ & $1$ \\
\hline
$\varkappa_V$ & $0$ \\
\hline
Condensate &$\left< \bar{d} d\right> $  \\
\hline
$\alpha_V$ & $-4H_2^r - 8L_8^r$ \\
\hline
$\xi_V$ & $0$ \\
\hline
$\varkappa_V$ & $1$ \\
\hline
\end{tabular}
\caption{NLO  for the  condensates in \soth theory.}
\label{tab:CondensatesSO4}
\end{table}

\section{Fit of the LECs for the \spth theory}\label{sec:fit}

In this section, we extract the NLO low-energy constants (LECs) together with one NNLO linear combination for the $Sp(N_c=4)$ gauge theory with $N_F=2$ fundamental quarks, as described by the \spth EFT at low energies. Our analysis combines lattice results for non-degenerate pion masses and decay constants from \cite{Kulkarni:2022bvh} with lattice determinations of pion scattering lengths in the mass-degenerate case from \cite{Dengler:2024maq}.

A first NLO fit to this dataset was carried out in \cite{Kolesova:2025ghl}. For comparison, a fit has also been performed for the $SU(N_c=2)\cong Sp(N_c=2)$ gauge theory with fermions in the fundamental (pseudoreal) representation in \cite{Arthur:2016dir}. In the case of real-world QCD with gauge group $SU(N_c=3)$, LEC extractions rely on a combination of experimental measurements and lattice inputs; see \cite{Bijnens:2014lea} for a  review.

\subsection{Lattice data}

\subsubsection{Pion masses and decay constants}
In \cite{Kulkarni:2022bvh}, lattice data are provided for three values of the  ratio $M_{\rho}/M_{\pi} \approx 1.14, 1.24, 1.46$ at the point of equal quark masses, where $M_{\rho}$ is the mass of vector meson. We use the dataset with $M_{\rho}/M_{\pi} \approx 1.46$, as it lies closest to the chiral limit. Each dataset includes pion and vector meson masses and their decay constants. In our notation, the singlet pion mass is $M_{\pi,3}$, while the four-plet pions have mass $M_{\pi,1}$, corresponding to $m_C$ and $m_A$ in~\cite{Kulkarni:2022bvh}.
Results are given at $\beta = 6.9, 7.05, 7.2$, with larger $\beta$ corresponding to smaller lattice spacing. Although no continuum extrapolation is performed, lattice systematics indicate that finite-spacing results approximate the continuum well. Since the NLO fit in \cite{Kolesova:2025ghl} shows weak $\beta$ dependence, we focus on $\beta = 7.2$, closest to the continuum. While \cite{Kulkarni:2022bvh} reports multiple lattice volumes without taking the infinite-volume limit, we use the extrapolated results from \cite{Kolesova:2025ghl} .

The data show that quark mass non-degeneracy induces splittings in both pion masses and decay constants. This effect is not captured at LO or NLO; accordingly, \cite{Kolesova:2025ghl} fits averaged decay constants with inflated uncertainties. At NNLO, however, non-degeneracy effects are fully included.
Quark mass splitting is parametrized using PCAC masses, extracted from lattice correlators via the partially conserved axial current relation. Since renormalization factors cancel in the ratio $r \equiv m_d^{\text{PCAC}}/m_u^{\text{PCAC}}$, we identify it with the parameter $R_q = (1 - r)/(1 + r)$ defined in Eq.~\eqref{eqquarkmassratioR}. For all $\beta$, when $r \gtrsim 5.5$, the lightest vector meson becomes nearly degenerate with the heavier pion, signaling a breakdown of the EFT. Following \cite{Kolesova:2025ghl}, we therefore restrict to $r \leq 5.5$, leaving five data points at $\beta = 7.2$ (see table \ref{tab:LECs_table_infinite_volume_limit}.)

\begin{table}[h]
\centering
\begin{tabular}{c c c c c c c c c c c}
\hline
$\beta$ & $am^0_u$ & $am^0_d$ & $r$  & $a^2(M_{\pi,1}^\infty)^2$   & $a^2(M_{\pi,3}^\infty)^2$ & $aF_{\pi,1}^\infty$ & $aF_{\pi,3}^\infty$ \\
\hline
7.2  & -0.794 &  -0.7 & 4.4(6)&  0.272(5)  & 0.204(6)    & 0.080(2)   &0.071(2) &\\
7.2  & -0.794 & -0.75 & 2.7(4)&  0.172(3)   & 0.153(4)   &  0.073(1) &0.070(2)  &\\
7.2  & -0.794 & -0.77 & 1.7(4)&  0.123(6)   & 0.112(5)   & 0.067(2)  & 0.066(1) &\\
7.2  & -0.794 & -0.78 & 1.4(3)&  0.105(22)   & 0.110(22) &  0.059(12) &  0.062(13)&\\ 
7.2  & -0.794 & -0.794 & 1.0(3)&  0.080(5)    & 0.079(5) & 0.058(2)  & 0.057(2) &\\
\hline
\end{tabular}
\caption{Infinite-volume extrapolation of the lattice data from ref.~\cite{Kulkarni:2022bvh}, performed in \cite{Kolesova:2025ghl}, in the region $M_{\rho}/M_{\pi} \approx 1.46$. Unlike the table in \cite{Kolesova:2025ghl}, the non-degenerate values of the decay constants are presented instead of the averaged value.}
\label{tab:LECs_table_infinite_volume_limit}
\end{table}

\subsubsection{Scattering length}
In \cite{Dengler:2024maq}, the scattering length for two mass-degenerate pions, 
$M_{\pi,1} = M_{\pi,3} \equiv M_{\pi}$, is computed using L\"{u}scher’s method 
\cite{Luscher:1986pf}. The two-pion operators are built from interpolating fields 
of the form $\bar{u}(x)\gamma_5 d(x)$. The analysis is carried out in the 
14-dimensional irreducible representation of the flavor $Sp(4)$ symmetry group, 
often referred to as the ``isospin-2'' channel by analogy with QCD. This 
channel does not couple to single vector-meson states, which instead transform in 
the 10-dimensional representation of $Sp(4)$.

The scattering length $a_0 M_\pi$ is given as a function of $M_\pi$ and $F_\pi$, 
and can be related to the NNLO results of \cite{Bijnens:2011fm}\footnote{In 
\cite{Dengler:2024maq}, the lattice results are presented both in terms of the 
ratio $M_\pi/F_\pi$ and as functions of the independent variables $M_\pi$ and 
$F_\pi$. In contrast to \cite{Kolesova:2025ghl}, where the ratio was used, we 
adopt the latter parametrization since we fit scale-dependent LECs, leading to an 
explicit $M_\pi$-dependence through the chiral logarithms.}.  Note that \cite{Bijnens:2011fm} employs a different convention for the scattering 
length than \cite{Dengler:2024maq}, namely $a_0 M_\pi = -a_0^{\text{MS}}$, where $a_0^{\text{MS}}$ is given by:
\begin{align}\label{scatteringLengthSpNNNLO}
\pi a_0^{\text{MS}}=&\  x_2 \left(-\frac{1}{32}\right) + x_2^2 \left( 2l_4^r - 2l_3^r + 2l_2^r + 2l_1^r - \frac{1}{128}\pi_{16}  +  \left(-\frac{5}{128}\right)L(M_\pi^2)\right) \nonumber  \\
&+  x_2^3\Bigg[ \Big( -256 (l_4^r)^2 + 256 l_3^r l_4^r - 64 (l_3^r)^2 + r_{A,0}^r 
+ \pi_{16} l_4^r - \pi_{16} l_3^r - 2\pi_{16} l_2^r + \pi_{16} l_1^r 
- \frac{43}{768}\pi_{16}^2 \Big) \nonumber\\
&+ L(M_\pi^2)
\left( -4l_4^r + \frac{3}{2}l_3^r - 6l_2^r - l_1^r + \frac{79}{768}\pi_{16} \right) \nonumber \\
&+ L^2(M_\pi^2)  \left(\frac{29}{384}\right)+ \pi^2 x_2^3 \left(\frac{17}{576}\pi_{16}^2\right)\Bigg],
\end{align}
where $x_2\equiv M_\pi^2/F_\pi^2$, $L(M_\pi^2)\equiv\pi_{16} \log \left( \frac{M^{2}_{\pi }}{\mu^{2} } \right)  $ and the following linear combination of NNLO LECs was introduced:
\begin{align}
r_{A,0}^r =& \ 2K_1^r + 16K_{10}^r + K_{11}^r + K_{13}^r + 4K_{14}^r + 4K_{15}^r + 16K_{16}^r - 3K_{17}^r - 12K_{18}^r\nonumber\\ & - 2K_{19}^r + 8K_2^r - 8K_{20}^r - 8K_{21}^r - 32K_{22}^r - 2K_{23}^r + 3K_{25}^r + 12K_{26}^r + 48K_{27}^r\nonumber\\
& + 4K_{28}^r + 8K_{29}^r - 2K_3^r - 2K_{31}^r - 2K_{33}^r - 8K_{35}^r + 2K_{37}^r + 3K_{39}^r \nonumber\\
&+ 12K_{40}^r - 4K_5^r + K_7^r + 4K_8^r + 4K_9^r.
\end{align}
In \cite{Dengler:2024maq}, several datasets are provided (see also 
\cite{dengler_2024_12920978} for the data release). Following 
\cite{Kolesova:2025ghl}, we use the scattering length data from the bottom panel 
of Fig.~3, corresponding to a subset of lattice ensembles selected to minimize 
discretization and finite-volume effects. As for the masses and decay constants, 
we restrict to the $\beta=7.2$ ensemble, which yields two data points, see table \ref{tab:scattering_data}. The infinite-volume limit is implemented within L\"{u}scher’s method, so the 
reported pion masses and scattering lengths are already extrapolated. While 
$F_\pi$ is given at finite volume, the lattices are sufficiently large that the 
associated systematic uncertainties are expected to be smaller than the 
statistical errors.
\begin{table}[h]
\centering
\begin{tabular}{c c c c c c}
\hline
$\beta$ & $T$ & $L$ & $a M_{\pi}^\infty $ &
 $aF_\pi$& $a_0M_{\pi}^\infty$  \\
\hline
7.2  & 36 & 24 & $0.3675(8)$ &  0.064(3) & 0.89(9)  \\
7.2  & 36 & 28 & $0.2852(4)$ &  0.057(1)  & 0.62(11) \\
\hline
\end{tabular}
\caption{
Lattice data determining the scattering length $a_0$ from \cite{Dengler:2024maq}, corresponding to the bottom panel of their figure~3. $M_{\pi}^\infty$ is the pion mass in the infinite volume limit, $a$ is the lattice spacing, $T$ and $L$ correspond to temporal and spacial extents of the largest lattice considered.}
\label{tab:scattering_data}
\end{table}

\subsection{Fitting the LECs}
We perform a global fit at NNLO precision using formulas~\eqref{massSpLO}–\eqref{massSpNNLO} for the pion masses, formulas~\eqref{decaySpLO}–\eqref{decaySpNNLO} for the pion decay constants, and formula~\eqref{scatteringLengthSpNNNLO} for the scattering length. In total, these formulas depend on 14 free parameters. We fit two LO parameters, $B_0m_u$ and $F$, five NLO LECs, $l_1^r,\dots,l_5^r$, and one NNLO linear combination, $r^r_{F,2}$, while setting $r^r_{M,0}$, $r^r_{M,1}$, $r^r_{M,2}$, $r^r_{F,0}$, $r^r_{F,1}$, and $r^r_{A,0}$ to zero due to strong degeneracies and limited data. The inclusion of $r^r_{F,2}$ provides sufficient freedom to fit the split decay constants, while the LEC $l_5^r$ is responsible for the mass splitting. 
The fit is performed for a fixed value of the renormalization scale $\mu=0.6a^{-1}$ which is chosen as the mass of the lighter vector meson at $r\approx4.4$ as presented in \cite{Kulkarni:2022bvh}.

We carried out a Bayesian analysis using Markov Chain Monte Carlo (MCMC) sampling to determine the posterior distributions of the model parameters. The sampling procedure was implemented with the \texttt{emcee} package~\cite{Foreman-Mackey:2012any}. Since the datasets of~\cite{Kulkarni:2022bvh} and~\cite{Dengler:2024maq} contain non-negligible uncertainties in the independent variables—namely the PCAC mass ratio $r$, the pion mass $M_\pi$, and the pion decay constant $F_\pi$—these quantities were treated as latent variables in the fit. Their true values were integrated out in the Bayesian inference procedure, thereby consistently propagating uncertainties from both dependent and independent observables into the posterior distributions.
The likelihood function for each dataset was assumed to be Gaussian. In the case of the scattering length data from Ref.~\cite{Dengler:2024maq}, where asymmetric errors are reported, we conservatively adopted the larger of the two uncertainties as an effective symmetric error. 
The prior distributions were chosen according to the expected chiral scaling of the low-energy constants. All NLO LECs were assigned Gaussian priors centered at zero with width $\pi_{16}$, whereas the NNLO parameter combination $r^r_{F,2}$ was given a Gaussian prior with variance of order $\pi_{16}^2$. These choices reflect the anticipated magnitude of higher-order corrections and are consistent with previous analyses, see for example Ref.~\cite{Kolesova:2025ghl}. For the LO parameters $F$ and $B_0m_u$, we employed uniform priors in the range $\left[0,0.1\right]$ in units of the inverse lattice spacing.

\begin{figure}[htbp]
\centering
\includegraphics[width=0.8\textwidth]{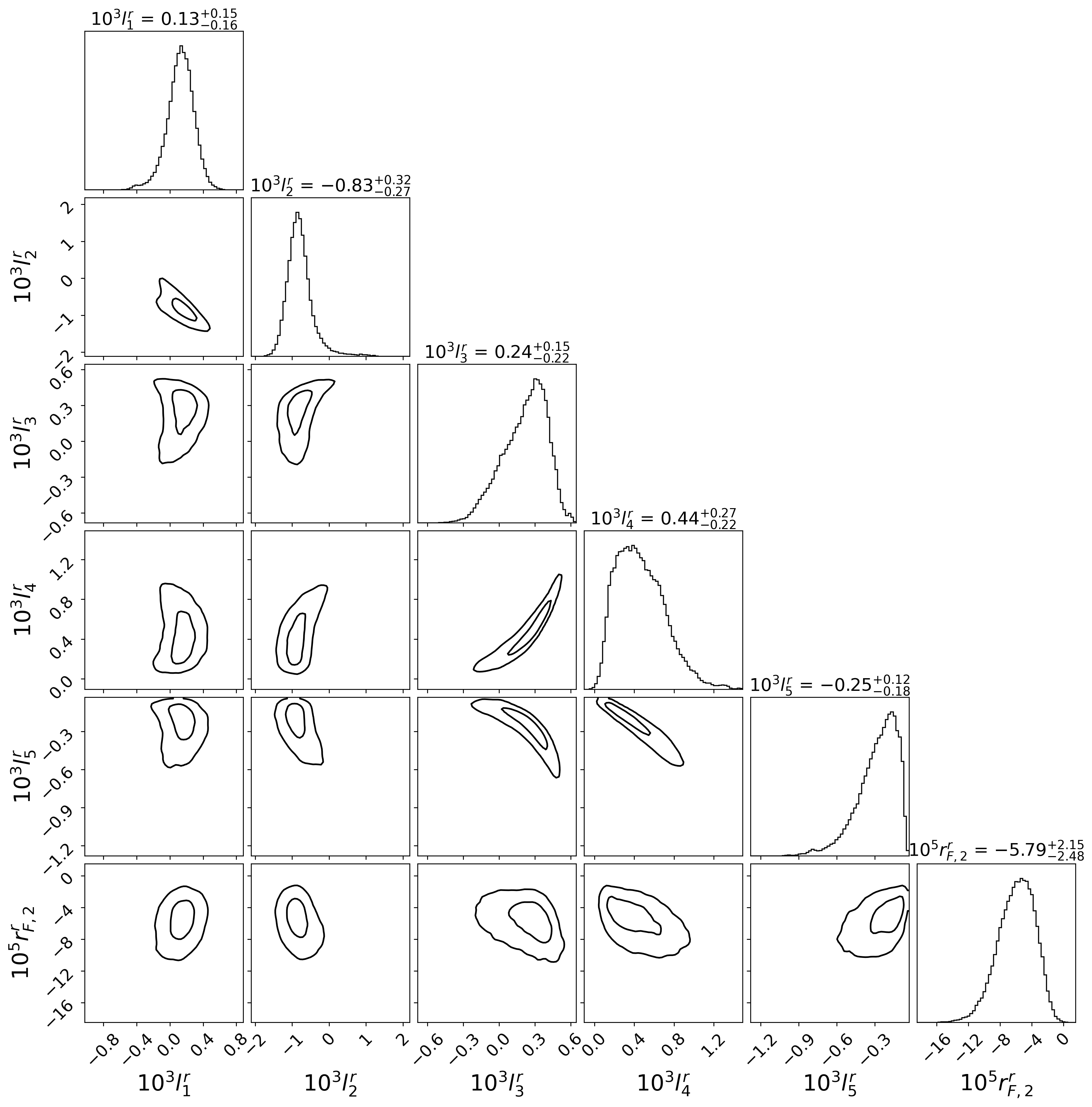}
\caption{
Triangle plot showing the posterior inference for NLO LECs $l^r_1, \dots l^r_5$ and the NNLO linear combination $r_{F,2}^r$ for $\beta=7.2$. The diagonal panels display the marginalized one-dimensional posteriors, with the median and 68\% credible intervals (16th–84th percentiles) indicated above each histogram. Off-diagonal panels show the marginalized two-dimensional posteriors, with contours enclosing 39.3\% and 86.5\% of the probability mass.
}
\label{fig:posteriors}
\end{figure}

The result of the fit is summarized in Fig.~\ref{fig:posteriors}, which shows the marginalized posterior distributions for each parameter, together with their median values and associated $1\sigma$ uncertainties.\footnote{$1\sigma$ interval corresponds to the 16th–84th percentiles (68\% credible interval) without assuming Gaussianity of the underlying distributions.} The fitted values of the dimensionful parameters are $B_0m_u=0.023(6)a^{-2}$ and $F=0.035(4)a^{-1}$. However, these values are not directly useful in practice because the lattice spacing $a$ is unknown.

\begin{figure}[htbp]
    \centering
\includegraphics[width=0.6\textwidth]{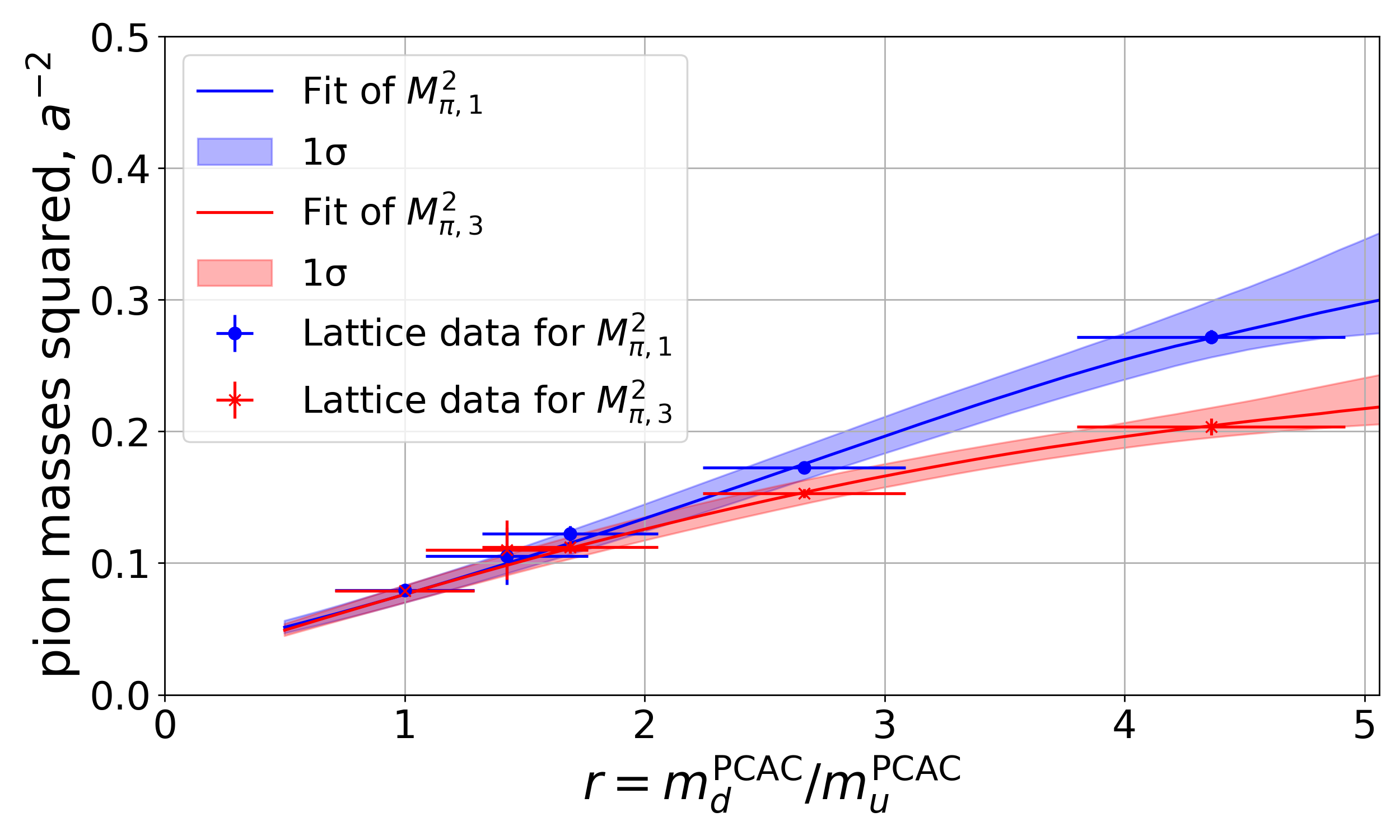}
    \caption{Lattice data \cite{Kulkarni:2022bvh} and fits for masses $M^2_{\pi,1}$ and $M^2_{\pi,3}$ as a function of $r$. The colored bands correspond to regions obtained when the LECs are varied within the 1$\sigma$ errors .  }\label{figFitMass}
\end{figure}

\begin{figure}[htbp]
    \centering
    \includegraphics[width=0.6\textwidth]{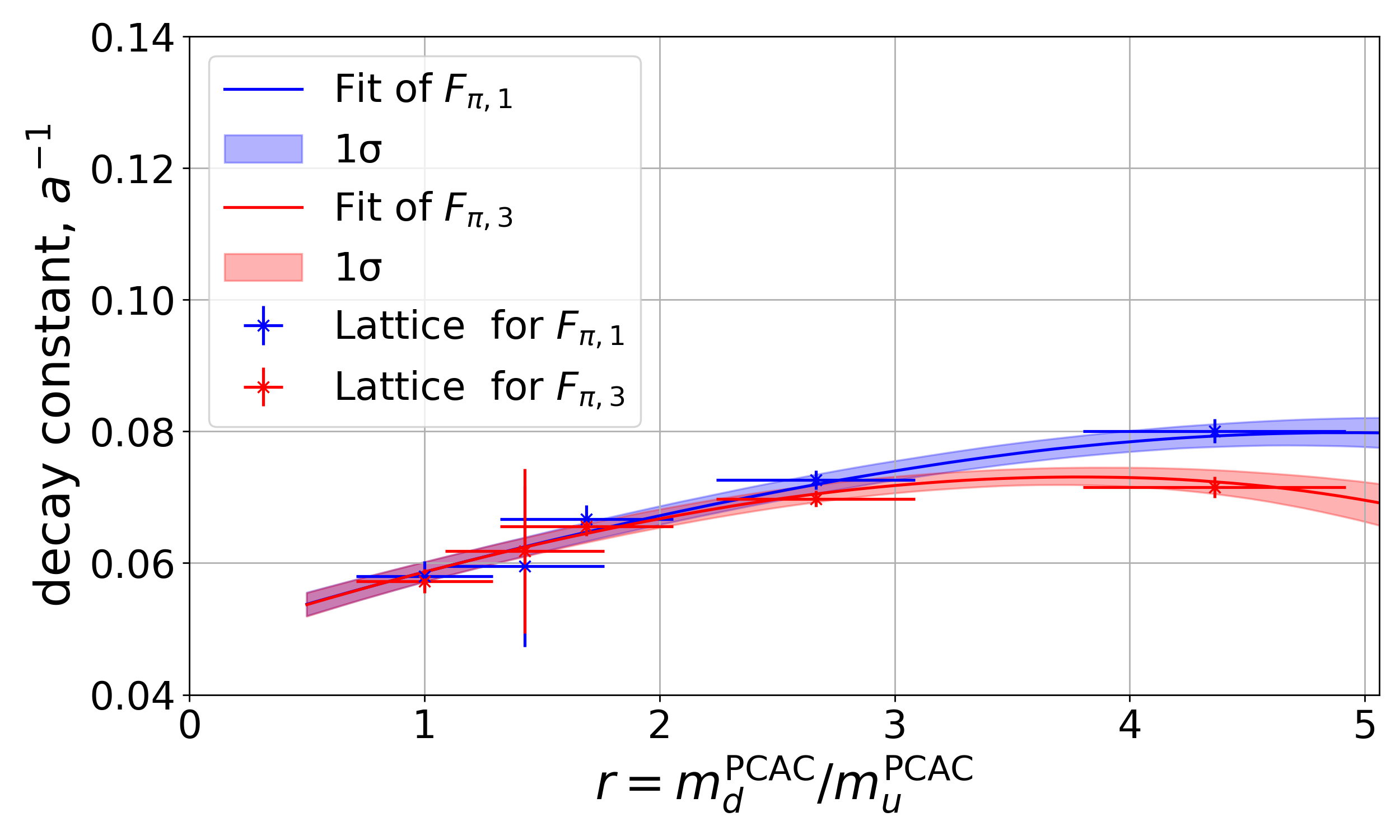}
    \caption{Lattice data \cite{Kulkarni:2022bvh} and fits for masses $F_{\pi,1}$ and $F_{\pi,3}$ as a function of $r$. The colored bands correspond to regions obtained when the LECs are varied within the 1$\sigma$ errors.
}\label{figFitDec}
\end{figure}

\begin{figure}[htbp]
    \centering
    \includegraphics[width=0.6\textwidth]{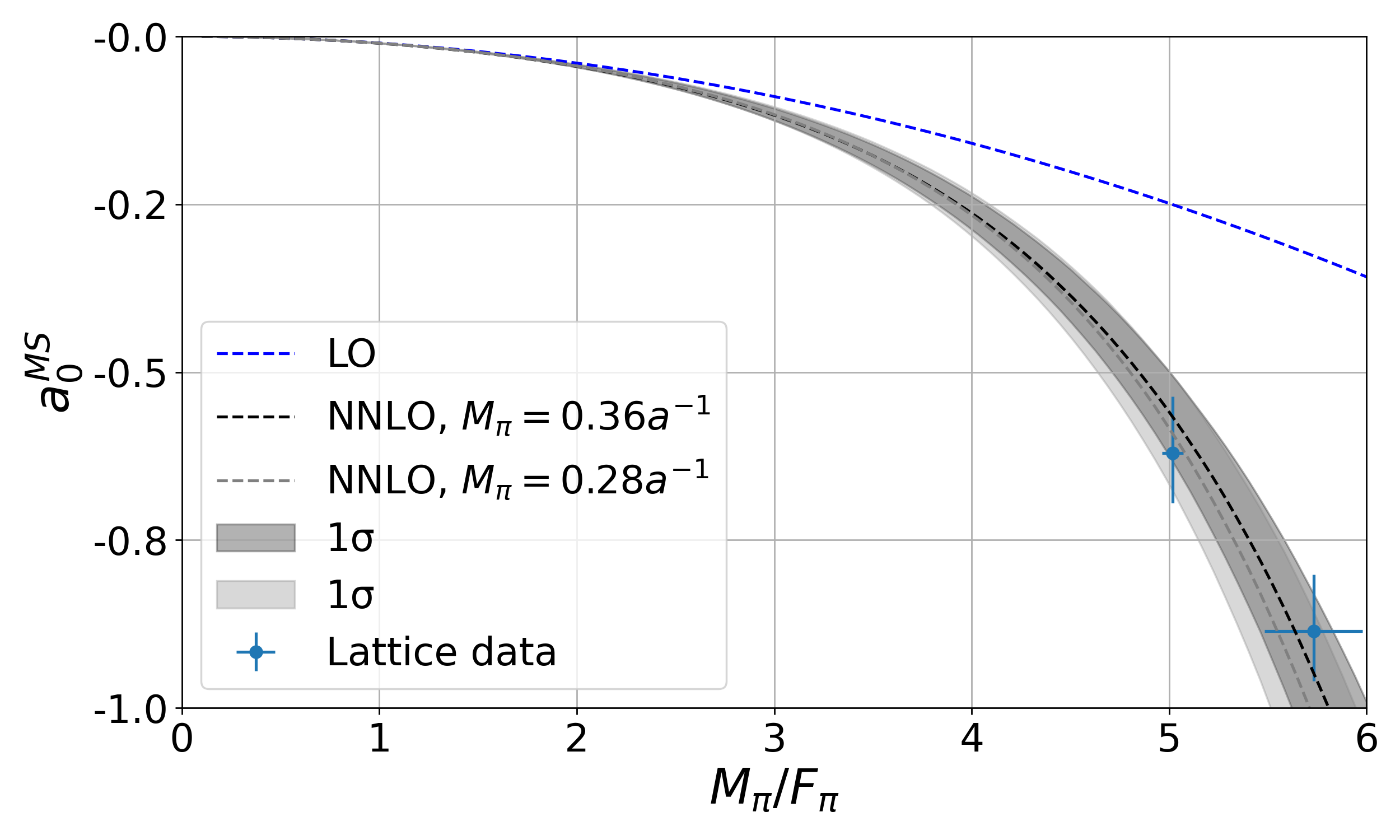}
    \caption{Lattice data \cite{dengler_2024_12920978} and a fit for the scattering length $a_0^\text{MS}$ as a function of $M_\pi/F_\pi$. Two NNLO curves are shown defined with masses corresponding to two data points. The LO result is shown as a reference.}
    \label{figFitScat}
\end{figure}

In figures \ref{figFitMass},~\ref{figFitDec} and~\ref{figFitScat}, the results of the fit is displayed for the pion masses, decay constant, and scattering length respectively.

\subsection{Application: self-interacting dark matter}

Dark pions have been proposed as dark matter candidates in a wide range of scenarios (see, e.g., ~\cite{Ryttov:2008xe,Essig:2009nc,Bai:2010qg,Buckley:2012ky,Frigerio:2012uc,Bhattacharya:2013kma,Cline:2013zca,Hochberg:2014kqa,Carmona:2015haa,Kopp:2016yji,Beauchesne:2018myj,Beauchesne:2019ato,Bernreuther:2019pfb,Contino:2020god,Davighi:2024zip,Alfano:2025non,Davighi:2025awm}). In most realizations, both the freeze-out process and present-day dark matter self-interactions take place in the non-relativistic regime. Furthermore, when the pion mass lies sufficiently below the cutoff scale $4\pi F_\pi$, chiral EFT provides a reliable framework for describing both phenomena. In this section, we concentrate on dark matter self-interactions.

Large dark matter self-interactions are a notable feature of pion dark matter models, as they can potentially alleviate the small-scale structure puzzles~\cite{Tulin:2017ara}. At the same time, overly strong self-interactions are constrained by observations of colliding galaxy clusters, including the Bullet Cluster~\cite{Randall:2008ppe,Robertson:2016xjh,Wittman:2017gxn}. These bounds are therefore crucial in defining the viable parameter space of strongly interacting dark matter scenarios~\cite{Hochberg:2014kqa,Hansen:2015yaa}. The relevance of higher-order corrections in this context was highlighted in~\cite{Hansen:2015yaa}. In \cite{Kolesova:2025ghl}, the NLO prediction for the self-interaction cross section was improved using LECs extracted from an NLO fit. Here, we extend this analysis by incorporating the results of the NNLO fit performed for the \spth theory into the study of self-scattering among degenerate pions.

To obtain the self-interaction cross section in the \spth theory we use the expression for the $2\to2$ amplitude for the degenerate case:
\begin{equation}\label{eq:pheno1}
\begin{split}
    \mathcal{M}^{ab\rightarrow cd} \left( s,t,u\right)  &=\xi^{abcd} B\left( s,t,u\right)  +\xi^{acdb} B\left( t,u,s\right)  +\xi^{adbc} B\left( u,s,t\right)  \\ &+\delta^{ab} \delta^{cd} C\left( s,t,u\right)  +\delta^{ac} \delta^{bd} C\left( t,u,s\right)  +\delta^{ad} \delta^{bc} C\left( u,s,t\right) 
\end{split}
\end{equation}
with the group-theoretical factor
\begin{equation}\label{eq:pheno2}
\xi^{abcd} = \left< X^{a}X^{b}X^{c}X^{d}\right>  +\left< X^{a}X^{d}X^{c}X^{b}\right>   =\frac{1}{2} \left( \delta^{ab} \delta^{cd} -\delta^{ac} \delta^{bd} +\delta^{ad} \delta^{bc} \right).
\end{equation}
The functions $B(u,s,t)$ and $C(u,s,t)$ at the NNLO are given in \cite{Bijnens:2011fm}. The non-relativistic self-scattering cross-section ($s\to4M_\pi^2,\  t\to0$) reads:
\begin{align}\label{eq:pheno11}
  \sigma_{2\to2} =\ & \frac{1}{128\pi N^{2}_{\pi }M^{2}_{\pi }} \sum^{N_{\pi }=5}_{a,b,c,d=1} \left| \mathcal{M}^{ab\rightarrow cd} \right|^{2} \nonumber \\& = 
\frac{x_2^2}{\pi M_\pi^2}\Bigg[
\frac{9}{512} + x_2 \left(
\frac{7}{20}l_4^r
+ \frac{19}{20}l_3^r
+ \frac{7}{20}l_2^r
+ \frac{7}{20}l_1^r
- \frac{451}{5120}L(M_\pi^2)
+ \frac{383}{5120}\pi_{16}
\right) \nonumber\\
&+ x_2^2 \Bigg(
r_\sigma^r
- \frac{56}{5}(l_4^r)^2
- \frac{304}{5}l_3^r l_4^r
+ \frac{256}{5}(l_3^r)^2 
+ \frac{336}{5}l_2^r l_4^r
- \frac{112}{5}l_2^r l_3^r
+ \frac{168}{5}(l_2^r)^2\nonumber \\
&\qquad
+ \frac{336}{5}l_1^r l_4^r
- \frac{112}{5}l_1^r l_3^r
+ \frac{336}{5}l_1^r l_2^r
+ \frac{168}{5}(l_1^r)^2\nonumber \\
&\qquad
- \frac{1113}{80}L(M_\pi^2)\,l_4^r
- \frac{15}{4}L(M_\pi^2)\,l_3^r
- \frac{803}{80}L(M_\pi^2)\,l_2^r
- \frac{131}{16}L(M_\pi^2)\,l_1^r \nonumber\\
&\qquad
+ \frac{64903}{122880}L(M_\pi^2)^2
+ \frac{567}{80}\pi_{16}l_4^r
+ \frac{291}{80}\pi_{16}l_3^r + \frac{131}{16}\pi_{16}l_2^r\nonumber\\
&\qquad
+ \frac{671}{80}\pi_{16}l_1^r
- \frac{49517}{61440}\pi_{16}L(M_\pi^2)
+ \frac{46459}{122880}\pi_{16}^2-\pi^{2} \pi^{2}_{16} \frac{427}{23040} 
\Bigg)
\Bigg],
\end{align}
where the following linear combination of NNLO LECs was introduced:
\begin{align}
\begin{aligned}
r_\sigma^r =\frac{1}{40}\Big(&
-38 K_1^r
+112 K_{10}^r
+7 K_{11}^r
+7 K_{13}^r
+28 K_{14}^r
+28 K_{15}^r
+112 K_{16}^r \\
&-21 K_{17}^r
-84 K_{18}^r
+38 K_{19}^r
-152 K_2^r
+152 K_{20}^r
+152 K_{21}^r
+608 K_{22}^r \\
&+38 K_{23}^r
+21 K_{25}^r
+84 K_{26}^r
+336 K_{27}^r
-76 K_{28}^r
-152 K_{29}^r \\
&+90 K_3^r
+90 K_{31}^r
+104 K_{32}^r
+38 K_{33}^r
+152 K_{35}^r
-90 K_{37}^r \\
&-104 K_{38}^r
+21 K_{39}^r
+104 K_4^r
+84 K_{40}^r
+180 K_5^r
+104 K_6^r \\
&+7 K_7^r
+28 K_8^r
+28 K_9^r
\Big).
\end{aligned}
\end{align}

\begin{figure}[t]
    \centering
    \includegraphics[width=0.6\textwidth]{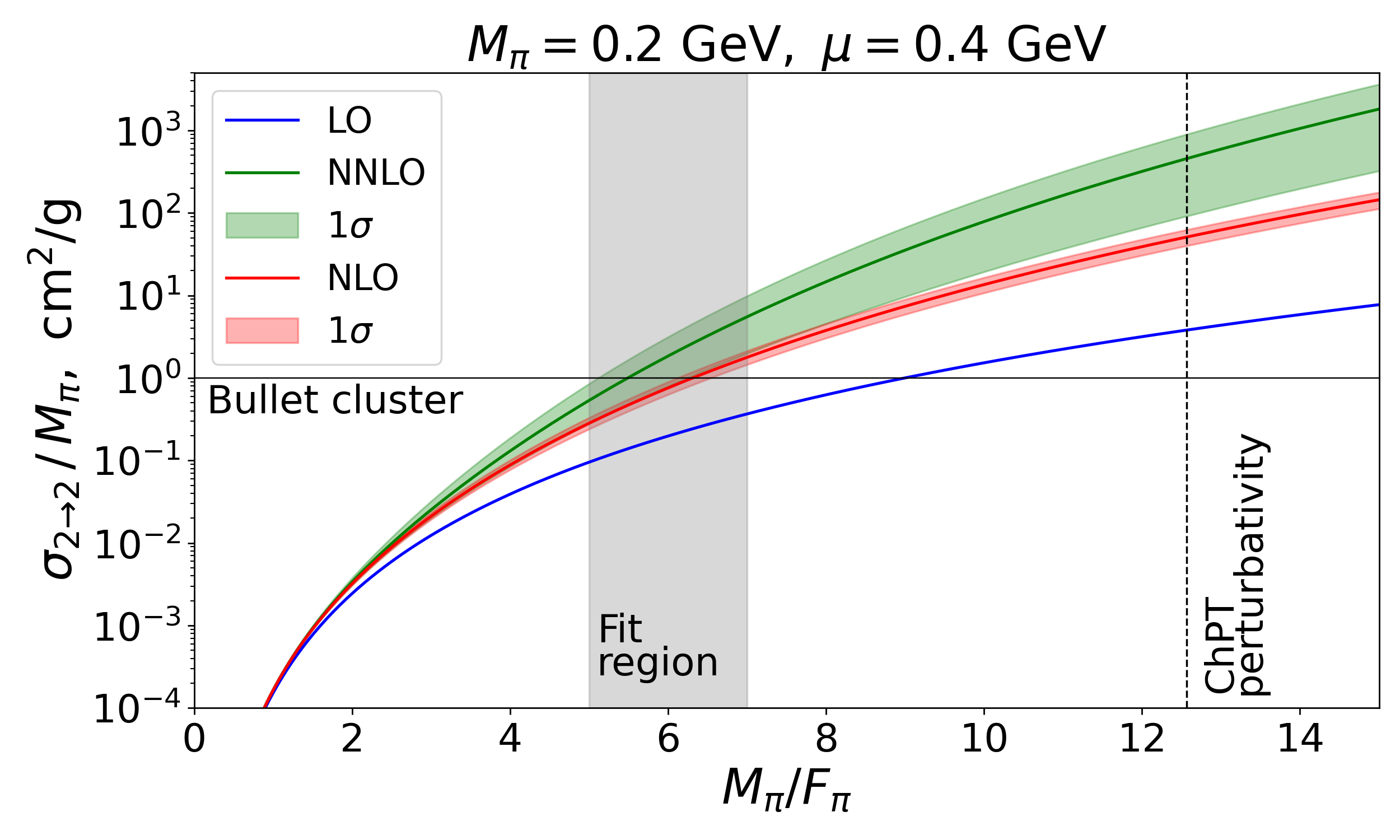} 
    \caption{Dependence of the non-relativistic $2\to 2$ pion cross section \eqref{eq:pheno11} on $M_\pi/F_\pi$ for a fixed (typical SIMP) mass $M_\pi = 0.2\,$GeV. To obtain a plot for a different pion mass, $M'_\pi$, the curves should be multiplied by a factor $(M_\pi/M'_\pi)^3$ (this assumes that the $\mu$ scale is correspondingly adjusted to preserve $M_\pi'/\mu\approx2$ in order to be consistent with the fit of the lattice data). We also depict the constraint on dark matter self-interaction by observation of the Bullet cluster and the bound on ChPT validity, $M_\pi/F_\pi = 4\pi$. Finally, we indicate the region of $M_\pi/F_\pi$ for which the lattice data are available, i.e., where our NNLO treatment should be most accurate. Note that the NNLO curve is plotted with $1\sigma$ error coming from the fits of LECs. 
    }
    \label{fig:Self-scattering}
\end{figure}

%%%%%%%%%%%%%%%%%%%%%%%%%%%%%%%%%%
We present the resulting non-relativistic LO, NLO, and NNLO $2\to2$ pion self-scattering cross sections in figure \ref{fig:Self-scattering} for a representative dark pion mass of $M_\pi = 0.2$ GeV. The renormalization scale is chosen as $\mu = 0.4$ GeV in order to approximately reproduce the ratio $M_\pi/\mu \approx 2$ employed in the fit. The NLO and NNLO predictions depend on the values of the LECs determined in section \ref{sec:fit}, and the corresponding $1\sigma$ uncertainty bands are obtained from the posterior distributions shown in figure \ref{fig:posteriors}. The particular NNLO combination $r_\sigma^r$ was not constrained by the fit; for this coefficient, we assume a normal distribution centered at zero with a characteristic width of $\pi_{16}^2$.

Consistent with the Adler zero principle~\cite{Brauner:2024juy}, the $2\to2$ cross section rapidly decreases in the limit $M_\pi/F_\pi \to 0$. In the considered regime, the scattering process is entirely non-relativistic, and the cross section is non-vanishing only because the pseudo-Goldstone pions acquire a finite mass. Taking the NNLO result as the most reliable prediction, the perturbative expansion indicates an estimated truncation uncertainty of roughly $30\%$ for the LO result at $x \approx 2.1$, and for the NLO result at $x \approx 3.8$. We compare our predictions with the constraint on dark matter self-interactions inferred from observations of the Bullet Cluster, $\sigma_{2\to2}/M_\pi \lesssim 1,\mathrm{cm}^2/\mathrm{g}$~\cite{Randall:2008ppe}. For the benchmark value $M_\pi = 0.2$ GeV, the LO, NLO, and NNLO self-scattering cross sections exceed this bound at approximately $M_\pi/F_\pi \approx 9.0$, $6.3$, and $5.5$, respectively.

Lattice studies indicate that the masses of dark vector mesons can become comparable to the dark pion mass when $M_\pi/F_\pi$ is large~\cite{Hietanen:2014xca,Bennett:2019jzz,Kulkarni:2022bvh}. Such states may substantially affect the freeze-out dynamics (see, e.g., \cite{Berlin:2018tvf,Arthur:2016dir,Bernreuther:2023kcg,SIMP_VM}). Nevertheless, the non-relativistic self-interaction cross section considered here is expected to remain largely unaffected away from the resonant regime $M_\rho \sim 2 M_\pi$, which corresponds roughly to $M_\pi/F_\pi \sim 4$ according to the lattice results of~\cite{Bennett:2019jzz}. In particular, when vector mesons are included explicitly in the effective theory, the tree-level vector-exchange contribution to pion scattering is momentum suppressed and vanishes in the limit $v\to0$~\cite{Choi:2018iit}. 

Note that  \eqref{eq:pheno11} is valid only in the mass-degenerate case. For $r \neq 1$, the singlet pion is the lighter state and is expected to dominate the dark matter abundance. In this case, the self-scattering cross section is reduced. However, in the presence of a mediator coupling to the SM, the singlet pion becomes unstable, making this scenario less viable.

%%%%%%%%%%%%%%%%%%%%%%%%%%%%%%%%%%

%%%%%%%%%%%%%%%%%%%%%%%%%%%%%%%%%%%%%%%%%%%%%%%%%%%%%%%%%%%%%%%%%%%%%%%%%%%%
\section{Conclusions}
In this paper, we considered QCD-like theories with $N_F=2$ fermion flavours in pseudoreal and real representations. For the pseudoreal case with the symmetry-breaking pattern \spth, we presented the reduced NLO Lagrangian. In both cases, we calculated the NLO and NNLO contributions to the pion masses, decay constants, and vacuum condensates for non-degenerate fermion masses. This analysis extends the previous work of \cite{Bijnens:2009qm}, where these expressions were derived for the case of degenerate masses, and \cite{Kolesova:2025ghl}, where the NLO formulas for the non-degenerate case were obtained. The results presented here are in agreement with these earlier works.

We made use of the spectroscopic lattice data from \cite{Kulkarni:2022bvh} and the scattering data from \cite{Dengler:2024maq} for the $Sp(N_c=4)$ gauge theory with two fermion flavours in order to fit the NLO LECs of the reduced \spth\ Lagrangian at NNLO precision. This refines the previous results of \cite{Kolesova:2025ghl}, where not all features of the lattice data could be reproduced at NLO. However, due to a large degeneracy among the NNLO contributions, the NNLO LECs remain poorly constrained given the currently available lattice data.

We further applied the fitted results to confirm the importance of higher-order corrections for the phenomenology of strongly interacting pionic dark matter, especially in the regime of relatively large $M_\pi/F_\pi$. NLO and NNLO corrections give significant contributions to the self-scattering cross section and can  affect the viable dark matter parameter space. 

A natural direction for future work is a more precise determination of the NNLO LECs once additional lattice data become available. Another important extension would be the inclusion of finite-volume effects for the non-degenerate case, extending the NNLO analysis of \cite{Bijnens:2015xba} performed for degenerate fermion masses. Such corrections are particularly relevant for precise comparisons with lattice simulations and for improving the extraction of low-energy constants from lattice data.

\acknowledgments

The authors would also like to thank Helena Kolesova, Yannick Dengler, Axel Maas, and Fabian Zierler for useful discussions. 

D.K. would like to thank Lund University for its hospitality during the completion of this work. 
The work of D.K.\ is supported by the Research Council of Norway under the FRIPRO Young Research Talent grant no.~335388. The stay of D.K.\ at Lund University is supported by the Research Council of Norway through the ``Funding for Research Stays Abroad for Doctoral and Postdoctoral Fellows’’ grant no.~362518.

\clearpage
\appendix
\section{NNLO masses in \soth theory}\label{appSO4Masses}
In this appendix, we present the NNLO contribution to masses in the \soth theory.
\begin{align}%M1 NNLO
&\frac{F^4}{M^2_{1}}M^2_{1,\text{NNLO}} =
M_1^2 \Bigg(
\frac{1}{4}H^F(M_6^2,M_6^2,M_1^2,M_1^2)
+ \frac{1}{4}(M_6^2,M_8^2,M_1^2,M_1^2)
+ \frac{1}{6}H^F(M_6^2,M_1^2,M_1^2,M_1^2)  \nonumber\\
&+ \frac{1}{4}H^F(M_8^2,M_8^2,M_1^2,M_1^2)
+ \frac{1}{6}H^F(M_8^2,M_1^2,M_1^2,M_1^2) + \frac{1}{3}H^F_1(M_6^2,M_1^2,M_1^2,M_1^2) \nonumber\\
&
+ \frac{1}{3}H^F_1(M_8^2,M_1^2,M_1^2,M_1^2) - \frac{1}{3}H^F_1(M_1^2,M_1^2,M_6^2,M_1^2)
- \frac{1}{3}H^F_1(M_1^2,M_1^2,M_8^2,M_1^2)  \nonumber\\
&+ \frac{3}{4}H^F_{21}(M_6^2,M_6^2,M_1^2,M_1^2)
+ \frac{3}{4}H^F_{21}(M_8^2,M_8^2,M_1^2,M_1^2) + \frac{3}{4}H^F_{21}(M_1^2,M_6^2,M_8^2,M_1^2) \nonumber\\
&
+ \frac{3}{4}H^F_{21}(M_1^2,M_6^2,M_1^2,M_1^2)
+ \frac{3}{4}H^F_{21}(M_1^2,M_8^2,M_1^2,M_1^2)
\Bigg)  \nonumber\\
&+ M_1^4 \Big(
-128 L^r_5 L^r_8 -512 L^r_5 L^r_6 + 64 (L^r_5)^2
-512 L^r_4 L^r_8 -2048 L^r_4 L^r_6  \nonumber\\
& +512 L^r_4 L^r_5 +1024 (L^r_4)^2 + r_{M,0}^r
\Big)  \nonumber\\
&+ M_1^4 \pi_{16} \Big(
-\frac{23}{768} -4 L^r_8 -16 L^r_6 +2 L^r_5 +8 L^r_4
+5 L^r_3 +20 L^r_2 +4 L^r_1 +9 L^r_0
\Big) + M_1^4 \pi_{16}^2 \Big(
-\frac{1745}{1152} 
\Big)  \nonumber\\
&+ R_q\,M_1^2 \Big(
\frac{1}{4}H^F(M_6^2,M_6^2,M_1^2,M_1^2)
-\frac{1}{4}H^F(M_8^2,M_8^2,M_1^2,M_1^2)
\Big) + R_q^2 M_1^4 r^r_{M,1}\nonumber \\
&+ R_q^2 M_1^4 \pi_{16} \Big(
-\frac{1}{192} +4 L^r_8 +16 L^r_7 +2 L^r_3 +8 L^r_2 +4 L^r_0
\Big) + R_q^2 M_1^4 \pi_{16}^2 \Big(
-\frac{7}{16} 
\Big) \nonumber\\& + \overline{A}(M_6^2) M_1^2 \Big(
16 L^r_8 +32 L^r_6 -8 L^r_5 -32 L^r_4
+10 L^r_3 +8 L^r_2 +32 L^r_1 +4 L^r_0
\Big) + \overline{A}(M_6^2) M_1^2 \pi_{16} 
\frac{5}{8}  \nonumber\\
&+ \overline{A}(M_6^2) R_q M_1^2 \Big(
16 L^r_8 +32 L^r_6 -8 L^r_5 -32 L^r_4
+10 L^r_3 +8 L^r_2 +32 L^r_1 +4 L^r_0
\Big)  \nonumber \\
&+ (\overline{A}(M_6^2))^2 \Big(
-\frac{1}{4} - \frac{1}{2}(1+R_q)^{-1}
\Big) \nonumber \\
&+ \overline{A}(M_6^2)\overline{A}(M_8^2)\left(-\frac{1}{4}\right) + \overline{A}(M_6^2)\overline{A}(M_1^2) R_q \left(-\frac{1}{8}\right) + \overline{A}(M_6^2)\overline{A}(M_1^2)\left(-\frac{5}{8}\right)  \nonumber\\
&+ \overline{A}(M_8^2) M_1^2 \Big(
16 L^r_8 +32 L^r_6 -8 L^r_5 -32 L^r_4
+10 L^r_3 +8 L^r_2 +32 L^r_1 +4 L^r_0
\Big)  \nonumber\\
&+ \overline{A}(M_8^2) M_1^2 \pi_{16} 
\frac{5}{8} + \overline{A}(M_8^2) R_q M_1^2 \Big(
-16 L^r_8 -32 L^r_6 +8 L^r_5 +32 L^r_4
-10 L^r_3 -8 L^r_2 -32 L^r_1 -4 L^r_0
\Big)  \nonumber\\
&+ (\overline{A}(M_8^2))^2 \Big(
-\frac{1}{4} + \frac{1}{2}( -1 + R_q )^{-1}
\Big) + \overline{A}(M_8^2)\overline{A}(M_1^2) R_q \frac{1}{8} +\overline{A}(M_8^2)\overline{A}(M_1^2)\left(-\frac{5}{8}\right)  \nonumber\\
&+ \overline{A}(M_1^2) M_1^2 \Big(
52 L^r_8 +144 L^r_6 -26 L^r_5 -104 L^r_4
+26 L^r_3 +40 L^r_2 +88 L^r_1 +14 L^r_0
\Big) \nonumber \\
&+ \overline{A}(M_1^2) M_1^2 \pi_{16} 
\frac{33}{16}  + \overline{A}(M_1^2) R_q^2 M_1^2 (-4 L^r_8 -16 L^r_7) + (\overline{A}(M_1^2))^2 \left(-\frac{19}{32}\right),
\end{align}

%M3 NNLO
\begin{align}
    &\frac{F^4}{M^2_{1}}M^2_{3,\text{NNLO}} =
M_1^2 \Bigg(
\frac{1}{4}H^F(M_6^2,M_6^2,M_1^2,M_1^2)
+ \frac{1}{3}H^F(M_6^2,M_1^2,M_1^2,M_1^2)
 \nonumber\\&+ \frac{1}{4}H^F(M_8^2,M_8^2,M_1^2,M_1^2)+ \frac{1}{3}H^F(M_8^2,M_1^2,M_1^2,M_1^2) - \frac{4}{3}H^F_1(M_6^2,M_1^2,M_1^2,M_1^2)
 \nonumber\\&- \frac{4}{3}H^F_1(M_8^2,M_1^2,M_1^2,M_1^2)
+ \frac{4}{3}H^F_1(M_1^2,M_1^2,M_6^2,M_1^2)
+ \frac{4}{3}H^F_1(M_1^2,M_1^2,M_8^2,M_1^2) \nonumber \\& + \frac{3}{2}H^F_{21}(M_6^2,M_1^2,M_1^2,M_1^2)
+ \frac{3}{2}H^F_{21}(M_8^2,M_1^2,M_1^2,M_1^2)
\Bigg) 
 \nonumber\\
&+ M_1^4 \Big(
-128 L^r_5 L^r_8 -512 L^r_5 L^r_6 + 64 (L^r_5)^2
-512 L^r_4 L^r_8 -2048 L^r_4 L^r_6  +512 L^r_4 L^r_5 +1024 (L^r_4)^2 + r^r_{M,0}
\Big)  \nonumber\\
&+ M_1^4 \pi_{16} \Big(
-\frac{11}{384} -4 L^r_8 -16 L^r_6 +2 L^r_5 +8 L^r_4
+5 L^r_3 +20 L^r_2 +4 L^r_1 +9 L^r_0
\Big) + M_1^4 \pi_{16}^2 \Big(
-\frac{179}{144} 
\Big) \nonumber \\
&+ R_q M_1^2 \Big(
\frac{1}{2}H^F(M_6^2,M_6^2,M_1^2,M_1^2)
+ \frac{5}{6}H^F(M_6^2,M_1^2,M_1^2,M_1^2)
- \frac{1}{2}H^F(M_8^2,M_8^2,M_1^2,M_1^2)
 \nonumber\\&- \frac{5}{6}H^F(M_8^2,M_1^2,M_1^2,M_1^2) - \frac{4}{3}H^F_1(M_6^2,M_1^2,M_1^2,M_1^2)
+ \frac{4}{3}H^F_1(M_8^2,M_1^2,M_1^2,M_1^2)\nonumber
\\&+ \frac{4}{3}H^F_1(M_1^2,M_1^2,M_6^2,M_1^2)
- \frac{4}{3}H^F_1(M_1^2,M_1^2,M_8^2,M_1^2)
\Big)  \nonumber\\
&+ R_q^2 M_1^2 \Big(
\frac{1}{4}H^F(M_6^2,M_6^2,M_1^2,M_1^2)
+ \frac{1}{2}H^F(M_6^2,M_1^2,M_1^2,M_1^2)
+ \frac{1}{4}H^F(M_8^2,M_8^2,M_1^2,M_1^2)
 \nonumber\\&+ \frac{1}{2}H^F(M_8^2,M_1^2,M_1^2,M_1^2)
\Big)  + R_q^2 M_1^4 \Big(
-128 L^r_5 L^r_8 -512 L^r_5 L^r_7 -512 L^r_4 L^r_8
-2048 L^r_4 L^r_7 + r^r_{M,2}
\Big)  \nonumber\\
&+ R_q^2 M_1^4 \pi_{16} \Big(
-\frac{19}{384} -52 L^r_8 -16 L^r_7 -64 L^r_6
+24 L^r_5 +32 L^r_4 +4 L^r_3 +8 L^r_2 +4 L^r_0
\Big) + R_q^2 M_1^4 \pi_{16}^2 \Big(
-\frac{37}{8} 
\Big)  \nonumber\\
&+ \overline{A}(M_6^2) M_1^2 \Big(
40 L^r_8 +96 L^r_6 -20 L^r_5 -80 L^r_4
+12 L^r_3 +8 L^r_2 +32 L^r_1 +12 L^r_0
\Big) + \overline{A}(M_6^2) M_1^2 \pi_{16} \Big(
-\frac{1}{8} 
\Big)  \nonumber\\
&+ \overline{A}(M_6^2) R_q M_1^2 \Big(
80 L^r_8 +64 L^r_7 +96 L^r_6 -32 L^r_5 -80 L^r_4
+12 L^r_3 +8 L^r_2 +32 L^r_1 +12 L^r_0
\Big)  \nonumber\\
&+ \overline{A}(M_6^2) R_q M_1^2 \pi_{16} \Big(
-\frac{1}{8} 
\Big) + \overline{A}(M_6^2) R_q^2 M_1^2 (40 L^r_8 +64 L^r_7 -12 L^r_5)  \nonumber\\
& + (\overline{A}(M_6^2))^2 R_q \left(-\frac{1}{2}\right)
+ (\overline{A}(M_6^2))^2 \left(-\frac{3}{4}\right) + \overline{A}(M_6^2)\overline{A}(M_1^2) R_q \left(-\frac{5}{8}\right)
+ \overline{A}(M_6^2)\overline{A}(M_1^2) \left(-\frac{5}{8}\right) \nonumber \\
&+ \overline{A}(M_8^2) M_1^2 \Big(
40 L^r_8 +96 L^r_6 -20 L^r_5 -80 L^r_4
+12 L^r_3 +8 L^r_2 +32 L^r_1 +12 L^r_0
\Big) + \overline{A}(M_8^2) M_1^2 \pi_{16} \Big(
-\frac{1}{8} 
\Big) \nonumber \\
&+ \overline{A}(M_8^2) R_q M_1^2 \Big(
-80 L^r_8 -64 L^r_7 -96 L^r_6 +32 L^r_5 +80 L^r_4
-12 L^r_3 -8 L^r_2 -32 L^r_1 -12 L^r_0
\Big)  \nonumber\\
&+ \overline{A}(M_8^2) R_q M_1^2 \pi_{16} 
\frac{1}{8}
 + \overline{A}(M_8^2) R_q^2 M_1^2 (40 L^r_8 +64 L^r_7 -12 L^r_5) +  (\overline{A}(M_8^2))^2 R_q \frac{1}{2} \nonumber\\
&+ (\overline{A}(M_8^2))^2 \left(-\frac{3}{4}\right)+ \overline{A}(M_8^2)\overline{A}(M_1^2) R_q \frac{5}{8}
+ \overline{A}(M_8^2)\overline{A}(M_1^2) \left(-\frac{5}{8}\right)\nonumber \\
&+ \overline{A}(M_1^2) M_1^2 \Big(
4 L^r_8 +16 L^r_6 -2 L^r_5 -8 L^r_4
+22 L^r_3 +40 L^r_2 +88 L^r_1 -2 L^r_0
\Big)\nonumber \\
&+ \overline{A}(M_1^2) M_1^2 \pi_{16} 
\frac{43}{12} 
 + \overline{A}(M_1^2) R_q^2 M_1^2 (20 L^r_8 +80 L^r_7) + \overline{A}(M_1^2) R_q^2 M_1^2 \pi_{16} \Big(
-\frac{1}{4} 
\Big) \nonumber\\
&+ (\overline{A}(M_1^2))^2 R_q^2 \left(-\frac{5}{4}\right)
+ (\overline{A}(M_1^2))^2 \left(-\frac{19}{32}\right),
\end{align}

%M6 NNLO
\begin{align}
&\frac{F^4}{M^2_{6}}M^2_{6,\text{NNLO}}=
M_1^2 \Bigg(
\frac{5}{8}H^F(M_6^2,M_1^2,M_1^2,M_6^2)
+ \frac{1}{4}H^F(M_8^2,M_1^2,M_1^2,M_6^2)
+ \frac{1}{6}H^F(M_1^2,M_1^2,M_1^2,M_6^2)\nonumber \\
&+ \frac{3}{4}H^F_{21}(M_8^2,M_1^2,M_1^2,M_6^2)
+ \frac{3}{2}H^F_{21}(M_1^2,M_6^2,M_1^2,M_6^2)
+ \frac{3}{4}H^F_{21}(M_1^2,M_1^2,M_1^2,M_6^2)
\Bigg)\nonumber \\
&+ M_1^4 \Big(
-128 L^r_5 L^r_8 -512 L^r_5 L^r_6 + 64 (L^r_5)^2
-512 L^r_4 L^r_8 -2048 L^r_4 L^r_6 \nonumber\\
& +512 L^r_4 L^r_5 +1024 (L^r_4)^2 + r_{M,0}^r
\Big) \nonumber\\
&+ M_1^4 \pi_{16} \Big(
-\frac{41}{1536} -4 L^r_8 -16 L^r_6 +2 L^r_5 +8 L^r_4
+5 L^r_3 +20 L^r_2 +4 L^r_1 +9 L^r_0
\Big) + M_1^4 \pi_{16}^2 \Big(
-\frac{581}{576} 
\Big) \nonumber\\
&+ R_q M_1^2 \Big(
\frac{1}{8}H^F(M_6^2,M_1^2,M_1^2,M_6^2)
- \frac{1}{4}H^F(M_8^2,M_1^2,M_1^2,M_6^2)
- \frac{1}{12}H^F(M_1^2,M_1^2,M_1^2,M_6^2) \nonumber\\
&+ \frac{3}{4}H^F_{21}(M_8^2,M_1^2,M_1^2,M_6^2)
+ \frac{3}{2}H^F_{21}(M_1^2,M_6^2,M_1^2,M_6^2)
+ \frac{3}{4}H^F_{21}(M_1^2,M_1^2,M_1^2,M_6^2)
\Big)\nonumber \\
&+ R_q M_1^4 \Big(
-256 L^r_5 L^r_8 -512 L^r_5 L^r_6 +128 (L^r_5)^2
-512 L^r_4 L^r_8 +512 L^r_4 L^r_5  + r_{M,3}^r
\Big) \nonumber\\
&+ R_q M_1^4 \pi_{16} \Big(
-\frac{3}{256} +4 L^r_3 +4 L^r_2 +8 L^r_1 +8 L^r_0
\Big) + R_q M_1^4 \pi_{16}^2 
\frac{103}{288} 
 \nonumber\\
&+ R_q^2 M_1^4 \Big(
-128 L^r_5 L^r_8 +64 (L^r_5)^2 + r_{M,4}^r
\Big)\nonumber \\
&+ R_q^2 M_1^4 \pi_{16} \Big(
-\frac{7}{1536} -4 L^r_8 -16 L^r_7 +2 L^r_3 +10 L^r_2
+4 L^r_1 +4 L^r_0
\Big) + R_q^2 M_1^4 \pi_{16}^2 
\frac{265}{576}  
\nonumber\\
&+ \overline{A}(M_6^2) M_1^2 \Big(
32 L^r_8 +64 L^r_6 -16 L^r_5 -32 L^r_4
+20 L^r_3 +28 L^r_2 +40 L^r_1 +8 L^r_0
\Big) + \overline{A}(M_6^2) M_1^2 \pi_{16} 
\frac{31}{12} 
 \nonumber\\
&+ \overline{A}(M_6^2) R_q M_1^2 \Big(
32 L^r_8 +64 L^r_6 -16 L^r_5 -32 L^r_4
+20 L^r_3 +28 L^r_2 +40 L^r_1 +8 L^r_0
\Big) + \overline{A}(M_6^2) R_q M_1^2 \pi_{16} 
\frac{25}{12}
 \nonumber\\
&+ (\overline{A}(M_6^2))^2 \Big(
\frac{7}{16} - \frac{1}{4}(1+R_q)^{-1}
\Big) + \overline{A}(M_6^2)\overline{A}(M_1^2) R_q \frac{1}{8}
+ \overline{A}(M_6^2)\overline{A}(M_1^2) \left(-\frac{3}{8}\right)\nonumber \\
&+ \overline{A}(M_8^2) M_1^2 \Big(
32 L^r_6 -32 L^r_4 +8 L^r_2 +32 L^r_1
\Big)+ \overline{A}(M_8^2) M_1^2 \pi_{16} \Big(
-\frac{1}{8} 
\Big)\nonumber \\
&+ \overline{A}(M_8^2) R_q M_1^2 \Big(
-32 L^r_6 +32 L^r_4 -8 L^r_2 -32 L^r_1
\Big) + \overline{A}(M_8^2) R_q M_1^2 \pi_{16} 
\frac{1}{8}\nonumber\\
&+ (\overline{A}(M_8^2))^2 \Big(
-\frac{1}{8} 
\Big) + \overline{A}(M_8^2)\overline{A}(M_1^2) R_q \left(-\frac{1}{8}\right)
+ \overline{A}(M_8^2)\overline{A}(M_1^2) \frac{1}{8}\nonumber\\
&+ \overline{A}(M_1^2) M_1^2 \Big(
52 L^r_8 +112 L^r_6 -26 L^r_5 -104 L^r_4
+26 L^r_3 +20 L^r_2 +80 L^r_1 +14 L^r_0
\Big)\nonumber \\
&+ \overline{A}(M_1^2) M_1^2 \pi_{16} 
\frac{19}{16} 
+ \overline{A}(M_1^2) R_q M_1^2 (16 L^r_8 +32 L^r_7 -4 L^r_5) \nonumber\\
&+ \overline{A}(M_1^2) R_q M_1^2 \pi_{16} \Big(
-\frac{1}{4} 
\Big) + \overline{A}(M_1^2) R_q^2 M_1^2 (4 L^r_8 +16 L^r_7) \nonumber\\
&+ (\overline{A}(M_1^2))^2 R_q \left(-\frac{1}{2}\right)
+ (\overline{A}(M_1^2))^2 \left(-\frac{89}{32}\right),
\end{align}

%M8 NNLO
\begin{align}
&\frac{F^4}{M^2_{8}}M^2_{8,\text{NNLO}}=
M_1^2 \Bigg(
\frac{1}{4}H^F(M_6^2,M_1^2,M_1^2,M_8^2)
+ \frac{5}{8}H^F(M_8^2,M_1^2,M_1^2,M_8^2)
+ \frac{1}{6}H^F(M_1^2,M_1^2,M_1^2,M_8^2) \nonumber\\
&+ \frac{3}{4}H^F_{21}(M_6^2,M_1^2,M_1^2,M_8^2)
+ \frac{3}{2}H^F_{21}(M_1^2,M_8^2,M_1^2,M_8^2)
+ \frac{3}{4}H^F_{21}(M_1^2,M_1^2,M_1^2,M_8^2)
\Bigg)\nonumber \\
&+ M_1^4 \Big(
-128 L^r_5 L^r_8 -512 L^r_5 L^r_6 + 64 (L^r_5)^2
-512 L^r_4 L^r_8 -2048 L^r_4 L^r_6  +512 L^r_4 L^r_5 +1024 (L^r_4)^2 + r_{M,0}^r
\Big) \nonumber\\
&+ M_1^4 \pi_{16} \Big(
-\frac{41}{1536} -4 L^r_8 -16 L^r_6 +2 L^r_5 +8 L^r_4
+5 L^r_3 +20 L^r_2 +4 L^r_1 +9 L^r_0
\Big) + M_1^4 \pi_{16}^2 \Big(
-\frac{581}{576} 
\Big) \nonumber\\
&+ R_q M_1^2 \Big(
\frac{1}{4}H^F(M_6^2,M_1^2,M_1^2,M_8^2)
- \frac{1}{8}H^F(M_8^2,M_1^2,M_1^2,M_8^2)
+ \frac{1}{12}H^F(M_1^2,M_1^2,M_1^2,M_8^2) \nonumber\\
&- \frac{3}{4}H^F_{21}(M_6^2,M_1^2,M_1^2,M_8^2)
- \frac{3}{2}H^F_{21}(M_1^2,M_8^2,M_1^2,M_8^2)
- \frac{3}{4}H^F_{21}(M_1^2,M_1^2,M_1^2,M_8^2)
\Big) \nonumber\\
&+ R_q M_1^4 \Big(
256 L^r_5 L^r_8 +512 L^r_5 L^r_6 -128 (L^r_5)^2
+512 L^r_4 L^r_8 -512 L^r_4 L^r_5 -  r_{M,3}^r
\Big) \nonumber\\
&+ R_q M_1^4 \pi_{16} \Big(
\frac{3}{256} -4 L^r_3 -4 L^r_2 -8 L^r_1 -8 L^r_0
\Big) + R_q M_1^4 \pi_{16}^2 \Big(
-\frac{103}{288} 
\Big) \nonumber\\
&+ R_q^2 M_1^4 \Big(
-128 L^r_5 L^r_8 +64 (L^r_5)^2 + r_{M,4}^r
\Big) \nonumber\\
&+ R_q^2 M_1^4 \pi_{16} \Big(
-\frac{7}{1536} -4 L^r_8 -16 L^r_7 +2 L^r_3 +10 L^r_2
+4 L^r_1 +4 L^r_0
\Big) + R_q^2 M_1^4 \pi_{16}^2 
\frac{265}{576} 
 \nonumber\\
&+ \overline{A}(M_6^2) M_1^2 (32 L^r_6 -32 L^r_4 +8 L^r_2 +32 L^r_1) + \overline{A}(M_6^2) M_1^2 \pi_{16} \Big(
-\frac{1}{8} 
\Big) \nonumber\\
&+ \overline{A}(M_6^2) R_q M_1^2 (32 L^r_6 -32 L^r_4 +8 L^r_2 +32 L^r_1) + \overline{A}(M_6^2) R_q M_1^2 \pi_{16} \Big(
-\frac{1}{8} 
\Big) \nonumber\\
&+ (\overline{A}(M_6^2))^2 \Big(
-\frac{1}{8}
\Big) + \overline{A}(M_6^2)\overline{A}(M_1^2) R_q \frac{1}{8}
+ \overline{A}(M_6^2)\overline{A}(M_1^2) \frac{1}{8}\nonumber\\
&+ \overline{A}(M_8^2) M_1^2 \Big(
32 L^r_8 +64 L^r_6 -16 L^r_5 -32 L^r_4
+20 L^r_3 +28 L^r_2 +40 L^r_1 +8 L^r_0
\Big) + \overline{A}(M_8^2) M_1^2 \pi_{16} 
\frac{31}{12}
\nonumber \\
&+ \overline{A}(M_8^2) R_q M_1^2 \Big(
-32 L^r_8 -64 L^r_6 +16 L^r_5 +32 L^r_4
-20 L^r_3 -28 L^r_2 -40 L^r_1 -8 L^r_0
\Big)\nonumber \\
&+ \overline{A}(M_8^2) R_q M_1^2 \pi_{16} \Big(
-\frac{25}{12}
\Big) + (\overline{A}(M_8^2))^2 \Big(
\frac{7}{16} + \frac{1}{4}(-1+R_q)^{-1}
\Big)\nonumber \\
&+ \overline{A}(M_8^2)\overline{A}(M_1^2) R_q \left(-\frac{1}{8}\right)
+ \overline{A}(M_8^2)\overline{A}(M_1^2) \left(-\frac{3}{8}\right)\nonumber \\
&+ \overline{A}(M_1^2) M_1^2 \Big(
52 L^r_8 +112 L^r_6 -26 L^r_5 -104 L^r_4
+26 L^r_3 +20 L^r_2 +80 L^r_1 +14 L^r_0
\Big) + \overline{A}(M_1^2) M_1^2 \pi_{16} 
\frac{19}{16} 
 \nonumber\\
&+ \overline{A}(M_1^2) R_q M_1^2 (-16 L^r_8 -32 L^r_7 +4 L^r_5) + \overline{A}(M_1^2) R_q M_1^2 \pi_{16} 
\frac{1}{4} 
\nonumber \\
&+ \overline{A}(M_1^2) R_q^2 M_1^2 (4 L^r_8 +16 L^r_7) + (\overline{A}(M_1^2))^2 R_q \frac{1}{2}
+ (\overline{A}(M_1^2))^2 \left(-\frac{89}{32}\right).
\end{align}

In these formulas, the linear combinations \eqref{rM0}–\eqref{rM2} of the NNLO LECs were used, together with the following additional combinations:
\begin{align}
r_{M,3}^r&=-32 \Big(
2 K_{17}^r + 4 K_{18}^r + K_{19}^r + 2 K_{20}^r + K_{23}^r
- 3 K_{25}^r - 4 K_{26}^r - 2 K_{39}^r - 4 K_{40}^r
\Big),\\
r_{M,4}^r&=-16 \Big(
2 K_{17}^r + K_{19}^r + 4 K_{21}^r + K_{23}^r
- 3 K_{25}^r - 4 K_{26}^r - 2 K_{39}^r
\Big).
\end{align}
%%%%%%%%%%%%%%%%%%%%%%%%%%%%%%%%%%%%%%%%  

\section{NNLO decay constants in \soth theory}\label{appSO4Decay}
In this appendix, we present the NNLO contribution to decay constants in the \soth theory.

%F1
\begin{align}
&F^3 F_{1,\text{NNLO}} =
M_1^2 \Big(
-\frac{1}{8}H^F(M_6^2,M_6^2,M_1^2,M_1^2)
-\frac{1}{8}H^F(M_6^2,M_8^2,M_1^2,M_1^2)
-\frac{1}{8}H^F(M_6^2,M_1^2,M_1^2,M_1^2)\nonumber \\
&
-\frac{1}{8}H^F(M_8^2,M_8^2,M_1^2,M_1^2)
-\frac{1}{8}H^F(M_8^2,M_1^2,M_1^2,M_1^2)
\Big) \nonumber\\
&+ M_1^4 \Big(
r_{F,0}^r - 8 (L^r_5)^2 - 64 L^r_4 L^r_5 - 128 (L^r_4)^2\nonumber \\
& + \frac{1}{8} H^{F\prime}(M_6^2,M_6^2,M_1^2,M_1^2)
+ \frac{1}{8} H^{F\prime}(M_6^2,M_8^2,M_1^2,M_1^2)
+ \frac{1}{12} H^{F\prime}(M_6^2,M_1^2,M_1^2,M_1^2) \nonumber\\
&+ \frac{1}{8} H^{F\prime}(M_8^2,M_8^2,M_1^2,M_1^2)
+ \frac{1}{12} H^{F\prime}(M_8^2,M_1^2,M_1^2,M_1^2)+ \frac{1}{6} H^{F\prime}_{1}(M_6^2,M_1^2,M_1^2,M_1^2) \nonumber \\
&
+ \frac{1}{6} H^{F\prime}_{1}(M_8^2,M_1^2,M_1^2,M_1^2)  - \frac{1}{6} H^{F\prime}_{1}(M_1^2,M_6^2,M_1^2,M_1^2)
- \frac{1}{6} H^{F\prime}_{1}(M_1^2,M_8^2,M_1^2,M_1^2)\nonumber \\
&+ \frac{3}{8} H^{F\prime}_{21}(M_6^2,M_6^2,M_1^2,M_1^2)
+ \frac{3}{8} H^{F\prime}_{21}(M_8^2,M_8^2,M_1^2,M_1^2)
+ \frac{3}{8} H^{F\prime}_{21}(M_1^2,M_6^2,M_8^2,M_1^2)\nonumber \\
&+ \frac{3}{8} H^{F\prime}_{21}(M_1^2,M_6^2,M_1^2,M_1^2)
+ \frac{3}{8} H^{F\prime}_{21}(M_1^2,M_8^2,M_1^2,M_1^2)
\Big)\nonumber \\
&+ M_1^4 \pi_{16} \Big(
\frac{5}{512} - 16 L^r_8 - 64 L^r_6 + 8 L^r_5 + 32 L^r_4
- \frac{5}{2}L^r_3 - 10 L^r_2 - 2 L^r_1 - \frac{9}{2}L^r_0
\Big)+ M_1^4 \pi_{16}^2 \frac{341}{768}\nonumber \\
&+ R_q M_1^2 \Big(
-\frac{1}{8}H^F(M_6^2,M_6^2,M_1^2,M_1^2)
+ \frac{1}{8}H^F(M_8^2,M_8^2,M_1^2,M_1^2)
\Big) \nonumber\\
&+ R_q M_1^4 \Big(
\frac{1}{8} H^{F\prime}(M_6^2,M_6^2,M_1^2,M_1^2)
- \frac{1}{8} H^{F\prime}(M_8^2,M_8^2,M_1^2,M_1^2)
\Big)\nonumber \\
&+ R_q^2 M_1^4 (r_{F,1}^r)
+ R_q^2 M_1^4 \pi_{16} \Big(
\frac{1}{384} - 12 L^r_8 - 16 L^r_7 + 4 L^r_5
- L^r_3 - 4 L^r_2 - 2 L^r_0
\Big) + R_q^2 M_1^4 \pi_{16}^2 \frac{7}{32} \nonumber\\
&+ \overline{A}(M_6^2) M_1^2 \Big(
4 L^r_8 + 16 L^r_6 - L^r_5 - 4 L^r_4
- 5 L^r_3 - 4 L^r_2 - 16 L^r_1 - 2 L^r_0
\Big) \nonumber\\
&+ \overline{A}(M_6^2) M_1^2 \pi_{16} \left(-\frac{7}{16}\right)
+ \overline{A}(M_6^2) R_q M_1^2 \Big(
4 L^r_8 - 2 L^r_5 + 8 L^r_4
- 5 L^r_3 - 4 L^r_2 - 16 L^r_1 - 2 L^r_0
\Big) \nonumber\\
&+ \overline{A}(M_6^2) R_q M_1^2 \pi_{16} \left(-\frac{1}{16}\right) + (\overline{A}(M_6^2))^2 \Big(
\frac{3}{32} + \frac{1}{8}(1+R_q)^{-1}
\Big)
+ \overline{A}(M_6^2)\overline{A}(M_8^2)\frac{1}{16}\nonumber\\
&+ \overline{A}(M_6^2)\overline{A}(M_1^2) R_q \frac{1}{8}
+ \overline{A}(M_6^2)\overline{A}(M_1^2) \frac{5}{16} \nonumber\\
&+ \overline{A}(M_8^2) M_1^2 \Big(
4 L^r_8 + 16 L^r_6 - L^r_5 - 4 L^r_4
- 5 L^r_3 - 4 L^r_2 - 16 L^r_1 - 2 L^r_0
\Big) \nonumber\\
&+ \overline{A}(M_8^2) M_1^2 \pi_{16} \left(-\frac{7}{16}\right)
+ \overline{A}(M_8^2) R_q M_1^2 \Big(
-4 L^r_8 + 2 L^r_5 - 8 L^r_4
+ 5 L^r_3 + 4 L^r_2 + 16 L^r_1 + 2 L^r_0
\Big)\nonumber \\
&+ \overline{A}(M_8^2) R_q M_1^2 \pi_{16} \frac{1}{16} + (\overline{A}(M_8^2))^2 \Big(
\frac{3}{32} - \frac{1}{8}(-1+R_q)^{-1}
\Big)
+ \overline{A}(M_8^2)\overline{A}(M_1^2) R_q \left(-\frac{1}{8}\right)
\nonumber\\
&+ \overline{A}(M_8^2)\overline{A}(M_1^2) \frac{5}{16} + \overline{A}(M_1^2) M_1^2 \Big(
8 L^r_8 + 32 L^r_6 - L^r_5 - 4 L^r_4
- 13 L^r_3 - 20 L^r_2 - 44 L^r_1 - 7 L^r_0
\Big) \nonumber\\
&+ \overline{A}(M_1^2) M_1^2 \pi_{16} \left(-\frac{13}{12}\right)
+ \overline{A}(M_1^2) R_q^2 M_1^2 (4 L^r_8 + 16 L^r_7)
+ (\overline{A}(M_1^2))^2 \frac{1}{8},
\end{align}

%F3  
\begin{align}
&F^3 F_{3,\text{NNLO}} =
M_1^2 \Big(
-\frac{1}{4}H^F(M_6^2,M_1^2,M_1^2,M_1^2)
-\frac{1}{4}H^F(M_8^2,M_1^2,M_1^2,M_1^2)
\Big) \nonumber\\
&+ M_1^4 \Big(
r_{F,0}^r - 8 (L^r_5)^2 - 64 L^r_4 L^r_5 - 128 (L^r_4)^2  + \frac{1}{8} H^{F\prime}(M_6^2,M_6^2,M_1^2,M_1^2)
+ \frac{1}{6} H^{F\prime}(M_6^2,M_1^2,M_1^2,M_1^2)\nonumber\\&
+ \frac{1}{8} H^{F\prime}(M_8^2,M_8^2,M_1^2,M_1^2)
+ \frac{1}{6} H^{F\prime}(M_8^2,M_1^2,M_1^2,M_1^2)  - \frac{2}{3} H^{F\prime}_{1}(M_6^2,M_1^2,M_1^2,M_1^2)\nonumber\\
&
- \frac{2}{3} H^{F\prime}_{1}(M_8^2,M_1^2,M_1^2,M_1^2)
+ \frac{2}{3} H^{F\prime}_{1}(M_1^2,M_6^2,M_1^2,M_1^2)
+ \frac{2}{3} H^{F\prime}_{1}(M_1^2,M_8^2,M_1^2,M_1^2) \nonumber\\
& + \frac{3}{4} H^{F\prime}_{21}(M_6^2,M_1^2,M_1^2,M_1^2)
+ \frac{3}{4} H^{F\prime}_{21}(M_8^2,M_1^2,M_1^2,M_1^2)
\Big) \nonumber\\
&+ M_1^4 \pi_{16} \Big(
\frac{1}{128} - 16 L^r_8 - 64 L^r_6 + 8 L^r_5 + 32 L^r_4
- \frac{5}{2}L^r_3 - 10 L^r_2 - 2 L^r_1 - \frac{9}{2}L^r_0
\Big) + M_1^4 \pi_{16}^2 \frac{11}{48} \nonumber\\
&+ R_q M_1^2 \Big(
-\frac{1}{4}H^F(M_6^2,M_1^2,M_1^2,M_1^2)
+ \frac{1}{4}H^F(M_8^2,M_1^2,M_1^2,M_1^2)
\Big) + R_q M_1^4 \Big(
\frac{1}{4} H^{F\prime}(M_6^2,M_6^2,M_1^2,M_1^2) \nonumber\\&+ \frac{5}{12} H^{F\prime}(M_6^2,M_1^2,M_1^2,M_1^2)  - \frac{1}{4} H^{F\prime}(M_8^2,M_8^2,M_1^2,M_1^2)
- \frac{5}{12} H^{F\prime}(M_8^2,M_1^2,M_1^2,M_1^2) \nonumber\\
& - \frac{2}{3} H^{F\prime}_{1}(M_6^2,M_1^2,M_1^2,M_1^2)
+ \frac{2}{3} H^{F\prime}_{1}(M_8^2,M_1^2,M_1^2,M_1^2)
+ \frac{2}{3} H^{F\prime}_{1}(M_1^2,M_6^2,M_1^2,M_1^2)
 \nonumber\\
& - \frac{2}{3} H^{F\prime}_{1}(M_1^2,M_8^2,M_1^2,M_1^2)
\Big)+ R_q^2 M_1^4 \Big(
r_{F,2}^r
+ \frac{1}{8} H^{F\prime}(M_6^2,M_6^2,M_1^2,M_1^2)
+ \frac{1}{4} H^{F\prime}(M_6^2,M_1^2,M_1^2,M_1^2)\nonumber \\
& + \frac{1}{8} H^{F\prime}(M_8^2,M_8^2,M_1^2,M_1^2)
+ \frac{1}{4} H^{F\prime}(M_8^2,M_1^2,M_1^2,M_1^2)
\Big) \nonumber\\
&+ R_q^2 M_1^4 \pi_{16} \Big(
\frac{1}{384} - 2 L^r_3 - 4 L^r_2 - 2 L^r_0
\Big)
+ R_q^2 M_1^4 \pi_{16}^2 \frac{3}{16}\nonumber \\
&+ \overline{A}(M_6^2) M_1^2 (2 L^r_5 + 8 L^r_4 - 6 L^r_3 - 4 L^r_2 - 16 L^r_1 - 6 L^r_0)\nonumber \\
&+ \overline{A}(M_6^2) R_q M_1^2 (2 L^r_5 + 8 L^r_4 - 6 L^r_3 - 4 L^r_2 - 16 L^r_1 - 6 L^r_0)
+ (\overline{A}(M_6^2))^2 \frac{1}{8} \nonumber\\
&+ \overline{A}(M_8^2) M_1^2 (2 L^r_5 + 8 L^r_4 - 6 L^r_3 - 4 L^r_2 - 16 L^r_1 - 6 L^r_0)\nonumber \\
&+ \overline{A}(M_8^2) R_q M_1^2 (-2 L^r_5 - 8 L^r_4 + 6 L^r_3 + 4 L^r_2 + 16 L^r_1 + 6 L^r_0)
+ (\overline{A}(M_8^2))^2 \frac{1}{8} \nonumber\\
&+ \overline{A}(M_1^2) M_1^2 \Big(
16 L^r_8 + 64 L^r_6 - 7 L^r_5 - 28 L^r_4
- 11 L^r_3 - 20 L^r_2 - 44 L^r_1 + L^r_0
\Big) \nonumber\\
&+ \overline{A}(M_1^2) M_1^2 \pi_{16} \left(-\frac{73}{32}\right)
+ (\overline{A}(M_1^2))^2 \frac{5}{8},
\end{align}

%F6
\begin{align}
&F^3 F_{6,\text{NNLO}} =
M_1^2 \Big(
-\frac{1}{4}H^F(M_6^2,M_1^2,M_1^2,M_6^2)
-\frac{1}{8}H^F(M_8^2,M_1^2,M_1^2,M_6^2)
-\frac{1}{8}H^F(M_1^2,M_1^2,M_1^2,M_6^2)
\Big) \nonumber\\
&+ M_1^4 \Big(
r_{F,0}^r - 8 (L^r_5)^2 - 64 L^r_4 L^r_5 - 128 (L^r_4)^2  + \frac{5}{16} H^{F\prime}(M_6^2,M_1^2,M_1^2,M_6^2)
+ \frac{1}{8} H^{F\prime}(M_8^2,M_1^2,M_1^2,M_6^2)\nonumber\\
&
+ \frac{1}{12} H^{F\prime}(M_1^2,M_1^2,M_1^2,M_6^2)  + \frac{3}{8} H^{F\prime}_{21}(M_8^2,M_1^2,M_1^2,M_6^2)
+ \frac{3}{4} H^{F\prime}_{21}(M_1^2,M_6^2,M_1^2,M_6^2)
\nonumber\\
&+ \frac{3}{8} H^{F\prime}_{21}(M_1^2,M_1^2,M_1^2,M_6^2)
\Big) + M_1^4 \pi_{16} \Big(
\frac{1}{128} - 16 L^r_8 - 64 L^r_6 + 8 L^r_5 + 32 L^r_4
- \frac{5}{2}L^r_3 - 10 L^r_2 - 2 L^r_1 - \frac{9}{2}L^r_0
\Big)\nonumber\\& + M_1^4 \pi_{16}^2 \frac{109}{384}+ R_q M_1^2 \frac{1}{8}H^F(M_8^2,M_1^2,M_1^2,M_6^2) + R_q M_1^4 \Big(
r_{F,3}^r - 16 (L^r_5)^2 - 64 L^r_4 L^r_5 \nonumber 
\end{align}
\begin{align}
&+ \frac{3}{8} H^{F\prime}(M_6^2,M_1^2,M_1^2,M_6^2)
+ \frac{1}{24} H^{F\prime}(M_1^2,M_1^2,M_1^2,M_6^2)  + \frac{3}{4} H^{F\prime}_{21}(M_8^2,M_1^2,M_1^2,M_6^2)
\nonumber\\
&+ \frac{3}{2} H^{F\prime}_{21}(M_1^2,M_6^2,M_1^2,M_6^2)
+ \frac{3}{4} H^{F\prime}_{21}(M_1^2,M_1^2,M_1^2,M_6^2)
\Big) \nonumber\\
&+ R_q M_1^4 \pi_{16} \Big(
-\frac{1}{768} - 16 L^r_8 - 32 L^r_6 + 8 L^r_5 + 16 L^r_4
- 2 L^r_3 - 2 L^r_2 - 4 L^r_1 - 4 L^r_0
\Big) + R_q M_1^4 \pi_{16}^2 \left(-\frac{17}{192}\right) \nonumber\\
&+ R_q^2 M_1^4 \Big(
r_{F,4}^r - 8 (L^r_5)^2
+ \frac{1}{16} H^{F\prime}(M_6^2,M_1^2,M_1^2,M_6^2)
- \frac{1}{8} H^{F\prime}(M_8^2,M_1^2,M_1^2,M_6^2)
\nonumber\\&- \frac{1}{24} H^{F\prime}(M_1^2,M_1^2,M_1^2,M_6^2) + \frac{3}{8} H^{F\prime}_{21}(M_8^2,M_1^2,M_1^2,M_6^2)
+ \frac{3}{4} H^{F\prime}_{21}(M_1^2,M_6^2,M_1^2,M_6^2)
\nonumber\\&+ \frac{3}{8} H^{F\prime}_{21}(M_1^2,M_1^2,M_1^2,M_6^2)
\Big) + R_q^2 M_1^4 \pi_{16} \Big(
\frac{1}{1536} - 8 L^r_8 + 4 L^r_5 - L^r_3 - 5 L^r_2 - 2 L^r_1 - 2 L^r_0
\Big) \nonumber\\
&+ R_q^2 M_1^4 \pi_{16}^2 \frac{31}{384} + \overline{A}(M_6^2) M_1^2 \Big(
8 L^r_8 + 32 L^r_6 - 2 L^r_5 - 16 L^r_4
- 10 L^r_3 - 14 L^r_2 - 20 L^r_1 - 4 L^r_0
\Big) \nonumber\\
&+ \overline{A}(M_6^2) M_1^2 \pi_{16} \left(-\frac{23}{16}\right) + \overline{A}(M_6^2) R_q M_1^2 \Big(
8 L^r_8 - 2 L^r_5 + 8 L^r_4
- 10 L^r_3 - 14 L^r_2 - 20 L^r_1 - 4 L^r_0
\Big) \nonumber\\
&+ \overline{A}(M_6^2) R_q M_1^2 \pi_{16} \left(-\frac{19}{16}\right) + (\overline{A}(M_6^2))^2 \Big(-\frac{1}{4} + \frac{1}{8}(1+R_q)^{-1}\Big)
+ \overline{A}(M_6^2)\overline{A}(M_1^2)\frac{1}{8}\nonumber\\
&+ \overline{A}(M_8^2) M_1^2 (8 L^r_4 - 4 L^r_2 - 16 L^r_1)
+ \overline{A}(M_8^2) R_q M_1^2 (-8 L^r_4 + 4 L^r_2 + 16 L^r_1)+ (\overline{A}(M_8^2))^2 \frac{1}{16} \nonumber\\
&+ \overline{A}(M_1^2) M_1^2 \Big(
8 L^r_8 + 32 L^r_6 - L^r_5 - 4 L^r_4
- 13 L^r_3 - 10 L^r_2 - 40 L^r_1 - 7 L^r_0
\Big)\nonumber \\
&+ \overline{A}(M_1^2) M_1^2 \pi_{16} \left(-\frac{7}{8}\right) + \overline{A}(M_1^2) R_q M_1^2 3 L^r_5
+ \overline{A}(M_1^2) R_q M_1^2 \pi_{16} \left(-\frac{1}{8}\right)\nonumber\\
& + (\overline{A}(M_1^2))^2 R_q \left(-\frac{1}{8}\right)
+ (\overline{A}(M_1^2))^2 \frac{13}{16},
\end{align}

%F8
\begin{align}
&F^3 F_{8,\text{NNLO}} =
M_1^2 \Big(
-\frac{1}{8}H^F(M_6^2,M_1^2,M_1^2,M_8^2)
-\frac{1}{4}H^F(M_8^2,M_1^2,M_1^2,M_8^2)\nonumber
\\&-\frac{1}{8}H^F(M_1^2,M_1^2,M_1^2,M_8^2)
\Big)+ M_1^4 \Big(
r_{F,0}^r - 8 (L^r_5)^2 - 64 L^r_4 L^r_5 - 128 (L^r_4)^2\nonumber \\
&+ \frac{1}{8}H^{F\prime}(M_6^2,M_1^2,M_1^2,M_8^2)
+ \frac{5}{16} H^{F\prime}(M_8^2,M_1^2,M_1^2,M_8^2)
+ \frac{1}{12} H^{F\prime}(M_1^2,M_1^2,M_1^2,M_8^2) \nonumber\\
& + \frac{3}{8} H^{F\prime}_{21}(M_6^2,M_1^2,M_1^2,M_8^2)
+ \frac{3}{4} H^{F\prime}_{21}(M_1^2,M_8^2,M_1^2,M_8^2)
+ \frac{3}{8} H^{F\prime}_{21}(M_1^2,M_1^2,M_1^2,M_8^2)
\Big)\nonumber \\
&+ M_1^4 \pi_{16} \Big(
\frac{1}{128} - 16 L^r_8 - 64 L^r_6 + 8 L^r_5 + 32 L^r_4
- \frac{5}{2}L^r_3 - 10 L^r_2 - 2 L^r_1 - \frac{9}{2}L^r_0
\Big) + M_1^4 \pi_{16}^2 \frac{109}{384} \nonumber\\
&+ R_q M_1^2 \left( -\frac{1}{8}H^F(M_6^2,M_1^2,M_1^2,M_8^2) \right)\nonumber \\
&+ R_q M_1^4 \Big(
- r_{F,3}^r + 16 (L^r_5)^2 + 64 L^r_4 L^r_5
- \frac{3}{8} H^{F\prime}(M_8^2,M_1^2,M_1^2,M_8^2)
- \frac{1}{24} H^{F\prime}(M_1^2,M_1^2,M_1^2,M_8^2) \nonumber\\
& - \frac{3}{4} H^{F\prime}_{21}(M_6^2,M_1^2,M_1^2,M_8^2)
- \frac{3}{2} H^{F\prime}_{21}(M_1^2,M_8^2,M_1^2,M_8^2)
- \frac{3}{4} H^{F\prime}_{21}(M_1^2,M_1^2,M_1^2,M_8^2)
\Big) \nonumber\\
&+ R_q M_1^4 \pi_{16} \Big(
\frac{1}{768} + 16 L^r_8 + 32 L^r_6 - 8 L^r_5 - 16 L^r_4
+ 2 L^r_3 + 2 L^r_2 + 4 L^r_1 + 4 L^r_0
\Big) + R_q M_1^4 \pi_{16}^2 \frac{17}{192}\nonumber \\
&+ R_q^2 M_1^4 \Big(
r_{F,4}^r - 8 (L^r_5)^2
- \frac{1}{8} H^{F\prime}(M_6^2,M_1^2,M_1^2,M_8^2)
+ \frac{1}{16} H^{F\prime}(M_8^2,M_1^2,M_1^2,M_8^2)
\nonumber\\&- \frac{1}{24} H^{F\prime}(M_1^2,M_1^2,M_1^2,M_8^2) + \frac{3}{8} H^{F\prime}_{21}(M_6^2,M_1^2,M_1^2,M_8^2)
+ \frac{3}{4} H^{F\prime}_{21}(M_1^2,M_8^2,M_1^2,M_8^2)\nonumber
\\
&+ \frac{3}{8} H^{F\prime}_{21}(M_1^2,M_1^2,M_1^2,M_8^2)
\Big) + R_q^2 M_1^4 \pi_{16} \Big(
\frac{1}{1536} - 8 L^r_8 + 4 L^r_5 - L^r_3 - 5 L^r_2 - 2 L^r_1 - 2 L^r_0
\Big) \nonumber\\
&+ R_q^2 M_1^4 \pi_{16}^2 \frac{31}{384} + \overline{A}(M_6^2) M_1^2 (8 L^r_4 - 4 L^r_2 - 16 L^r_1)
+ \overline{A}(M_6^2) R_q M_1^2 (8 L^r_4 - 4 L^r_2 - 16 L^r_1) \nonumber\\
&+ (\overline{A}(M_6^2))^2 \frac{1}{16} + \overline{A}(M_8^2) M_1^2 \Big(
8 L^r_8 + 32 L^r_6 - 2 L^r_5 - 16 L^r_4
- 10 L^r_3 - 14 L^r_2 - 20 L^r_1 - 4 L^r_0
\Big)\nonumber \\
&+ \overline{A}(M_8^2) M_1^2 \pi_{16} \left(-\frac{23}{16}\right)+ \overline{A}(M_8^2) R_q M_1^2 \Big(
-8 L^r_8 + 2 L^r_5 - 8 L^r_4
+ 10 L^r_3 + 14 L^r_2 + 20 L^r_1 + 4 L^r_0
\Big) \nonumber \\&+ \overline{A}(M_8^2) R_q M_1^2 \pi_{16} \frac{19}{16} + (\overline{A}(M_8^2))^2 \Big(
-\frac{1}{4} - \frac{1}{8}(-1+R_q)^{-1}
\Big)
+ \overline{A}(M_8^2)\overline{A}(M_1^2)\frac{1}{8} \nonumber\\
&+ \overline{A}(M_1^2) M_1^2 \Big(
8 L^r_8 + 32 L^r_6 - L^r_5 - 4 L^r_4
- 13 L^r_3 - 10 L^r_2 - 40 L^r_1 - 7 L^r_0
\Big) + \overline{A}(M_1^2) M_1^2 \pi_{16} \left(-\frac{7}{8}\right) \nonumber\\
&+ \overline{A}(M_1^2) R_q M_1^2 (-3 L^r_5)
+ \overline{A}(M_1^2) R_q M_1^2 \pi_{16} \frac{1}{8}+ (\overline{A}(M_1^2))^2 R_q \frac{1}{8}
+ (\overline{A}(M_1^2))^2 \frac{13}{16}.
\end{align}
In these formulas, the linear combinations \eqref{rF0}–\eqref{rF2} of the NNLO LECs were used, together with the following additional combinations:
\begin{align}
r_{F,3}^r &= 16 \Big( K_{19}^r + 2 K_{20}^r + K_{23}^r \Big), \\
r_{F,4}^r &= 8 \Big( K_{19}^r + 4 K_{21}^r + K_{23}^r \Big).
\end{align}

\section{NNLO condensates in \soth theory}\label{appSO4Vacuum}
In this appendix, we present the NNLO contribution to condensates  in the \soth theory.

%uBar*u 

\begin{align}
&\frac{F^{4}}{\left< \bar{u} u\right>_{\text{LO} }  } \left< \bar{u} u\right>_{\text{NNLO} }  = M_1^4 \,r_{V,0}^r + M_1^4 \pi_{16} \Big(
-36 L^r_8 - 144 L^r_6 + 18 L^r_5 + 72 L^r_4
\Big) \nonumber\\
&+ R_q\, M_1^4 \frac{1}{2} r_{V,1}^r
+ R_q\, M_1^4 \pi_{16} \Big(
-32 L^r_8 - 64 L^r_6 + 16 L^r_5 + 32 L^r_4
\Big)\nonumber \\
&+ R_q^2 M_1^4 \frac{1}{4} r_{V,1}^r
+ R_q^2 M_1^4 \pi_{16} \Big(
-20 L^r_8 - 16 L^r_7 + 8 L^r_5
\Big) \nonumber\\
&+ \overline{A}(M_6^2) M_1^2 \Big(
48 L^r_8 + 160 L^r_6 - 24 L^r_5 - 80 L^r_4
\Big)
+ \overline{A}(M_6^2) M_1^2 \pi_{16} \left(-\frac{1}{8}\right)\nonumber \\
&+ \overline{A}(M_6^2) R_q M_1^2 \Big(
48 L^r_8 + 32 L^r_6 - 24 L^r_5 - 16 L^r_4
\Big)
+ \overline{A}(M_6^2) R_q M_1^2 \pi_{16} \left(-\frac{1}{8}\right)\nonumber \\
&+ \overline{A}(M_6^2)\overline{A}(M_1^2) R_q \frac{1}{8}
+ \overline{A}(M_6^2)\overline{A}(M_1^2) \frac{5}{8} + \overline{A}(M_8^2) M_1^2 (32 L^r_6 - 16 L^r_4)\nonumber
\\
&+ \overline{A}(M_8^2) M_1^2 \pi_{16} \left(-\frac{1}{8}\right) + \overline{A}(M_8^2) R_q M_1^2 (-32 L^r_6 + 16 L^r_4)
+ \overline{A}(M_8^2) R_q M_1^2 \pi_{16} \frac{1}{8}\nonumber \\
&+ \overline{A}(M_8^2)\overline{A}(M_1^2) R_q \left(-\frac{1}{8}\right)
+ \overline{A}(M_8^2)\overline{A}(M_1^2) \frac{1}{8} + \overline{A}(M_1^2) M_1^2 \Big(
60 L^r_8 + 240 L^r_6 - 30 L^r_5 - 120 L^r_4
\Big) \nonumber\\
&+ \overline{A}(M_1^2) M_1^2 \pi_{16} \left(-\frac{5}{16}\right) + \overline{A}(M_1^2) R_q M_1^2 (8 L^r_8 + 32 L^r_7)
+ \overline{A}(M_1^2) R_q M_1^2 \pi_{16} \left(-\frac{1}{4}\right) \nonumber\\
&+ \overline{A}(M_1^2) R_q^2 M_1^2 (4 L^r_8 + 16 L^r_7)
+ (\overline{A}(M_1^2))^2 \frac{3}{32},
\end{align}

%%%%%%%

%dBar*d 

\begin{align}
&\frac{F^{4}}{\left< \bar{d} d\right>_{\text{LO} }  } \left< \bar{d} d\right>_{\text{NNLO} } =  M_1^4 \,r_{V,0}^r + M_1^4 \pi_{16} \Big(
-36 L^r_8 - 144 L^r_6 + 18 L^r_5 + 72 L^r_4
\Big) \nonumber\\
&+ R_q\, M_1^4 \left(-\frac{1}{2} r_{V,1}^r\right)
+ R_q\, M_1^4 \pi_{16} \Big(
32 L^r_8 + 64 L^r_6 - 16 L^r_5 - 32 L^r_4
\Big) \nonumber\\
&+ R_q^2 M_1^4 \frac{1}{4} r_{V,1}^r
+ R_q^2 M_1^4 \pi_{16} \Big(
-20 L^r_8 - 16 L^r_7 + 8 L^r_5
\Big) + \overline{A}(M_6^2) M_1^2 (32 L^r_6 - 16 L^r_4)
\nonumber\\&+ \overline{A}(M_6^2) M_1^2 \pi_{16} \left(-\frac{1}{8}\right) + \overline{A}(M_6^2) R_q M_1^2 (32 L^r_6 - 16 L^r_4)
+ \overline{A}(M_6^2) R_q M_1^2 \pi_{16} \left(-\frac{1}{8}\right) \nonumber\\
&+ \overline{A}(M_6^2)\overline{A}(M_1^2) R_q \frac{1}{8}
+ \overline{A}(M_6^2)\overline{A}(M_1^2) \frac{1}{8} + \overline{A}(M_8^2) M_1^2 \Big(
48 L^r_8 + 160 L^r_6 - 24 L^r_5 - 80 L^r_4
\Big)
\nonumber\\&+ \overline{A}(M_8^2) M_1^2 \pi_{16} \left(-\frac{1}{8}\right) + \overline{A}(M_8^2) R_q M_1^2 \Big(
-48 L^r_8 - 32 L^r_6 + 24 L^r_5 + 16 L^r_4
\Big)\nonumber\\
&
+ \overline{A}(M_8^2) R_q M_1^2 \pi_{16} \frac{1}{8}+ \overline{A}(M_8^2)\overline{A}(M_1^2) R_q \left(-\frac{1}{8}\right)
+ \overline{A}(M_8^2)\overline{A}(M_1^2) \frac{5}{8} \nonumber\\
&+ \overline{A}(M_1^2) M_1^2 \Big(
60 L^r_8 + 240 L^r_6 - 30 L^r_5 - 120 L^r_4
\Big) + \overline{A}(M_1^2) M_1^2 \pi_{16} \left(-\frac{5}{16}\right) \nonumber\\
&+ \overline{A}(M_1^2) R_q M_1^2 (-8 L^r_8 - 32 L^r_7)
+ \overline{A}(M_1^2) R_q M_1^2 \pi_{16} \frac{1}{4} \nonumber\\
&+ \overline{A}(M_1^2) R_q^2 M_1^2 (4 L^r_8 + 16 L^r_7)
+ (\overline{A}(M_1^2))^2 \frac{3}{32}.
\end{align}

In these formulas, the linear combinations \eqref{rV0} and \eqref{rV1} of the NNLO LECs were used.

\section{Generators of $SU(4)$}\label{appendixSU4gen}

The same set of Hermitian generators for $SU(4)$ is used as in \cite{Kolesova:2025ghl}:
\begin{align}
&T^1 = \begin{pmatrix}
\frac{1}{2} & 0 & 0 & 0 \\
0 & \frac{1}{2} & 0 & 0 \\
0 & 0 & -\frac{1}{2} & 0 \\
0 & 0 & 0 & -\frac{1}{2}
\end{pmatrix},\quad
T^2 = \begin{pmatrix}
0 & \frac{1}{2} & 0 & 0 \\
\frac{1}{2} & 0 & 0 & 0 \\
0 & 0 & 0 & \frac{1}{2} \\
0 & 0 & \frac{1}{2} & 0
\end{pmatrix},\quad
T^3 = \begin{pmatrix}
0 & \frac{1}{2} & 0 & 0 \\
\frac{1}{2} & 0 & 0 & 0 \\
0 & 0 & 0 & -\frac{1}{2} \\
0 & 0 & -\frac{1}{2} & 0
\end{pmatrix},
\nonumber\\[10pt]
&T^4 = \begin{pmatrix}
0 & -\frac{i}{2} & 0 & 0 \\
\frac{i}{2} & 0 & 0 & 0 \\
0 & 0 & 0 & \frac{i}{2} \\
0 & 0 & -\frac{i}{2} & 0
\end{pmatrix},\quad
T^5 = \begin{pmatrix}
0 & -\frac{i}{2} & 0 & 0 \\
\frac{i}{2} & 0 & 0 & 0 \\
0 & 0 & 0 & -\frac{i}{2} \\
0 & 0 & \frac{i}{2} & 0
\end{pmatrix},\quad
T^6 = \begin{pmatrix}
\frac{1}{2} & 0 & 0 & 0 \\
0 & -\frac{1}{2} & 0 & 0 \\
0 & 0 & \frac{1}{2}  & 0 \\
0 & 0 & 0 & -\frac{1}{2}
\end{pmatrix},
\nonumber\\[10pt]
&T^7 = \begin{pmatrix}
\frac{1}{2} & 0 & 0 & 0 \\
0 & -\frac{1}{2} & 0 & 0 \\
0 & 0 & -\frac{1}{2} & 0 \\
0 & 0 & 0 & \frac{1}{2}
\end{pmatrix},\quad
T^8 = \begin{pmatrix}
0 & 0 & 0 & \frac{1}{2} \\
0 & 0 & \frac{1}{2} & 0 \\
0 & \frac{1}{2} & 0 & 0 \\
\frac{1}{2} & 0 & 0 & 0
\end{pmatrix},\quad
T^9 = \begin{pmatrix}
0 & 0 & 0 & \frac{i}{2} \\
0 & 0 & \frac{i}{2} & 0 \\
0 & -\frac{i}{2} & 0 & 0 \\
-\frac{i}{2} & 0 & 0 & 0
\end{pmatrix}, \nonumber\\[10pt]
&T^{10} = \begin{pmatrix}
0 & 0 & \frac{1}{\sqrt{2}} & 0 \\
0 & 0 & 0 & 0 \\
\frac{1}{\sqrt{2}} & 0 & 0 & 0 \\
0 & 0 & 0 & 0
\end{pmatrix},\quad
T^{11} = \begin{pmatrix}
0 & 0 & \frac{i}{\sqrt{2}} & 0 \\
0 & 0 & 0 & 0\\
-\frac{i}{\sqrt{2}} & 0 & 0 & 0 \\
0 & 0& 0 & 0
\end{pmatrix},\quad
T^{12} = \begin{pmatrix}
0 & 0 & 0 & 0 \\
0 & 0 & 0 & \frac{1}{\sqrt{2}} \\
0 & 0 & 0 & 0 \\
0 & \frac{1}{\sqrt{2}} & 0 & 0
\end{pmatrix},\nonumber\\[10pt]
&T^{13} = \begin{pmatrix}
0 & 0 & 0 & 0 \\
0 & 0 & 0 & \frac{i}{\sqrt{2}} \\
0 & 0 & 0 & 0 \\
0 & -\frac{i}{\sqrt{2}} & 0 & 0
\end{pmatrix},\quad
T^{14} = \begin{pmatrix}
0 & 0& 0 & \frac{1}{2}  \\
0 & 0 & -\frac{1}{2}  & 0 \\
0& -\frac{1}{2}  & 0 & 0 \\
\frac{1}{2} & 0 & 0 & 0
\end{pmatrix},\quad
T^{15} = \begin{pmatrix}
0 & 0& 0 & \frac{i}{2}  \\
0 & 0 & -\frac{i}{2}  & 0 \\ 0& \frac{i}{2}  & 0 & 0 \\ -\frac{i}{2} & 0 & 0 & 0 
\end{pmatrix}.
\end{align}
The generators are normalized as $\left< T^{a}T^{b}\right>  =\delta^{ab}$. For the \spth theory the broken generators corresponding to the pion fields read
\begin{equation}
 X^{1}\equiv T^{2},\quad X^{2}\equiv T^{4},\quad X^{3}\equiv T^{6},\quad X^{4}\equiv T^{14},\quad X^{5}\equiv T^{15},
\end{equation}
while for the \soth theory they are identified as
\begin{align}
X^{1}&\equiv T^{2},\quad X^{2}\equiv T^{4},\quad X^{3}\equiv T^{6},\quad X^{4}\equiv T^{8},\quad X^{5}\equiv T^{9},\quad \nonumber \\ X^{6}&\equiv T^{10},\quad X^{7}\equiv T^{11},\quad X^{8}\equiv T^{12},\quad X^{9}\equiv T^{13}. 
\end{align}
%%%%%%%%%% 

%%%%%%%%%% Bibliography

\addcontentsline{toc}{section}{References}
\bibliographystyle{JHEP}
\bibliography{biblio}

@article{Bijnens:2014gsa,
    author = "Bijnens, Johan",
    title = "{CHIRON: a package for ChPT numerical results at two loops}",
    eprint = "1412.0887",
    archivePrefix = "arXiv",
    primaryClass = "hep-ph",
    reportNumber = "LU-TP-14-41",
    doi = "10.1140/epjc/s10052-014-3249-9",
    journal = "Eur. Phys. J. C",
    volume = "75",
    number = "1",
    pages = "27",
    year = "2015"
}

@article{Bijnens:2006zp,
    author = "Bijnens, Johan",
    title = "{Chiral perturbation theory beyond one loop}",
    eprint = "hep-ph/0604043",
    archivePrefix = "arXiv",
    reportNumber = "LU-TP-06-16",
    doi = "10.1016/j.ppnp.2006.08.002",
    journal = "Prog. Part. Nucl. Phys.",
    volume = "58",
    pages = "521--586",
    year = "2007"
}

@article{Bijnens:1999hw,
    author = "Bijnens, J. and Colangelo, G. and Ecker, G.",
    title = "{Renormalization of chiral perturbation theory to order $p^6$}",
    eprint = "hep-ph/9907333",
    archivePrefix = "arXiv",
    reportNumber = "LU-TP-99-14, ZU-TH-18-99, UWTHPH-1999-42",
    doi = "10.1006/aphy.1999.5982",
    journal = "Annals Phys.",
    volume = "280",
    pages = "100--139",
    year = "2000"
}

@article{Peskin:1980gc,
    author = "Peskin, Michael E.",
    title = "{The Alignment of the Vacuum in Theories of Technicolor}",
    reportNumber = "SACLAY-DPH-T-80-46",
    doi = "10.1016/0550-3213(80)90051-6",
    journal = "Nucl. Phys. B",
    volume = "175",
    pages = "197--233",
    year = "1980"
}

@article{Preskill:1980mz,
    author = "Preskill, John",
    title = "{Subgroup Alignment in Hypercolor Theories}",
    reportNumber = "HUTP-80-A033",
    doi = "10.1016/0550-3213(81)90265-0",
    journal = "Nucl. Phys. B",
    volume = "177",
    pages = "21--59",
    year = "1981"
}

@article{Dimopoulos:1979sp,
    author = "Dimopoulos, Savas",
    editor = "Zichichi, A.",
    title = "{Technicolored Signatures}",
    reportNumber = "PRINT-79-0962 (STANFORD)",
    doi = "10.1016/0550-3213(80)90277-1",
    journal = "Nucl. Phys. B",
    volume = "168",
    pages = "69--92",
    year = "1980"
}

@misc{SIMP_VM,
      title={ The path to realistic SIMP dark matter (In preparation)}, 
      author={Torsten Bringmann and Helena Kole\v{s}ov\'{a} and Daniil Krichevskiy and Halvor Melkild and Joachim Pomper},
      year={2026}, 
}

@article{Cline:2021itd,
    author = "Cline, James M.",
    title = "{Dark atoms and composite dark matter}",
    eprint = "2108.10314",
    archivePrefix = "arXiv",
    primaryClass = "hep-ph",
    doi = "10.21468/SciPostPhysLectNotes.52",
    journal = "SciPost Phys. Lect. Notes",
    volume = "52",
    pages = "1",
    year = "2022"
}

@article{Peskin:2025lsg,
    author = "Peskin, Michael E.",
    title = "{What is the Hierarchy Problem?}",
    eprint = "2505.00694",
    archivePrefix = "arXiv",
    primaryClass = "hep-ph",
    doi = "10.1016/j.nuclphysb.2025.116971",
    journal = "Nucl. Phys. B",
    volume = "1018",
    pages = "116971",
    year = "2025"
}

@article{Balazs:2024uyj,
    author = "Balazs, Csaba and Bringmann, Torsten and Kahlhoefer, Felix and White, Martin",
    title = "{A Primer on Dark Matter}",
    eprint = "2411.05062",
    archivePrefix = "arXiv",
    primaryClass = "astro-ph.CO",
    doi = "10.1016/B978-0-443-21439-4.00070-5",
    journal = "Astrophysics",
    volume = "5",
    pages = "17",
    year = "2026"
}

@book{Peskin:1995ev,
    author = "Peskin, Michael E. and Schroeder, Daniel V.",
    title = "{An Introduction to quantum field theory}",
    doi = "10.1201/9780429503559",
    isbn = "978-0-201-50397-5, 978-0-429-50355-9, 978-0-429-49417-8",
    publisher = "Addison-Wesley",
    address = "Reading, USA",
    year = "1995"
}

@article{Kolesova:2025ghl,
    author = "Kole{\v{s}}ov{\'a}, Helena and Krichevskiy, Daniil and Kulkarni, Suchita",
    title = "{NLO observables for QCD-like theories and application to pion dark matter}",
    eprint = "2509.07102",
    archivePrefix = "arXiv",
    primaryClass = "hep-ph",
    doi = "10.1007/JHEP05(2026)042",
    journal = "JHEP",
    volume = "05",
    pages = "042",
    year = "2026"
}

@article{Gasser:1984gg,
    author = "Gasser, J. and Leutwyler, H.",
    title = "{Chiral Perturbation Theory: Expansions in the Mass of the Strange Quark}",
    reportNumber = "CERN-TH-3798",
    doi = "10.1016/0550-3213(85)90492-4",
    journal = "Nucl. Phys. B",
    volume = "250",
    pages = "465--516",
    year = "1985"
}

@article{Bijnens:2015xba,
    author = {Bijnens, Johan and R{\"o}ssler, Thomas},
    title = "{Finite Volume and Partially Quenched QCD-like Effective Field Theories}",
    eprint = "1509.04082",
    archivePrefix = "arXiv",
    primaryClass = "hep-lat",
    reportNumber = "LU-TP-15-34",
    doi = "10.1007/JHEP11(2015)017",
    journal = "JHEP",
    volume = "11",
    pages = "017",
    year = "2015"
}

@article{Weinberg:1978kz,
    author = "Weinberg, Steven",
    editor = "Deser, S.",
    title = "{Phenomenological Lagrangians}",
    reportNumber = "HUTP-78-A051A",
    doi = "10.1016/0378-4371(79)90223-1",
    journal = "Physica A",
    volume = "96",
    number = "1-2",
    pages = "327--340",
    year = "1979"
}

@article{Bijnens:1999sh,
    author = "Bijnens, Johan and Colangelo, Gilberto and Ecker, Gerhard",
    title = "{The Mesonic chiral Lagrangian of order p$^{6}$}",
    eprint = "hep-ph/9902437",
    archivePrefix = "arXiv",
    reportNumber = "LU-TP-99-02, UWTHPH-1999-02, ZU-TH-9-99",
    doi = "10.1088/1126-6708/1999/02/020",
    journal = "JHEP",
    volume = "02",
    pages = "020",
    year = "1999"
}

@article{Davighi:2024zip,
    author = "Davighi, Joe and Greljo, Admir and Selimovic, Nudzeim",
    title = "{Topological Portal to the Dark Sector}",
    eprint = "2401.09528",
    archivePrefix = "arXiv",
    primaryClass = "hep-ph",
    reportNumber = "CERN-TH-2024-010, CERN-TH-2024-010",
    doi = "10.1103/PhysRevLett.134.111804",
    journal = "Phys. Rev. Lett.",
    volume = "134",
    number = "11",
    pages = "111804",
    year = "2025"
}

@article{Davighi:2025awm,
    author = "Davighi, Joe and Moldovsky, Serah and Murayama, Hitoshi and Scherb, Christiane and Selimovic, Nudzeim",
    title = "{Topological Freeze-out by Semi-Annihilation}",
    eprint = "2506.05468",
    archivePrefix = "arXiv",
    primaryClass = "hep-ph",
    month = "6",
    year = "2025"
}

@article{Alfano:2025non,
    author = "Alfano, Anja and Evans, Nick and Kulkarni, Suchita and Porod, Werner",
    title = "{Surveying the theory space of pion dark matter}",
    eprint = "2509.04892",
    archivePrefix = "arXiv",
    primaryClass = "hep-ph",
    month = "9",
    year = "2025"
}

@article{Carmona:2015haa,
    author = "Carmona, Adrian and Chala, Mikael",
    title = "{Composite Dark Sectors}",
    eprint = "1504.00332",
    archivePrefix = "arXiv",
    primaryClass = "hep-ph",
    reportNumber = "DESY-15-047",
    doi = "10.1007/JHEP06(2015)105",
    journal = "JHEP",
    volume = "06",
    pages = "105",
    year = "2015"
}

@article{Frigerio:2012uc,
    author = "Frigerio, Michele and Pomarol, Alex and Riva, Francesco and Urbano, Alfredo",
    title = "{Composite Scalar Dark Matter}",
    eprint = "1204.2808",
    archivePrefix = "arXiv",
    primaryClass = "hep-ph",
    doi = "10.1007/JHEP07(2012)015",
    journal = "JHEP",
    volume = "07",
    pages = "015",
    year = "2012"
}

@article{Ryttov:2008xe,
    author = "Ryttov, Thomas A. and Sannino, Francesco",
    title = "{Ultra Minimal Technicolor and its Dark Matter TIMP}",
    eprint = "0809.0713",
    archivePrefix = "arXiv",
    primaryClass = "hep-ph",
    reportNumber = "CERN-PH-TH-2008-188",
    doi = "10.1103/PhysRevD.78.115010",
    journal = "Phys. Rev. D",
    volume = "78",
    pages = "115010",
    year = "2008"
}

@article{Contino:2020god,
    author = "Contino, Roberto and Podo, Alessandro and Revello, Filippo",
    title = "{Composite Dark Matter from Strongly-Interacting Chiral Dynamics}",
    eprint = "2008.10607",
    archivePrefix = "arXiv",
    primaryClass = "hep-ph",
    doi = "10.1007/JHEP02(2021)091",
    journal = "JHEP",
    volume = "02",
    pages = "091",
    year = "2021"
}

@article{Bernreuther:2019pfb,
    author = {Bernreuther, Elias and Kahlhoefer, Felix and Kr{\"a}mer, Michael and Tunney, Patrick},
    title = "{Strongly interacting dark sectors in the early Universe and at the LHC through a simplified portal}",
    eprint = "1907.04346",
    archivePrefix = "arXiv",
    primaryClass = "hep-ph",
    reportNumber = "TTK-19-25, P3H-19-019",
    doi = "10.1007/JHEP01(2020)162",
    journal = "JHEP",
    volume = "01",
    pages = "162",
    year = "2020"
}

@article{Kopp:2016yji,
    author = "Kopp, Joachim and Liu, Jia and Slatyer, Tracy R. and Wang, Xiao-Ping and Xue, Wei",
    title = "{Impeded Dark Matter}",
    eprint = "1609.02147",
    archivePrefix = "arXiv",
    primaryClass = "hep-ph",
    reportNumber = "MIT-CTP-4832, MITP-16-092",
    doi = "10.1007/JHEP12(2016)033",
    journal = "JHEP",
    volume = "12",
    pages = "033",
    year = "2016"
}

@article{Beauchesne:2018myj,
    author = "Beauchesne, Hugues and Bertuzzo, Enrico and Grilli Di Cortona, Giovanni",
    title = "{Dark matter in Hidden Valley models with stable and unstable light dark mesons}",
    eprint = "1809.10152",
    archivePrefix = "arXiv",
    primaryClass = "hep-ph",
    doi = "10.1007/JHEP04(2019)118",
    journal = "JHEP",
    volume = "04",
    pages = "118",
    year = "2019"
}

@article{Essig:2009nc,
    author = "Essig, Rouven and Schuster, Philip and Toro, Natalia",
    title = "{Probing Dark Forces and Light Hidden Sectors at Low-Energy $e^{+}e^{-}$ Colliders}",
    eprint = "0903.3941",
    archivePrefix = "arXiv",
    primaryClass = "hep-ph",
    reportNumber = "SLAC-PUB-13561, SU-ITP-09-12",
    doi = "10.1103/PhysRevD.80.015003",
    journal = "Phys. Rev. D",
    volume = "80",
    pages = "015003",
    year = "2009"
}

@article{Cline:2013zca,
    author = "Cline, James M. and Liu, Zuowei and Moore, Guy D. and Xue, Wei",
    title = "{Composite strongly interacting dark matter}",
    eprint = "1312.3325",
    archivePrefix = "arXiv",
    primaryClass = "hep-ph",
    doi = "10.1103/PhysRevD.90.015023",
    journal = "Phys. Rev. D",
    volume = "90",
    number = "1",
    pages = "015023",
    year = "2014"
}

@article{Bhattacharya:2013kma,
    author = "Bhattacharya, Subhaditya and Meli{\'c}, Bla{\v{z}}enka and Wudka, Jos{\'e}",
    title = "{Pionic Dark Matter}",
    eprint = "1307.2647",
    archivePrefix = "arXiv",
    primaryClass = "hep-ph",
    doi = "10.1007/JHEP02(2014)115",
    journal = "JHEP",
    volume = "02",
    pages = "115",
    year = "2014"
}

@article{Bai:2010qg,
    author = "Bai, Yang and Hill, Richard J.",
    title = "{Weakly Interacting Stable Pions}",
    eprint = "1005.0008",
    archivePrefix = "arXiv",
    primaryClass = "hep-ph",
    reportNumber = "FERMILAB-PUB-10-001-T, EFI-PREPRINT-10-9",
    doi = "10.1103/PhysRevD.82.111701",
    journal = "Phys. Rev. D",
    volume = "82",
    pages = "111701",
    year = "2010"
}

@article{Tulin:2017ara,
    author = "Tulin, Sean and Yu, Hai-Bo",
    title = "{Dark Matter Self-interactions and Small Scale Structure}",
    eprint = "1705.02358",
    archivePrefix = "arXiv",
    primaryClass = "hep-ph",
    doi = "10.1016/j.physrep.2017.11.004",
    journal = "Phys. Rept.",
    volume = "730",
    pages = "1--57",
    year = "2018"
}

@article{Nagata:2021ugx,
    author = "Nagata, Keitaro",
    title = "{Finite-density lattice QCD and sign problem: Current status and open problems}",
    eprint = "2108.12423",
    archivePrefix = "arXiv",
    primaryClass = "hep-lat",
    doi = "10.1016/j.ppnp.2022.103991",
    journal = "Prog. Part. Nucl. Phys.",
    volume = "127",
    pages = "103991",
    year = "2022"
}

@article{Beauchesne:2019ato,
    author = "Beauchesne, Hugues and Grilli di Cortona, Giovanni",
    title = "{Classification of dark pion multiplets as dark matter candidates and collider phenomenology}",
    eprint = "1910.10724",
    archivePrefix = "arXiv",
    primaryClass = "hep-ph",
    doi = "10.1007/JHEP02(2020)196",
    journal = "JHEP",
    volume = "02",
    pages = "196",
    year = "2020"
}

@article{Berlin:2018tvf,
    author = "Berlin, Asher and Blinov, Nikita and Gori, Stefania and Schuster, Philip and Toro, Natalia",
    title = "{Cosmology and Accelerator Tests of Strongly Interacting Dark Matter}",
    eprint = "1801.05805",
    archivePrefix = "arXiv",
    primaryClass = "hep-ph",
    reportNumber = "SLAC-PUB-17135",
    doi = "10.1103/PhysRevD.97.055033",
    journal = "Phys. Rev. D",
    volume = "97",
    number = "5",
    pages = "055033",
    year = "2018"
}

@article{Choi:2018iit,
    author = "Choi, Soo-Min and Lee, Hyun Min and Ko, Pyungwon and Natale, Alexander",
    title = "{Resolving phenomenological problems with strongly-interacting-massive-particle models with dark vector resonances}",
    eprint = "1801.07726",
    archivePrefix = "arXiv",
    primaryClass = "hep-ph",
    doi = "10.1103/PhysRevD.98.015034",
    journal = "Phys. Rev. D",
    volume = "98",
    number = "1",
    pages = "015034",
    year = "2018"
}

@article{Randall:2008ppe,
    author = "Randall, Scott W. and Markevitch, Maxim and Clowe, Douglas and Gonzalez, Anthony H. and Bradac, Marusa",
    title = "{Constraints on the Self-Interaction Cross-Section of Dark Matter from Numerical Simulations of the Merging Galaxy Cluster 1E 0657-56}",
    eprint = "0704.0261",
    archivePrefix = "arXiv",
    primaryClass = "astro-ph",
    doi = "10.1086/587859",
    journal = "Astrophys. J.",
    volume = "679",
    pages = "1173--1180",
    year = "2008"
}

@article{Robertson:2016xjh,
    author = "Robertson, Andrew and Massey, Richard and Eke, Vincent",
    title = "{What does the Bullet Cluster tell us about self-interacting dark matter?}",
    eprint = "1605.04307",
    archivePrefix = "arXiv",
    primaryClass = "astro-ph.CO",
    doi = "10.1093/mnras/stw2670",
    journal = "Mon. Not. Roy. Astron. Soc.",
    volume = "465",
    number = "1",
    pages = "569--587",
    year = "2017"
}

@article{Wittman:2017gxn,
    author = "Wittman, David and Golovich, Nathan and Dawson, William A.",
    title = "{The Mismeasure of Mergers: Revised Limits on Self-interacting Dark Matter in Merging Galaxy Clusters}",
    eprint = "1701.05877",
    archivePrefix = "arXiv",
    primaryClass = "astro-ph.CO",
    doi = "10.3847/1538-4357/aaee77",
    journal = "Astrophys. J.",
    volume = "869",
    number = "2",
    pages = "104",
    year = "2018"
}

@article{Buckley:2012ky,
    author = "Buckley, Matthew R. and Neil, Ethan T.",
    title = "{Thermal Dark Matter from a Confining Sector}",
    eprint = "1209.6054",
    archivePrefix = "arXiv",
    primaryClass = "hep-ph",
    reportNumber = "FERMILAB-PUB-12-533-A-T",
    doi = "10.1103/PhysRevD.87.043510",
    journal = "Phys. Rev. D",
    volume = "87",
    number = "4",
    pages = "043510",
    year = "2013"
}

@article{Foreman-Mackey:2012any,
    author = "Foreman-Mackey, Daniel and Hogg, David W. and Lang, Dustin and Goodman, Jonathan",
    title = "{emcee: The MCMC Hammer}",
    eprint = "1202.3665",
    archivePrefix = "arXiv",
    primaryClass = "astro-ph.IM",
    doi = "10.1086/670067",
    journal = "Publ. Astron. Soc. Pac.",
    volume = "125",
    pages = "306--312",
    year = "2013"
}

@misc{dengler_2024_12920978,
  author       = {Dengler, Yannick and
                  Maas, Axel and
                  Zierler, Fabian},
  title        = "{Scattering of dark pions in Sp(4) gauge theory -
                   Data release
                  }",
  month        = aug,
  year         = 2024,
  publisher    = {Zenodo},
  doi          = {10.5281/zenodo.12920978},
  url          = {https://doi.org/10.5281/zenodo.12920978},
}

@article{Luscher:1986pf,
    author = "Luscher, M.",
    title = "{Volume Dependence of the Energy Spectrum in Massive Quantum Field Theories. 2. Scattering States}",
    reportNumber = "DESY-86-034",
    doi = "10.1007/BF01211097",
    journal = "Commun. Math. Phys.",
    volume = "105",
    pages = "153--188",
    year = "1986"
}

@article{Kamada:2022zwb,
    author = "Kamada, Ayuki and Kobayashi, Shin and Kuwahara, Takumi",
    title = "{Perturbative unitarity of strongly interacting massive particle models}",
    eprint = "2210.01393",
    archivePrefix = "arXiv",
    primaryClass = "hep-ph",
    doi = "10.1007/JHEP02(2023)217",
    journal = "JHEP",
    volume = "02",
    pages = "217",
    year = "2023"
}

@Article{Coleman:1969sm,
  author  = {Coleman, Sidney R. and Wess, J. and Zumino, Bruno},
  journal = {Phys. Rev.},
  title   = {{Structure of phenomenological Lagrangians. 1.}},
  year    = {1969},
  pages   = {2239--2247},
  volume  = {177},
  doi     = {10.1103/PhysRev.177.2239},
}

@Article{Callan:1969sn,
  author  = {Callan, Jr., Curtis G. and Coleman, Sidney R. and Wess, J. and Zumino, Bruno},
  journal = {Phys. Rev.},
  title   = {{Structure of phenomenological Lagrangians. 2.}},
  year    = {1969},
  pages   = {2247--2250},
  volume  = {177},
  doi     = {10.1103/PhysRev.177.2247},
}

@article{Brauner:2024juy,
    author = "Brauner, Tom\'a\v{s}",
    title = "{Effective Field Theory for Spontaneously Broken Symmetry}",
    eprint = "2404.14518",
    archivePrefix = "arXiv",
    primaryClass = "hep-th",
    doi = "10.1007/978-3-031-48378-3",
    journal = "Lect. Notes Phys.",
    volume = "1023",
    pages = "pp.",
    year = "2024"
}

@article{Hietanen:2014xca,
    author = "Hietanen, Ari and Lewis, Randy and Pica, Claudio and Sannino, Francesco",
    title = "{Fundamental Composite Higgs Dynamics on the Lattice: SU(2) with Two Flavors}",
    eprint = "1404.2794",
    archivePrefix = "arXiv",
    primaryClass = "hep-lat",
    reportNumber = "CP3-ORIGINS-2014-012, DIAS-2014-12",
    doi = "10.1007/JHEP07(2014)116",
    journal = "JHEP",
    volume = "07",
    pages = "116",
    year = "2014"
}

@article{Arthur:2016dir,
    author = "Arthur, Rudy and Drach, Vincent and Hansen, Martin and Hietanen, Ari and Pica, Claudio and Sannino, Francesco",
    title = "{SU(2) gauge theory with two fundamental flavors: A minimal template for model building}",
    eprint = "1602.06559",
    archivePrefix = "arXiv",
    primaryClass = "hep-lat",
    reportNumber = "CP3-ORIGINS-2016-006, DIAS-2016-6, CERN-TH-2016-037",
    doi = "10.1103/PhysRevD.94.094507",
    journal = "Phys. Rev. D",
    volume = "94",
    number = "9",
    pages = "094507",
    year = "2016"
}

@article{Bijnens:2014lea,
    author = "Bijnens, Johan and Ecker, Gerhard",
    title = "{Mesonic low-energy constants}",
    eprint = "1405.6488",
    archivePrefix = "arXiv",
    primaryClass = "hep-ph",
    reportNumber = "LU-TP-14-16, UWTHPH-2014-10",
    doi = "10.1146/annurev-nucl-102313-025528",
    journal = "Ann. Rev. Nucl. Part. Sci.",
    volume = "64",
    pages = "149--174",
    year = "2014"
}

@Article{Hochberg:2014dra,
  author        = {Hochberg, Yonit and Kuflik, Eric and Volansky, Tomer and Wacker, Jay G.},
  journal       = {Phys. Rev. Lett.},
  title         = {{Mechanism for Thermal Relic Dark Matter of Strongly Interacting Massive Particles}},
  year          = {2014},
  pages         = {171301},
  volume        = {113},
  archiveprefix = {arXiv},
  doi           = {10.1103/PhysRevLett.113.171301},
  eprint        = {1402.5143},
  primaryclass  = {hep-ph},
}

@Article{Hochberg:2014kqa,
  author        = {Hochberg, Yonit and Kuflik, Eric and Murayama, Hitoshi and Volansky, Tomer and Wacker, Jay G.},
  journal       = {Phys. Rev. Lett.},
  title         = {{Model for Thermal Relic Dark Matter of Strongly Interacting Massive Particles}},
  year          = {2015},
  number        = {2},
  pages         = {021301},
  volume        = {115},
  archiveprefix = {arXiv},
  doi           = {10.1103/PhysRevLett.115.021301},
  eprint        = {1411.3727},
  primaryclass  = {hep-ph},
}

@Book{Scherer:2012xha,
  author = {Scherer, Stefan and Schindler, Matthias R.},
  title  = {{A Primer for Chiral Perturbation Theory}},
  year   = {2012},
  isbn   = {978-3-642-19253-1},
  volume = {830},
  doi    = {10.1007/978-3-642-19254-8},
  publisher = {Springer Berlin, Heidelberg},
}

@Article{Kulkarni:2022bvh,
  author        = {Kulkarni, Suchita and Maas, Axel and Mee, Se\'an and Nikolic, Marco and Pradler, Josef and Zierler, Fabian},
  journal       = {SciPost Phys.},
  title         = {{Low-energy effective description of dark $Sp(4)$ theories}},
  year          = {2023},
  number        = {3},
  pages         = {044},
  volume        = {14},
  archiveprefix = {arXiv},
  doi           = {10.21468/SciPostPhys.14.3.044},
  eprint        = {2202.05191},
  primaryclass  = {hep-ph},
}

@Article{Bijnens:2009qm,
  author        = {Bijnens, Johan and Lu, Jie},
  journal       = {JHEP},
  title         = {{Technicolor and other QCD-like theories at next-to-next-to-leading order}},
  year          = {2009},
  pages         = {116},
  volume        = {11},
  archiveprefix = {arXiv},
  doi           = {10.1088/1126-6708/2009/11/116},
  eprint        = {0910.5424},
  primaryclass  = {hep-ph},
  reportnumber  = {LU-TP-09-27},
}

@Article{Bijnens:2011fm,
  author        = {Bijnens, Johan and Lu, Jie},
  journal       = {JHEP},
  title         = {{Meson-meson Scattering in QCD-like Theories}},
  year          = {2011},
  pages         = {028},
  volume        = {03},
  archiveprefix = {arXiv},
  doi           = {10.1007/JHEP03(2011)028},
  eprint        = {1102.0172},
  primaryclass  = {hep-ph},
  reportnumber  = {LU-TP-11-07},
}

@Article{Pomper:2024otb,
  author        = {Pomper, Joachim and Kulkarni, Suchita},
  title         = {{Low energy effective theories of composite dark matter with real representations}},
  year          = {2026},
  month         = {2},
  archiveprefix = {arXiv},
  eprint        = {2402.04176},
  primaryclass  = {hep-ph},
  journal={SciPost Phys. Core},
  volume={9},
  pages={007},
  publisher={SciPost},
  doi={10.21468/SciPostPhysCore.9.1.007},
  url={https://scipost.org/10.21468/SciPostPhysCore.9.1.007}
}

@Article{Hansen:2015yaa,
  author        = {Hansen, Martin and Lang\ae{}ble, Kasper and Sannino, Francesco},
  journal       = {Phys. Rev. D},
  title         = {{SIMP model at NNLO in chiral perturbation theory}},
  year          = {2015},
  number        = {7},
  pages         = {075036},
  volume        = {92},
  archiveprefix = {arXiv},
  doi           = {10.1103/PhysRevD.92.075036},
  eprint        = {1507.01590},
  primaryclass  = {hep-ph},
  reportnumber  = {CP3-ORIGINS-2015-025, DIAS-2015-25},
}

@Article{Amoros:1999dp,
  author        = {Amoros, Gabriel and Bijnens, Johan and Talavera, P.},
  journal       = {Nucl. Phys. B},
  title         = {{Two point functions at two loops in three flavor chiral perturbation theory}},
  year          = {2000},
  pages         = {319--363},
  volume        = {568},
  archiveprefix = {arXiv},
  doi           = {10.1016/S0550-3213(99)00674-4},
  eprint        = {hep-ph/9907264},
  reportnumber  = {LU-TP-99-15},
}

@Article{Dengler:2024maq,
  author        = {Dengler, Yannick and Maas, Axel and Zierler, Fabian},
  journal       = {Phys. Rev. D},
  title         = {{Scattering of dark pions in Sp(4) gauge theory}},
  year          = {2024},
  number        = {5},
  pages         = {054513},
  volume        = {110},
  archiveprefix = {arXiv},
  doi           = {10.1103/PhysRevD.110.054513},
  eprint        = {2405.06506},
  primaryclass  = {hep-lat},
}

@Article{Gasser:1983yg,
  author       = {Gasser, J. and Leutwyler, H.},
  journal      = {Annals Phys.},
  title        = {{Chiral Perturbation Theory to One Loop}},
  year         = {1984},
  pages        = {142},
  volume       = {158},
  doi          = {10.1016/0003-4916(84)90242-2},
  reportnumber = {CERN-TH-3689},
}

@Article{Kogut:2000ek,
  author        = {Kogut, J. B. and Stephanov, Misha A. and Toublan, D. and Verbaarschot, J. J. M. and Zhitnitsky, A.},
  journal       = {Nucl. Phys. B},
  title         = {{QCD - like theories at finite baryon density}},
  year          = {2000},
  pages         = {477--513},
  volume        = {582},
  archiveprefix = {arXiv},
  doi           = {10.1016/S0550-3213(00)00242-X},
  eprint        = {hep-ph/0001171},
  reportnumber  = {SUNY-NTG-00-11},
}

@Article{Bernreuther:2023kcg,
  author        = {Bernreuther, Elias and Hemme, Nicoline and Kahlhoefer, Felix and Kulkarni, Suchita},
  journal       = {Phys. Rev. D},
  title         = {{Dark matter relic density in strongly interacting dark sectors with light vector mesons}},
  year          = {2024},
  number        = {3},
  pages         = {035009},
  volume        = {110},
  archiveprefix = {arXiv},
  doi           = {10.1103/PhysRevD.110.035009},
  eprint        = {2311.17157},
  primaryclass  = {hep-ph},
  reportnumber  = {FERMILAB-PUB-23-744-T, TTP23-057, P3H-23-096},
}

@article{Bennett:2019jzz,
    author = "Bennett, Ed and Hong, Deog Ki and Lee, Jong-Wan and Lin, C. -J. David and Lucini, Biagio and Piai, Maurizio and Vadacchino, Davide",
    title = "{Sp(4) gauge theories on the lattice: $N_f=2$ dynamical fundamental fermions}",
    eprint = "1909.12662",
    archivePrefix = "arXiv",
    primaryClass = "hep-lat",
    reportNumber = "PNUTP-19/A01",
    doi = "10.1007/JHEP12(2019)053",
    journal = "JHEP",
    volume = "12",
    pages = "053",
    year = "2019"
}

@article{Splittorff:2001fy,
    author = "Splittorff, K. and Toublan, D. and Verbaarschot, J. J. M.",
    title = "{Diquark condensate in QCD with two colors at next-to-leading order}",
    eprint = "hep-ph/0108040",
    archivePrefix = "arXiv",
    reportNumber = "SUNY-NTG-01-40",
    doi = "10.1016/S0550-3213(01)00536-3",
    journal = "Nucl. Phys. B",
    volume = "620",
    pages = "290--314",
    year = "2002"
}

@article{Cacciapaglia:2020kgq,
    author = "Cacciapaglia, Giacomo and Pica, Claudio and Sannino, Francesco",
    title = "{Fundamental Composite Dynamics: A Review}",
    eprint = "2002.04914",
    archivePrefix = "arXiv",
    primaryClass = "hep-ph",
    doi = "10.1016/j.physrep.2020.07.002",
    journal = "Phys. Rept.",
    volume = "877",
    pages = "1--70",
    year = "2020"
}

@article{Hill:2002ap,
    author = "Hill, Christopher T. and Simmons, Elizabeth H.",
    title = "{Strong Dynamics and Electroweak Symmetry Breaking}",
    eprint = "hep-ph/0203079",
    archivePrefix = "arXiv",
    reportNumber = "FERMILAB-PUB-02-045-T, BUHEP-01-09",
    doi = "10.1016/S0370-1573(03)00140-6",
    journal = "Phys. Rept.",
    volume = "381",
    pages = "235--402",
    year = "2003",
    note = "[Erratum: Phys.Rept. 390, 553--554 (2004)]"
}

@article{Smilga:1994tb,
    author = "Smilga, Andrei V. and Verbaarschot, J. J. M.",
    title = "{Spectral sum rules and finite volume partition function in gauge theories with real and pseudoreal fermions}",
    eprint = "hep-th/9404031",
    archivePrefix = "arXiv",
    reportNumber = "SUNY-NTG-94-18, TPI-MINN-94-10-T",
    doi = "10.1103/PhysRevD.51.829",
    journal = "Phys. Rev. D",
    volume = "51",
    pages = "829--837",
    year = "1995"
}
\end{document}